\documentclass[a4paper,11pt]{article}
\pdfoutput=1 

\usepackage{jheppub}     

\usepackage[T1]{fontenc} 
\usepackage{graphicx}    
\usepackage{subfigure}   
\usepackage{xcolor}

\def\beq{\begin{equation}}
\def\eeq{\end{equation}}
\def\bea{\begin{eqnarray}}
\def\eea{\end{eqnarray}}
\def\beqn{\begin{eqnarray}} 
\def\eeqn{\end{eqnarray}}
\def\nn{\nonumber}
\def\as{\alpha_{\rm S}}
\def\aem{\alpha}

\pdfoutput=1

\title{\boldmath Reconstructing partonic kinematics at colliders with Machine Learning}

\preprint{DESY 21-211}

\author[a]{D. F. Renter\'ia-Estrada,}
\author[a]{R. J. Hern\'andez-Pinto,}
\author[b,c]{G. F. R. Sborlini}
\author[d]{and P. Zurita}

\affiliation[a]{Facultad de Ciencias F\'isico-Matem\'aticas, Universidad Aut\'onoma de Sinaloa, Ciudad Universitaria, CP 80000 Culiac\'an, Mexico}
\affiliation[b]{Instituto de F\'{\i}sica Corpuscular, Universitat de Val\`{e}ncia -- Consejo Superior de Investigaciones Cient\'{\i}ficas, Parc Cient\'{\i}fic, E-46980 Paterna, Valencia, Spain}
\affiliation[c]{Deutsches Elektronen-Synchrotron DESY, Platanenallee 6, 15738 Zeuthen, Germany}
\affiliation[d]{Institut f\"ur Theoretische Physik, Universit\"at Regensburg, 93040 Regensburg, Germany, Universit\"at Regensburg, Germany}

\emailAdd{davidrenteria.fcfm@uas.edu.mx}
\emailAdd{roger@uas.edu.mx}
\emailAdd{german.sborlini@desy.de}
\emailAdd{maria.zurita@ur.de}

\abstract{In the context of high-energy physics, a reliable description of the parton-level kinematics plays a crucial role for understanding the internal structure of hadrons and improving the precision of the calculations. Here, we study the production of one hadron and a direct photon, including up to Next-to-Leading Order Quantum Chromodynamics and Leading-Order Quantum Electrodynamics corrections. Using a code based on Monte-Carlo integration, we simulate the collisions and analyze the events to determine the correlations among measurable and partonic quantities. Then, we use these results to feed three different Machine Learning algorithms that allow us to find the momentum fractions of the partons involved in the process, in terms of suitable combinations of the final state momenta. Our results are compatible with previous findings and suggest a powerful application of Machine-Learning to model high-energy collisions at the partonic-level with high-precision.}

\begin{document} 
\maketitle
\flushbottom

\section{Introduction}
\label{sec:Introduction}
Thanks to recent technological advances and increased computational power, Machine Learning (ML) has taken by storm our everyday life. Applications of ML cover fields as diverse as image and speech recognition, automatic language translation, product recommendation, stock market prediction and medical diagnosis, to mention some examples. High-energy physics has not remained indifferent to the opportunities offered by these techniques. In the last years several applications have been developed, particularly in regards to data analysis. Novel jet clustering algorithms that use improved classification to identify structures \cite{Nachman:2021yvi}, reconstruction of the Monte-Carlo (MC) parton shower variables \cite{Cranmer:2021gdt}, and reconstruction of the kinematics \cite{Arratia:2021tsq} are just some of the explored uses. In particular, the high luminosity upgrade of the Large Hadron Collider (LHC) and the upcoming Electron-Ion Collider (EIC) are feeding the interest of the community in ML\footnote{For more information concerning the LHC upgrade, we refer the reader to \href{https://home.cern/science/accelerators/high-luminosity-lhc}{https://home.cern/science/accelerators/high-luminosity-lhc}. Details about ML developments for the upcoming EIC were presented at the workshop \emph{AI4EIC - Experimental Applications of Artificial Intelligence for the Electron Ion collider} (\href{https://indico.bnl.gov/event/10699/}{https://indico.bnl.gov/event/10699/}).}. From a theoretical perspective there has been progress in the calculation of higher-order scattering amplitudes assisted by ML algorithms \cite{Aylett-Bullock:2021hmo} and, in phenomenology the NNPDF collaboration has pioneered the determination of the partonic structure of hadrons \cite{Forte:2002fg,Rojo:2004iq,DelDebbio:2007ee,Ball:2008by,Ball:2012cx,Ball:2014uwa,AbdulKhalek:2019bux,Ball:2021leu}. 

The successes of the perturbative expansion of Quantum Chromodynamics (QCD) to describe processes involving hadrons lies in the factorisation of the physical observables into hard (perturbative, process-dependent) and soft (non-perturbative, universal) terms \cite{Collins:1989gx}. The former describe the interaction between elementary particles while the latter encode all the information concerning non-perturbative physics, i.e., the description of the partons inside the hadrons before the interaction and their posterior hadronisation into detected particles. For these, only the scale evolution can be determined once they are known at some other scale, and thus must be obtained from data through global fits\footnote{Significant progress in the ab-initio calculation of parton densities is being carried out in the field of Lattice QCD \cite{Cichy:2018mum}.}. 

The simplest description of a hadron is that of a collection of partons moving in the same direction. The probability of finding a particular parton $a$ in a hadron $H$ carrying a fraction $x$ of its momentum is given by the parton distribution function (PDF) $f_{H/a}(x,\mu)$, when the hadron is explored at scale $\mu$. After the hard interaction all outgoing coloured particles will hadronise; the probability of a parton $a$ to fragment into a hadron $H$ with a fraction $z$ of its original momentum is described by the fragmentation function (FF) $D_{a/H}(z,\mu)$. This collinear picture is the best explored and in this framework several sets of PDFs and FFs have been extracted using standard regression techniques (e.g. \cite{Bailey:2020ooq,H1:2015ubc,PDFSMASSA,deFlorian:2014xna, Albino:2008fy}), MC sampling (e.g. \cite{Borsa:2021ran,Moffat:2021dji}) and MC sampling with neural networks (e.g. \cite{Forte:2002fg}). 

In order to perform a meaningful calculation, the hard cross-section must be convoluted with the PDFs and/or FFs, over the corresponding momentum fractions of the partons. In the inclusive deep inelastic scattering (DIS) process, where a lepton and a parton inside a hadron interact by exchanging momentum $Q^{2}\geq 1 \text{ GeV}^{2}$, measuring the scattered lepton (and/or final hadrons) provides the full kinematics of the event. Unfortunately, in proton-proton ($p+p$) collisions the situation is not so simple. One has to estimate the momenta of the initial partons (that enter in the evaluation of the PDFs) using the measured momenta and scattering angles of the final state particles. Depending on the process and the characteristic of the detectors, it can become a complicated task. Despite its inherent complexity, it is of the utmost importance in some situations. For example in the case of asymmetric proton-nucleus ($p+A$) collisions, particles created in the the backward (nucleus-going) direction are linked to initial partons in the nucleus with low-$x$, and those in the forward (proton-going) direction are related to partons in the nucleus with large-$x$. Depending on its exact value, one could have an enhancement or a suppression of the nuclear PDF w.r.t. the free proton one. Knowing the region of the detector associated with the kinematics of interest for a given process is also relevant for the efficient design and construction of the detectors \cite{AbdulKhalek:2021gbh}. The proper mapping of the measured kinematics onto the partonic level is crucial for a correct evaluation of the cross-sections and interpretation of the perturbative calculations. This can be done analytically at leading order (LO) for processes involving few particles, but as one considers higher orders the emission of real particles makes it hard to fully determine the kinematics, and normally phenomenological approximations are used. 

In the present work, we aim to use ML to determine the relation between the measurable four-momenta of the final particles and the parton-level kinematics. In particular, we focus on $p+p$ collisions with one photon plus one hadron in the final state, computed using QCD and Quantum Electrodynamics (QED) corrections. This process has already been identified as an interesting observable at the Relativistic Heavy-Ion Collider (RHIC) \cite{deFlorian:2010vy}. Our goal is to obtain the functions that, depending on the four-momenta of the photon and hadron, give $x_{i}$ (the fraction of momentum of the proton $i$ carried by the parton coming from it, $i=1,2$) and $z$, the fraction of energy of the parton coming from the hard interaction that is taken by the hadron (in our analysis a pion). 

This article is organised as follows. In Sec. \ref{sec:ComputationalSetup} we describe the framework used to implement the MC simulation of hadron-photon production, with special emphasis on the isolation prescription (Sec. \ref{ssec:CutsIsolation}). Relevant phenomenological aspects of the process are discussed in Sec. \ref{sec:Phenomenology}. The distributions w.r.t. different variables are presented in Sec. \ref{ssec:Distributions}, with the purpose of identifying the most probable configurations. We also explore the correlations between different measurable variables and the partonic momentum fractions in Sec. \ref{ssec:Correlation}. In Sec. \ref{sec:Reconstruction}, we detail the implementation of reconstruction algorithms based on ML to approximate the partonic momentum fractions using only measurable quantities. Finally, we discuss the results and comment on potential future applications of our methodology in Sec. \ref{sec:Conclusions}.


\section{Computational setup}
\label{sec:ComputationalSetup}
From the theoretical point of view, the calculation relies on the factorization theorem to separate the low-energy hadron dynamics (i.e. the non-perturbative component embodied into the PDFs and FFs) from the perturbative interactions of the fundamental particles. This approach is valid in the high-energy regime, under the assumption that the typical energy scale of the process is much larger than $\Lambda_{\rm QCD} \approx 900 {\rm MeV}$. The process under consideration is described by
\beq
H_1(P_1) + H_2(P_2) \to h(P^h) + \gamma(P^\gamma) \ ,
\label{eq:Proceso}
\eeq
and the differential cross-section is given by
\beqn
\nonumber d\sigma_{H_1 H_2  \to h \gamma} &=& \sum_{a_1 a_2 a_3 a_4} \int\, dx_1 \, dx_2 \, dz_1 \, dz_2 \, f_{H_1/a_1}(x_1,\mu_I) \,  f_{H_2/a_2}(x_2,\mu_I) \, D_{a_3/h}(z_1,\mu_F) \,  
\\ && \times  D_{a_4/\gamma}(z_2,\mu_F) \, d\hat\sigma_{a_1\, a_2 \to a_3 \, a_4}(x_1 P_1,x_2 P_2,P^h/z_1,P^\gamma/z_2;\mu_I,\mu_F,\mu_R) \, ,
\label{eq:DIFCrossSectionTOTAL}
\eeqn
where $\{a_i\}$ denote the possible flavours of the partons entering into the fundamental high-energy collision. $f_{H_i/a_j}(x,\mu_I)$ is the PDF of the parton at the initial state factorization scale $\mu_I$, and $D_{a_j/h}(z,\mu_F)$ is the FF of the parton at the final state factorization scale $\mu_F$. The partonic cross-section, $d\hat\sigma$, depends on the kinematics of the partons as well on the factorization and renormalization scales ($\mu_R$) and can be computed using perturbation theory. It is worth appreciating that we consider all the partons to be massless.

In Eq. (\ref{eq:DIFCrossSectionTOTAL}) we consider the photon as a parton, i.e. $a_i \in\{q,g,\gamma\}$. Namely, we rely on the extended parton model to include mixed QCD-QED corrections in a consistent way \cite{Roth:2004ti,Carrazza:2015dea,deFlorian:2015ujt,deFlorian:2016gvk,Sborlini:2016dfn}. However, we will assume that the fragmentation of a photon into any hadron is highly suppressed w.r.t. the same process initiated by a QCD parton. This implies that we neglect $D_{\gamma/h}$ and $a_3$ is always a QCD parton (quark or gluon). Also, since we are looking for a photon in the final state, we can write
\beq 
D_{a_4 / \gamma}(z_2,\mu_F) = \delta_{a_4 , \gamma} \delta(z_2-1) + (1-\delta_{a_4 , \gamma}) \tilde{D}_{a_4 / \gamma}(z_2,\mu_F) \, ,
\label{eq:DgammaDEF}
\eeq
which leads to
\beqn
\nonumber d\sigma_{H_1 H_2  \to h \gamma} &=& \sum_{a_1 a_2 a_3} \int\, dx_1 \, dx_2 \, dz \, f_{H_1/a_1}(x_1,\mu_I) \, f_{H_2/a_2}(x_2,\mu_I) \, D_{a_3/h}(z,\mu_F) \, 
\\ \nonumber && \times d\hat\sigma_{a_1\, a_2 \to a_3 \gamma}(x_1 P_1,x_2 P_2,P^h/z,P^\gamma;\mu_I,\mu_F,\mu_R) 
\\ \nonumber &+& \sum_{a_1 a_2 a_3} \sum_{a_4 \in {\rm QCD}} \int\, dx_1 \, dx_2 \, dz_1\, dz_2 \, f_{H_1/a_1}(x_1,\mu_I) \,  f_{H_2/a_2}(x_2,\mu_I) \, D_{a_3/h}(z_1,\mu_F) \, 
\\ && \times \tilde{D}_{a_4/\gamma}(z_2,\mu_F)  \,  d\hat\sigma_{a_1\, a_2 \to a_3 a_4}(x_1 P_1,x_2 P_2,P^h/z_1,P^\gamma/z_2;\mu_I,\mu_F,\mu_R) \, ,
\label{eq:DIFCrossSectionSTANDARD}
\eeqn
where $a_4$ is a QCD parton. By rewriting Eq. (\ref{eq:DIFCrossSectionTOTAL}) in this way, it is possible to separate two different mechanisms originating photons in the final state\footnote{A third mechanism is related to the presence of fracture functions, $M_{a_3,a_4/h,\gamma}$, which do not completely separate the non-perturbative interactions in the final state. Since we are interested in the high-energy limit of this process, such contributions will be suppressed by the same reasons supporting the validity of the factorization theorem.}. 
The first term describes the \emph{direct} production of an observed photon in the partonic collision; in the second term the observed \emph{resolved} photon is generated from a non-perturbative process initiated by the parton $a_4$. It is worth appreciating that these contributions are not individually distinguishable; however the latter can be suppressed by applying adequate prescriptions. By realising that the resolved component appears in the context of hadronisation, the photon being produced together with a bunch of hadrons, one can exploit this signature to enhance the direct photon: it is the motivation for introducing \emph{isolation prescriptions}. By selecting mainly those events that contain photons isolated from hadronic energy, the total cross-section can be approximated to
\beqn
\nonumber d\sigma_{H_1 H_2  \to h \gamma} &\approx& \sum_{a_1 a_2 a_3} \int\, dx_1 \, dx_2 \, dz \, f_{H_1/a_1}(x_1,\mu_I) \, f_{H_2/a_2}(x_2,\mu_I) \, D_{a_3/h}(z,\mu_F) \, 
\\ && \times d\hat\sigma^{(ISO)}_{a_1\, a_2 \to a_3 \gamma}(x_1 P_1,x_2 P_2,P^h/z,P^\gamma;\mu_I,\mu_F,\mu_R) \, ,
\label{eq:ISOLATED}
\eeqn
i.e. neglecting the resolved component and summing over all QCD-QED partons. The partonic cross-section $d\hat\sigma^{(ISO)}_{a_1\, a_2 \to a_3 \gamma}$ incorporates the isolation prescription and is described in greater detail in Sec. \ref{ssec:CutsIsolation}.

We can now move to the discussion of how to include the QED corrections. The next-to-leading order (NLO) pure QCD corrections for this process were computed in Refs. \cite{Arleo:2006xb,deFlorian:2010vy}. Since in this case we are dealing with mixed QCD-QED corrections, we have to consider the two couplings involved in the perturbative expansion. From the computational point of view, we can profit from the Abelianization techniques to directly obtain QED contributions from the QCD ones \cite{deFlorian:2015ujt,deFlorian:2016gvk,Cieri:2018sfk,Cieri:2020ikq,Cieri:2021fdb}. Given that the energy scale of the process is roughly ${\cal O}(10 \, {\rm GeV})$, we have $\as \approx 0.12$ and $\aem \approx 1/129$. This means $\aem \approx \as^2$, indicating that the LO QED corrections have the same weight as the NLO QCD ones. Therefore, the dominant contribution is given by the partonic channels $q\bar q \to g \gamma$ and $q g \to q \gamma$ at ${\cal O}(\as \aem)$, i.e.
\beqn  
d\hat\sigma^{{\rm ISO},(0)}_{a_1\,a_2 \to a_3 \, \gamma} &=& \frac{\as}{2\pi} \frac{\aem}{2\pi}\, \int d{\rm PS}^{2\to 2} \,  \frac{|{\cal M}^{(0)}|^2(x_1 P_1, x_2 P_2, P^h_3/z, P^\gamma)}{2 \hat s} \, {\cal S}_2 \, \, ,
\label{eq:xsISOLATEDLOQCD}
\eeqn
with ${\cal S}_2$ the measure function containing the definition of the kinematical selection cuts for the $2\to 2$ sub-processes. We have then to include ${\cal O}(\as^2 \aem)$ and ${\cal O}(\as \aem^2)$ contributions, associated to the partonic channels
\beq
q \bar q \to g \gamma g \, , \quad q g \to q \gamma g \, , \quad g g  \to q \gamma \bar q \, , \quad q \bar q \to Q \gamma \bar Q \, , \quad q Q \to q \gamma Q \, , 
\label{eq:PartonicChannelsNLOQCD}
\eeq
and
\beq 
q \gamma \to q \gamma \, , \quad  q\bar q \to \gamma \gamma \, ,
\label{eq:PartonicChannelsLOQED}
\eeq
respectively. In this way, the corrections to the partonic cross-section are given by \cite{Renteria-Estrada:2021rqp}
\beqn 
\nn d\hat\sigma^{{\rm ISO},(1)}_{a_1\,a_2 \to a_3 \, \gamma} &=& \frac{\aem^2}{4\pi^2}\, \int d{\rm PS}^{2\to 2} \,  \frac{|{\cal M}^{(0)}_{QED}|^2(x_1 P_1, x_2 P_2, P^h/z, P^\gamma)}{2 \hat s} \, {\cal S}_2 
\\ \nn &+& \frac{\as^2 }{4\pi^2} \frac{\aem}{2\pi}\, \int d{\rm PS}^{2\to 2} \, \frac{|{\cal M}^{(1)}|^2(x_1 P_1, x_2 P_2, P^h/z, P^\gamma)}{2 \hat s} \,{\cal S}_2 \,
\\ &+& \frac{\as^2 }{4\pi^2} \frac{\aem}{2\pi} \sum_{a_r} \int d{\rm PS}^{2\to 3} \, \frac{|{\cal M}^{(0)}|^2(x_1 P_1, x_2 P_2, P^h/z, P^\gamma, k_r)}{2 \hat s} \, {\cal S}_3 \,  ,
\label{eq:xsISOLATEDNLO}
\eeqn  
where $\hat s$ is the partonic center-of-mass energy and $r$ denotes the extra parton associated to the real radiation correction. $|{\cal M}^{(0)}|^2$ and $|{\cal M}^{(1)}|^2$ are the squared matrix-elements for the tree-level and one-loop corrections, respectively. In these expressions, ${\cal S}_3$ represents the measure function that implements the experimental cuts and the isolation prescription for the $2 \to 3$ sub-processes.

Since we are dealing with higher-order corrections, singularities will appear in the calculation. The LO QED is given by a (finite) Born level process. However, the NLO QCD corrections involve both ultraviolet (UV) and infrared (IR) singularities that must be regularized and cancelled to get a physical result. The regularization was done using Dimensional Regularization (DREG) \cite{Ashmore:1972uj,Cicuta:1972jf,tHooft:1972tcz,Bollini:1972ui}. The virtual corrections were computed starting from the one-loop QCD amplitude for the process $0 \to q \bar q g \gamma$, removing the UV poles through the renormalization in the $\overline{\rm MS}$ scheme. In order to cancel the IR singularities, we relied on the subtraction formalism \cite{Ellis:1990ek,Kunszt:1992tn,Frixione:1995ms,Catani:1996jh,Catani:1996vz}, splitting the real phase-space in regions containing only one kind of IR singularity. When combining the real and the virtual corrections, some of the IR divergences associated to final state radiation (FSR) cancel by virtue of the KLN theorem \cite{Kinoshita:1962ur,Lee:1964is}. But to achieve a full cancellation, counter-terms were added to remove the remaining initial-state and final-state contributions absorbed into the PDFs and FFs, respectively. In this way, the master formula for the partonic cross-section at NLO QCD + LO QED accuracy is symbolically given by
\beqn
\nn d\hat\sigma^{{\rm ISO},(1),{\rm finite}}_{a_1\,a_2 \to a_3 \, \gamma} &=& d\hat\sigma^{{\rm ISO},(1),{\rm ren.}}_{a_1\,a_2 \to a_3 \, \gamma} - \frac{C^{\rm UV}_{a_1\,a_2 \to a_3 \,  \gamma}}{\epsilon} \, \times \, d\hat\sigma^{{\rm ISO},(0)}_{a_1\,a_2 \to a_3 \, \gamma} 
\\ &-& d\hat\sigma^{{\rm ISO},{\rm cnt},(I)}_{a_1\,a_2 \to a_3 \, \gamma} - d\hat\sigma^{{\rm ISO},{\rm cnt},(F)}_{a_1\,a_2 \to a_3 \, \gamma} \, ,
\label{eq:CROSSSECTIONMASTER}
\eeqn
where $d\hat\sigma^{{\rm ISO},{\rm cnt},(I)}_{a_1\,a_2 \to a_3 \, \gamma}$ and $d\hat\sigma^{{\rm ISO},{\rm cnt},(F)}_{a_1\,a_2 \to a_3 \, \gamma}$ are the initial and final-state IR counter-terms, respectively. Here, $C^{\rm UV}_{a_1\,a_2 \to a_3 \,  \gamma}$ is the renormalization counter-term for the partonic process $a_1\,a_2 \to a_3 \,  \gamma$ in the $\overline{\rm MS}$ scheme\footnote{Explicit formulae for all the ingredients in this expression can be found in Refs. \cite{Frixione:1995ms,Sborlini:2009gsj}.}.

\subsection{Isolation prescription and other assumptions}
\label{ssec:CutsIsolation}
In order to suppress events with photons originated from the decay of hadrons, it is necessary to implement an isolation prescription. The idea behind most of the strategies available in the literature consists in quantifying the amount of hadronic energy surrounding a well-identified photon, and rejecting events with more hadronic energy than a certain threshold. Whilst most of the prescriptions work nicely at LO, not all of them are infrared safe. For instance, it is known that choosing a fixed cone eliminates events that play a crucial role in the cancellation of IR singularities. Thus, special care is needed in the implementation of these methods\footnote{An extensive study of different methods and their impact on the calculations is available in Refs. \cite{Cieri:2015wwa,Catani:2018krb,Gehrmann:2020oec}.}.

In this work, we rely on the smooth cone prescription introduced in Ref. \cite{Frixione:1998jh}. Its main advantage is that it suppresses the resolved component without preventing the emission of soft/collinear QCD radiation, which makes it IR-safe and fully suitable for higher-order calculations.
In the first place, we fix a reference point in the rapidity-azimuthal plane $(\eta_0,\phi_0)$, and define the distance
\beq 
r(j) = \sqrt{(\eta_j-\eta_0)^2+(\phi_j-\phi_0)^2} \, ,
\label{eq:DISTANCIA}
\eeq
with $(\eta_j,\phi_j)$ the angular coordinates of the parton $j$. Once we identify a photon in the detector, we trace a cone of radius $R$ around it and look for QCD partons inside. If no QCD radiation lays inside the cone, the photon is isolated. If not, we identify the QCD partons inside the cone, $\{a_j\}$, and measure their distance to the photon following Eq. (\ref{eq:DISTANCIA}). Then, for a fixed $r \leq R$, we calculate the sum of the hadronic transverse energy according to 
\beq 
E_T(r) = \sum_{r_j \leq r} E_{T_j} \, .
\label{eq:TransverseEnergy}
\eeq
We want to restrict $E_T$ by imposing an upper bound, thus limiting the amount of hadronic energy surrounding the photon. In the fixed cone prescription, this limit is a constant. However, for the smooth prescription, we introduce an arbitrary smooth function $\xi(r)$ satisfying $\xi(r) \to 0$ for $r\to 0$, and require $E_T(r)<\xi(r)$ for every $r<r_0$. Only if this condition is fulfilled, the photon is isolated; otherwise, the event is rejected.

The experimental implementation of this criterion requires a very high angular resolution, something that is usually not achievable in practise. This is one of the reasons because most of the current experiments still rely (mainly) on the fixed cone prescription. Fortunately, the difference between both approaches can be neglected for several relevant observables \cite{Cieri:2015wwa,Catani:2018krb}. In any case, technological improvements in detector science will certainly reduce the experimental limitations in the near future.

Finally, let us mention one further detail about the implementation. We will neglect the partonic channel $q \bar q \to \gamma \gamma$ in Eq. (\ref{eq:ISOLATED}), which would imply the introduction of the fragmentation $D_{\gamma/h}$. From the point of view of perturbation theory, this fragmentation can be interpreted as a collinear electromagnetic splitting $\gamma \to a + X$, with $a$ a QCD-parton that undergoes hadronization to generate the observed hadron $h$. Performing a naive counting, this contribution is ${\cal O}(\alpha^3)$ and turns out to be sub-leading w.r.t. the NLO QCD + LO QED terms studied in this work\footnote{This topic deserves attention, specially because non-perturbative contributions could enhance the production rate of hadrons from highly-energetic photons. Unfortunately, we were unable to find in the literature studies or a proper definition of $D_{\gamma/h}$ to be included within our simulations.}. 


\section{Phenomenological results}
\label{sec:Phenomenology}
Using the formalism explained in the previous Section, we calculated the unpolarized cross-section via a code that uses adaptive MC integration. In this program, the different contributions to $2 \to 2$ and $2 \to 3$ processes are computed independently, and kinematic cuts can be imposed. In particular, we reproduced the experimental cuts corresponding to the PHENIX detector, i.e.
\beq
|\eta^h| \leq 0.35 \, , \quad |\eta^\gamma| \leq 0.35 \, , \quad p_T^h \geq 2 \, {\rm GeV} \, , \quad 5 \, {\rm GeV} \leq p_T^\gamma \leq 15 \, {\rm GeV}  \, ,
\label{eq:CUTSexperimentales}
\eeq
with $\eta$ the rapidity of the particles measured in the hadronic center-of-mass frame. On top of that, we require $|\phi^h-\phi^\gamma|>2$ to retain those events with the photon and hadron produced almost back-to-back. We perform the simulations at centre-of-mass (c.m.) energy ($\sqrt{S_{CM}}$) $200$ GeV for RHIC and at $\sqrt{S_{CM}}=13$ TeV for LHC Run II, keeping in this case the same cuts described in Eq. (\ref{eq:CUTSexperimentales}). Since the pion is the lightest hadron and is produced more copiously, we restrict our attention to the case $h=\pi^+$. Additionally, we considered the scenario for Tevatron at $\sqrt{S_{CM}}=1.96$ TeV because it involves proton-antiproton ($p+\bar{p}$) collisions. In principle, this process might exhibit a different dependence on PDFs and FFs, compared to $p+p$ collisions.

Regarding the non-perturbative ingredients of the calculation, we used the LHAPDF package \cite{Buckley:2014ana,Andersen:2014efa} to have a unified framework for the PDF implementation. We relied on the \texttt{NNPDF4.0NLO} \cite{Ball:2021leu} and \texttt{NNPDF3.1luxQEDNLO} \cite{Campbell:2018wfu,Bertone:2017bme,Manohar:2017eqh,Manohar:2016nzj} parton distributions for the pure QCD and mixed QCD-QED calculations, respectively. In both cases, we use the set \texttt{0}, which corresponds to an average over the different replicas. For the fragmentation functions, we used the \texttt{DSS2014} set at NLO accuracy~\cite{deFlorian:2014xna,deFlorian:2007ekg}. Also, we evolve the QCD and QED couplings using the one-loop RGE with the initial conditions $\as(m_Z)=0.118$ and $\aem(m_Z)=1/128$.

Finally, we fixed the factorization and renormalization scales to be equal to the average transverse momenta of the hadron and the photon, i.e.
\beq
\mu_F = \mu_I = \mu_R = \frac{p_T^\pi+p_T^\gamma}{2} \, .
\label{eq:MUDEF}
\eeq
Regarding the implementation of the smooth isolation criteria, we used the function
\beq 
\xi(r) = E_T^\gamma \, \left(\frac{1-\cos(r)}{1-\cos(r_0)}\right)^4 \, ,
\label{eq:XIR}
\eeq
where $E_T^\gamma$ is the transverse energy of the photon and $r_0=0.4$. As mentioned before, the only requirement for $\xi(r)$ is that $\xi(r) \to 0$ smoothly, and Eq. (\ref{eq:XIR}) fulfils this condition.


\subsection{One-dimensional distributions}
\label{ssec:Distributions}
Since we are looking at the process $p + p \to \pi + \gamma + X$, 
\beq
{\cal V}_{\rm Exp} = \{p_T^\gamma,p_T^\pi,\eta^\gamma,\eta^\pi,\cos(\phi^\pi-\phi^\gamma)\} \, ,
\label{eq:VARIABLES}
\eeq
are the experimentally accessible variables measured in the c.m. system. Notice that we consider only the difference of the azimuthal angles, because the problem has rotational symmetry around the collision axis. Moreover, it turns out that $\cos(\phi^\pi-\phi^\gamma)$ is a variable often used by experimental collaborations \cite{PHENIX:2008osq}.

\begin{figure}[h!]
    \centering
    \subfigure{\includegraphics[width=75mm,height=6cm]{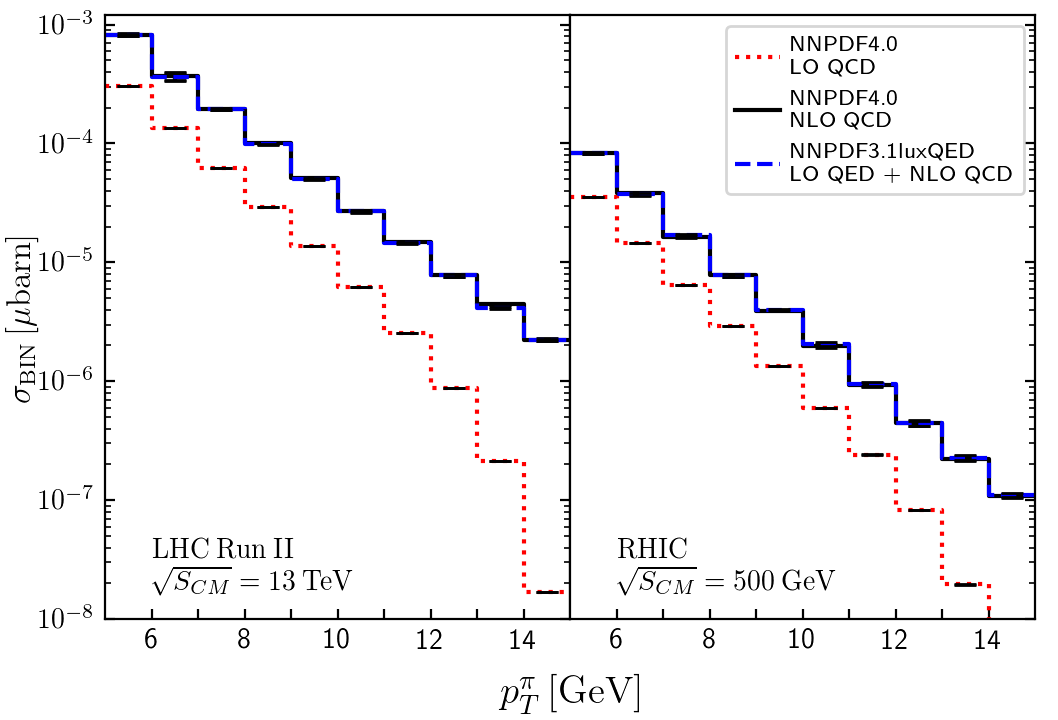}}
    \subfigure{\includegraphics[width=75mm,height=6cm]{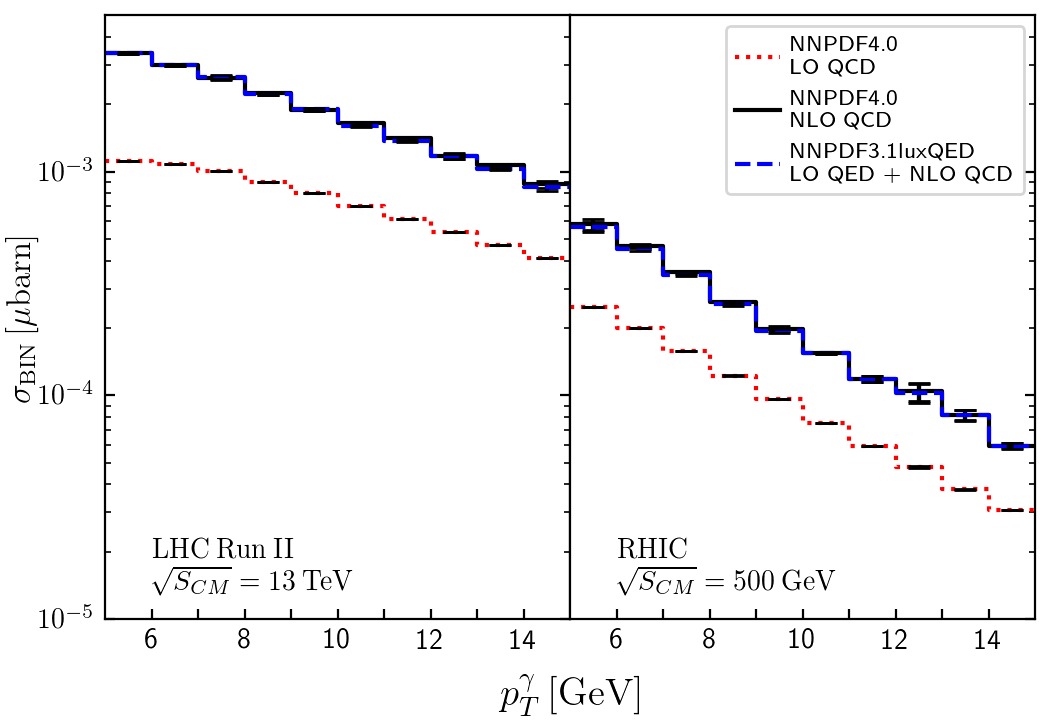}}
    \caption{Unpolarized cross-section for the production of one photon plus one pion as a function of the transverse momentum of the pion (left) and the photon (right), respectively. We considered the selection cuts described in the previous section, for LHC Run II and RHIC, respectively.}
    \label{fig:ONEdimensionPT}
\end{figure} 

\begin{figure}[h!]
    \centering
    \subfigure{\includegraphics[width=75mm,height=6cm]{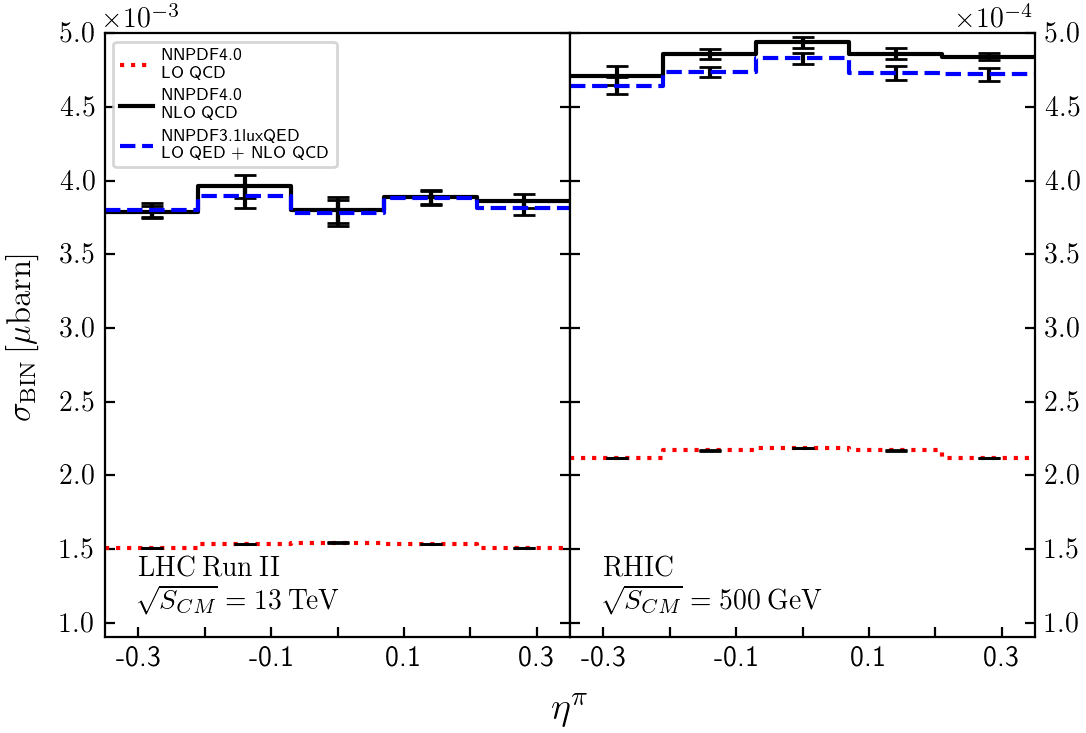}}
    \subfigure{\includegraphics[width=75mm,height=6cm]{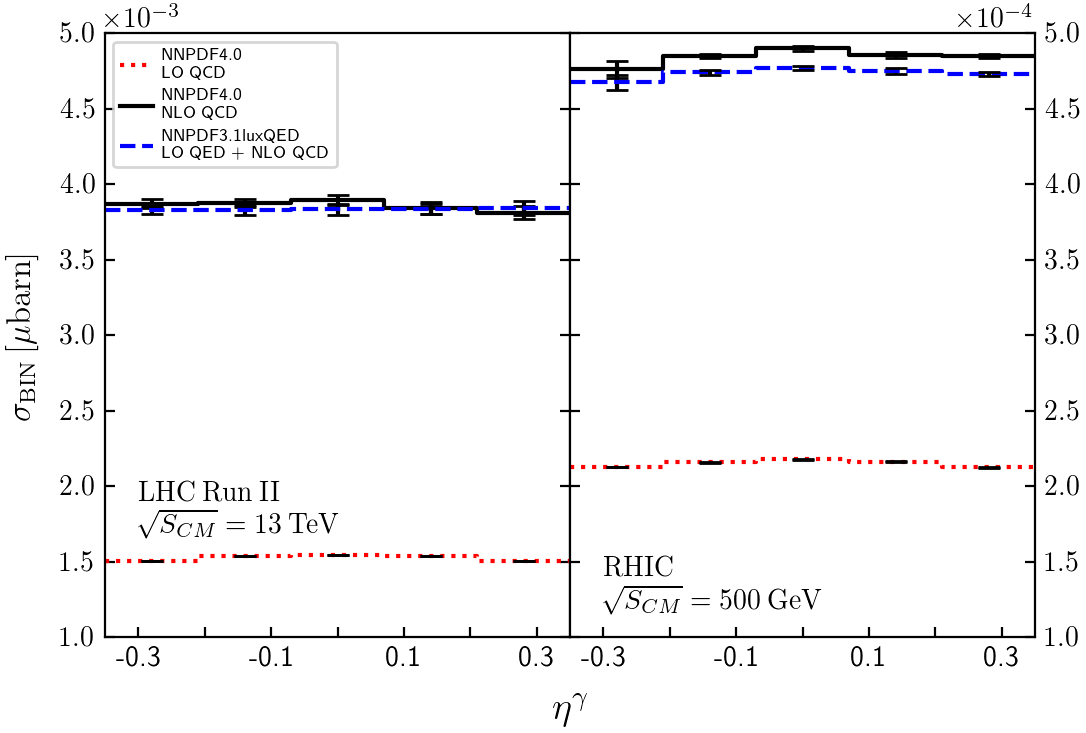}}
    \caption{Same as Fig. \ref{fig:ONEdimensionPT}, but now as a function of the rapidity of the pion (left) and the photon (right), respectively. }
    \label{fig:ONEdimensionETA}
\end{figure}

\begin{figure}[h!]
    \centering
    \includegraphics[width=75mm,height=6cm]{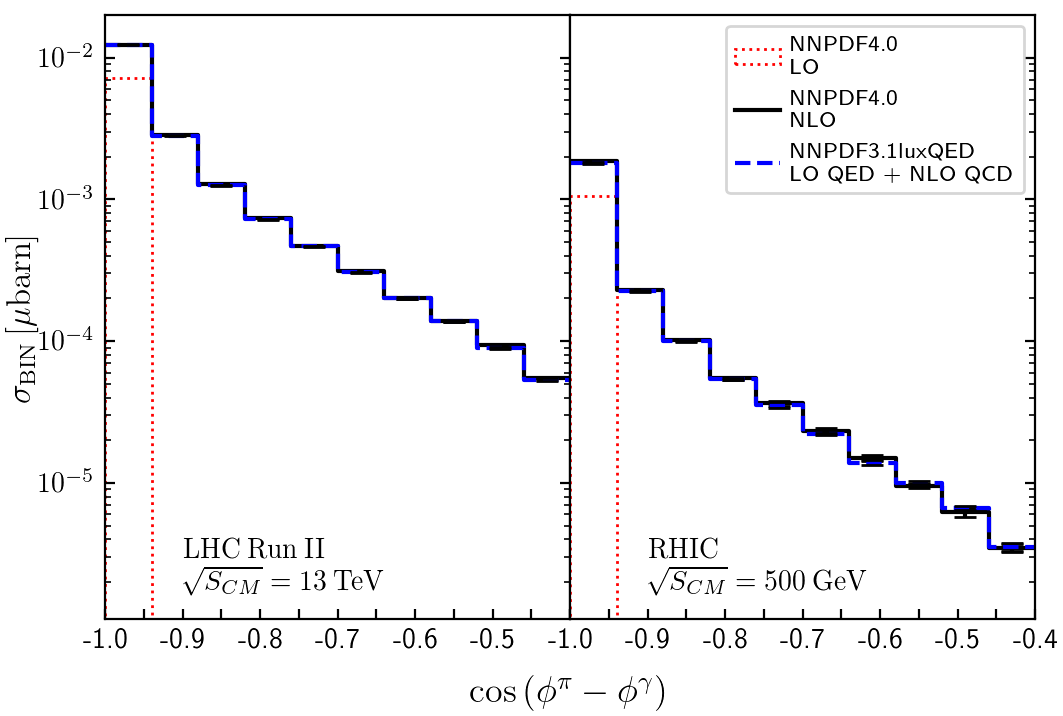}
    \caption{Dependence on $\cos{(\phi^\pi-\phi^{\gamma})}$ for LHC Run II (left) and RHIC (right).}
    \label{fig:ONEdimensionCOSPHI}
\end{figure}

\begin{figure}[h!]
    \centering
    \subfigure{\includegraphics[width=75mm,height=6cm]{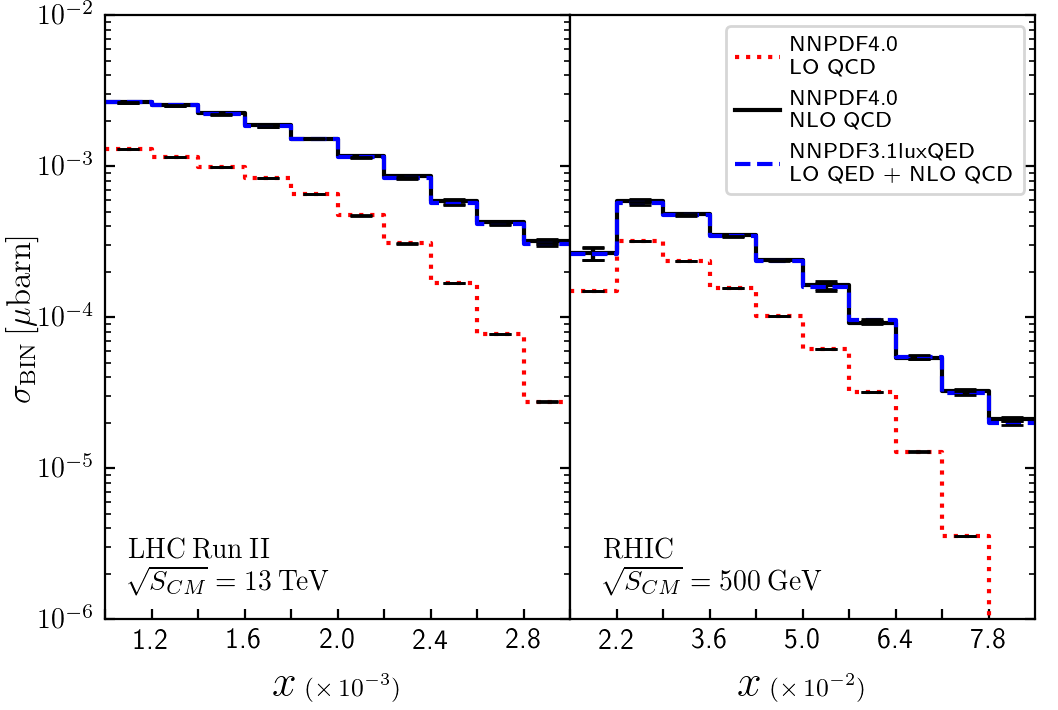}}
    \subfigure{\includegraphics[width=75mm,height=6cm]{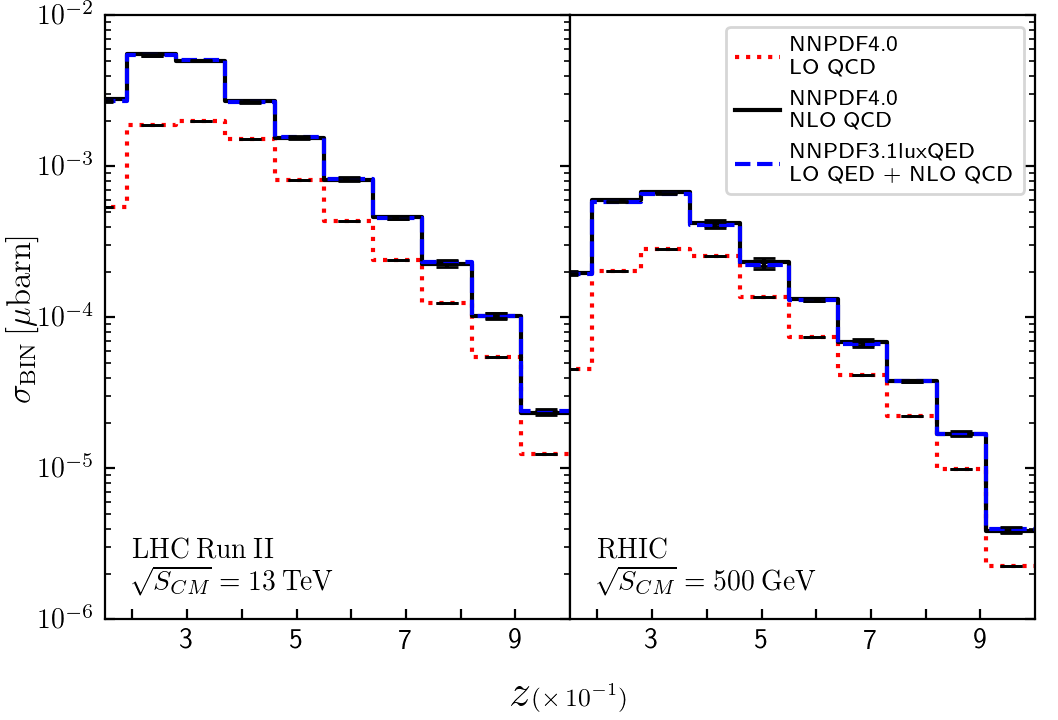}}
    \caption{Cross-section as a function of the partonic momentum fractions $x$ (left) and $z$ (right), for RHIC  and LHC Run II . Since these experiments involve $p+p$ collisions, we consider $x=x_1$ as given by Eq. (\ref{eq:DIFCrossSectionTOTAL}).}
    \label{fig:ONEdimensionX} 
\end{figure}

\begin{figure}[h!]
    \centering
    \includegraphics[width=75mm,height=6cm]{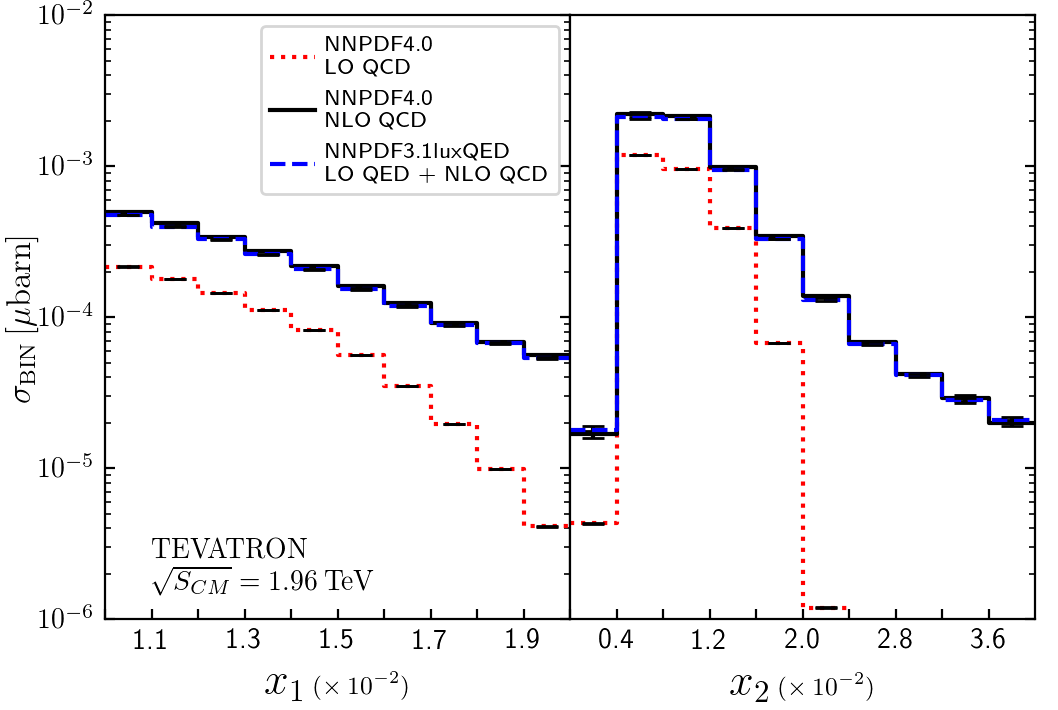}
    \caption{Cross-section as a function of the partonic momentum fraction $x_1$ (left) and $x_2$ (right), for Tevatron. $x_1$ corresponds to the momentum fraction associated to the proton, whilst $x_2$ to the antiproton.}
    \label{fig:ONEdimensionXteva}
\end{figure}

In Figures \ref{fig:ONEdimensionPT}, \ref{fig:ONEdimensionETA}, \ref{fig:ONEdimensionCOSPHI}, \ref{fig:ONEdimensionX} and \ref{fig:ONEdimensionXteva} we present the single differential cross-section as a function of the variables ${\cal V}_{\rm Exp}$ for RHIC and LHC Run II. Our predictions are shown for LO QCD (dotted red), NLO QCD (solid black) and NLO QCD + LO QED (dashed blue), considering the default scale choice defined in Eq. (\ref{eq:MUDEF}). In first place, we study the pion ($p_T^\pi$) and photon ($p_T^\gamma$) transverse-momentum spectrum in Fig. \ref{fig:ONEdimensionPT}. The cross-section increases for higher c.m. energies and the impact of the QED corrections also becomes more sizable. The distribution in $p_T^\pi$ falls faster than the $p_T^\gamma$-spectrum, mainly because of the convolution with the FFs. In fact, the experimental cuts imposed ensure an important contribution of events with close-to-Born kinematics. In this case, $p_T^\gamma$ is associated to the transverse momentum of the parton $c$ which fragments into a pion with momentum fraction $z$. Since the FFs tend to favour the region with $z \leq 0.2$ \cite{Abdolmaleki:2021yjf}, the suppression observed in Fig. \ref{fig:ONEdimensionPT} can be understood.

Next, we present the distributions in the rapidities (Fig. \ref{fig:ONEdimensionETA}) and the azimuthal variable $\cos(\phi^\pi-\phi^\gamma)$ (Fig. \ref{fig:ONEdimensionCOSPHI}). In both cases, we show a comparison between RHIC and LHC Run II. For the rapidity distribution, we appreciate a significant NLO QCD correction, although the added LO QED effects are very small. Regarding the azimuthal spectrum, we can observe in Fig. \ref{fig:ONEdimensionCOSPHI} a peak in the back-to-back region (i.e. $\cos(\phi^\pi-\phi^\gamma)=-1$), with a fast suppression for configurations beyond Born-level kinematics.

Besides the distributions w.r.t. the experimentally-accessible quantities, we can compute the differential cross-section as a function of the partonic momentum fractions, $x_1$, $x_2$ and $z$. For $p+p$ collisions we consider only the distributions w.r.t. $x_1$ due to the symmetry of the system. In what follows, $x$ and $x_1$ will be used interchangeably. The corresponding plots are shown in Fig. \ref{fig:ONEdimensionX}, for $x=x_1$ (left) and $z$ (right). We notice that the experimental cut in $p_{T}^{\gamma}$ induces a restriction on the maximum value of $x$ involved in the collision. In fact, using a LO approximation, we get
\beq 
x_{\rm Max} \approx \frac{p_T^\gamma}{\sqrt{S_{CM}}} \, ,
\label{eq:XMAXcut}
\eeq
beyond this value, the cross-section is drastically suppressed. For RHIC and LHC Run II, it translates to $x_{\rm Max} \approx 0.03$ and $x_{\rm Max} \approx 0.001$, respectively. Thus, we will use this information to restrict the $x$-range in the correlation analysis presented in the next section. In this way, we will avoid dealing with regions with a negligible amount of events. Notice that the higher the energy of the process, the lower the $x$-range accessible by the experiment.

Regarding the dependence in $z$ (right panel of Fig. \ref{fig:ONEdimensionX}), it reaches almost the endpoint region (i.e. $z=1$) with a reasonable amount of events. The fact that we impose $p_T^\pi \geq 2$ GeV translates into a lower bound for $z$ given by  
\beq 
z_{\rm Min} \approx \frac{p_T^\pi}{\sqrt{S_{CM}}} \, ,
\label{eq:ZMINcut}
\eeq
which corresponds to $z_{\rm Min}\approx 0.004$ and $z_{\rm Min} \approx 0.0001$ for RHIC and LHC Run II, respectively. Opposite to the case of the $x$-distribution, here the higher the energy of the process, the wider the accessible $z$-range. It is worthwhile noticing that the FFs used in this work do not include in the fit data with $z\leq 0.05$ and extrapolations into that region are most likely unreliable. The distribution present a peak, located at $z_{\rm Peak} \approx 0.35$ for RHIC ($z_{\rm Peak} \approx 0.25$ for LHC Run II). The position of the peaks depends on the explicit functional form of the PDFs and the FFs.
 
To conclude this section, we study the case of $p+\bar{p}$ collisions at Tevatron, with $\sqrt{S_{ CM}} = 1.96$ TeV. In this case, the symmetry between $x_1$ and $x_2$ is broken, since $x_1$ ($x_2$) corresponds to the momentum fraction of a parton inside a proton (antiproton). In Fig. \ref{fig:ONEdimensionXteva} we present the distribution for $x_1$ (left) and $x_2$ (right). We can appreciate that the distribution in $x_2$ reaches a peak around $x_2 \approx 0.01$ and then falls faster than the $x_1$-distribution. We know from previous studies that the partonic channel $gg$ is dominant \cite{deFlorian:2010vy}, and thus we expect the differences to take place in the $q \bar q$ and $q Q$ channels. This also has an impact when studying the $x_1$ vs $x_2$ correlations, as we will show in the next subsection.


\subsection{Correlations with the partonic momentum fractions}
\label{ssec:Correlation}
Since one of the main goals of this work is to reconstruct the partonic kinematics starting from experimentally accessible quantities, it is useful to first study the correlations among the different variables. This helps us to prioritize certain ansatzes depending on their functional form, in such a way that we capture the leading behaviour when exploring linear models. In the following, we restrict the discussion to RHIC kinematics (with the cuts defined in the previous section).

\begin{figure}[h!]
    \centering
    \subfigure{\includegraphics[width=75mm,height=5.9cm]{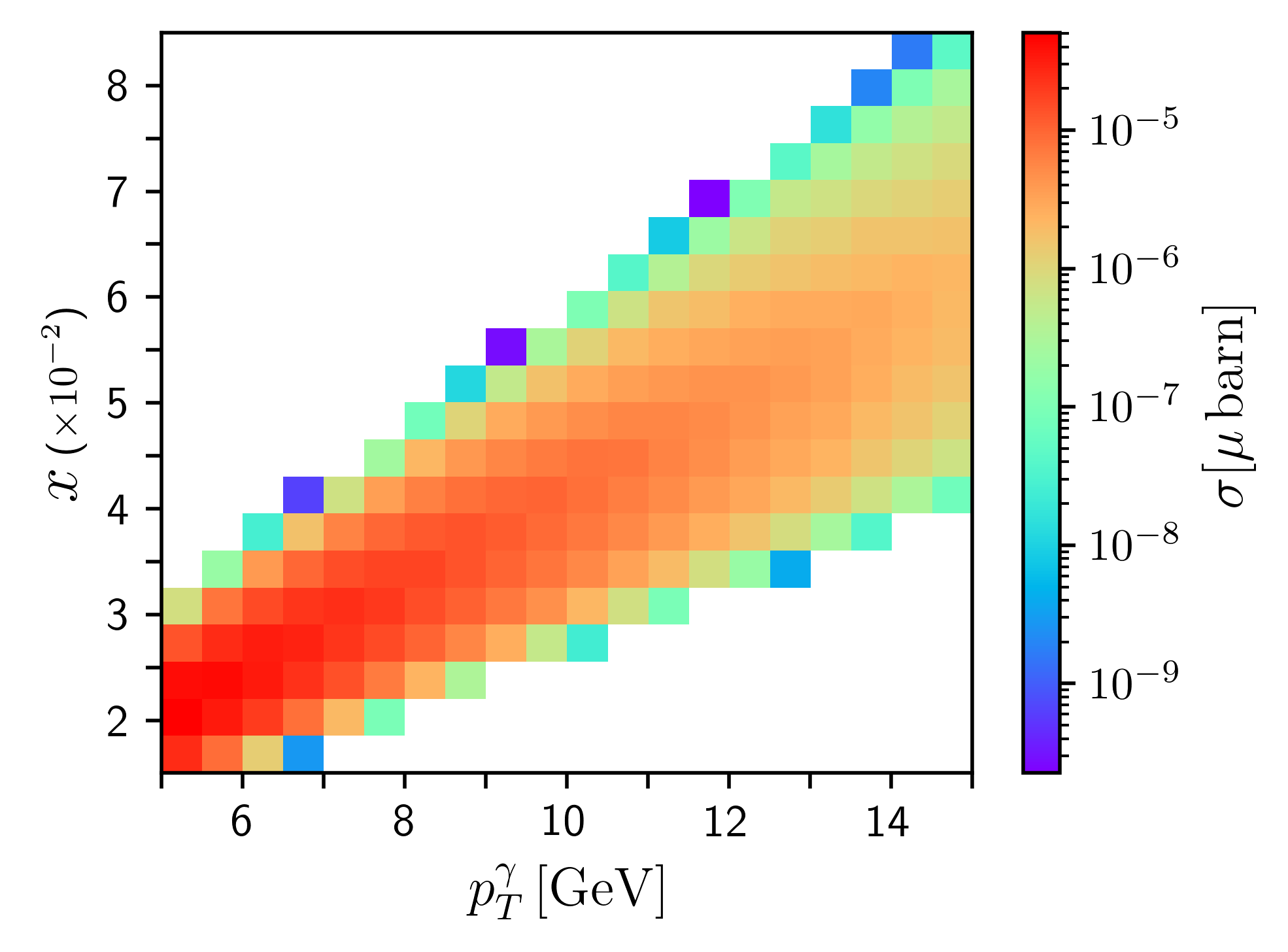}}
    \subfigure{\includegraphics[width=75mm,height=5.9cm]{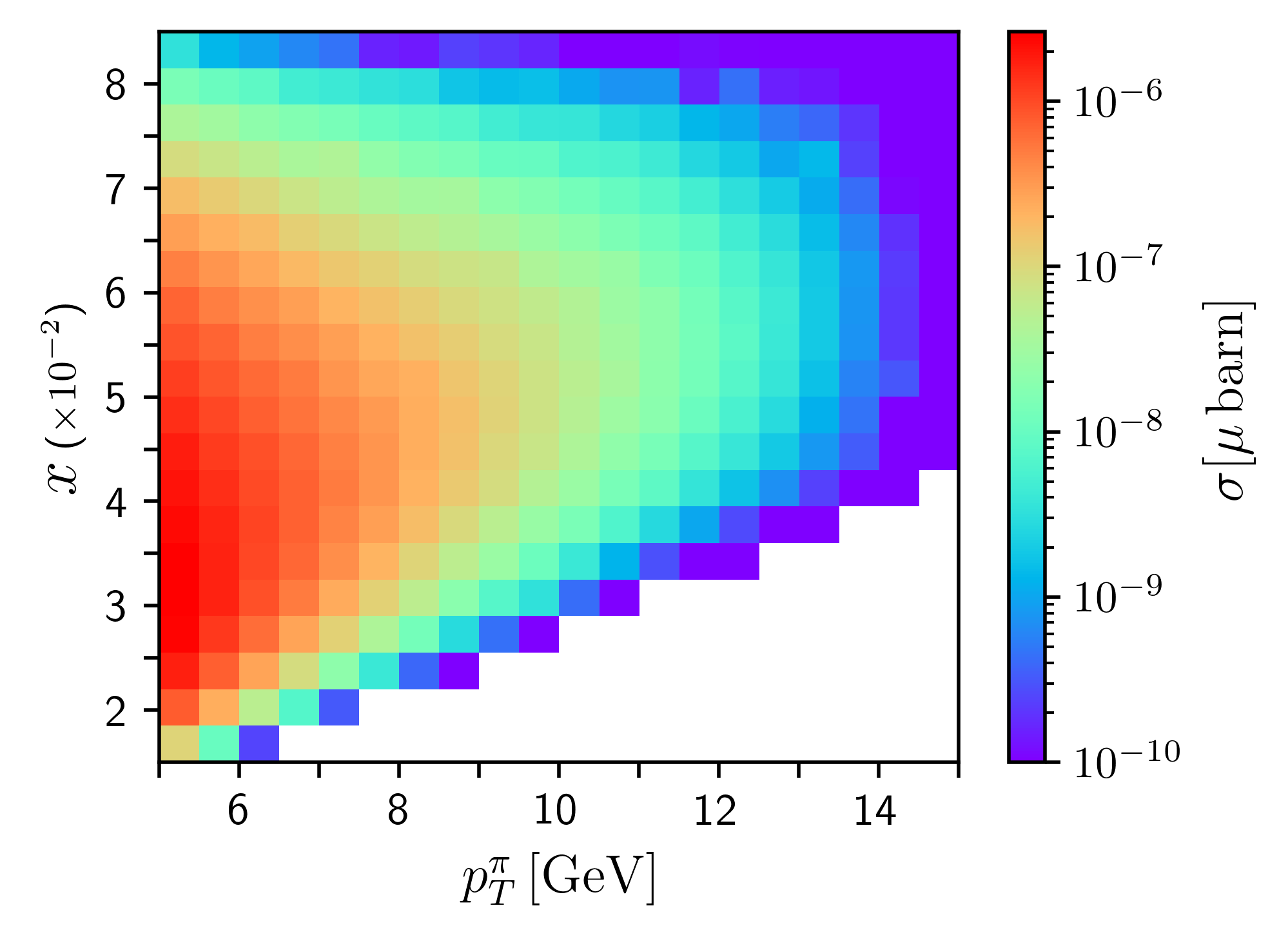}}
    \subfigure{\includegraphics[width=75mm,height=5.9cm]{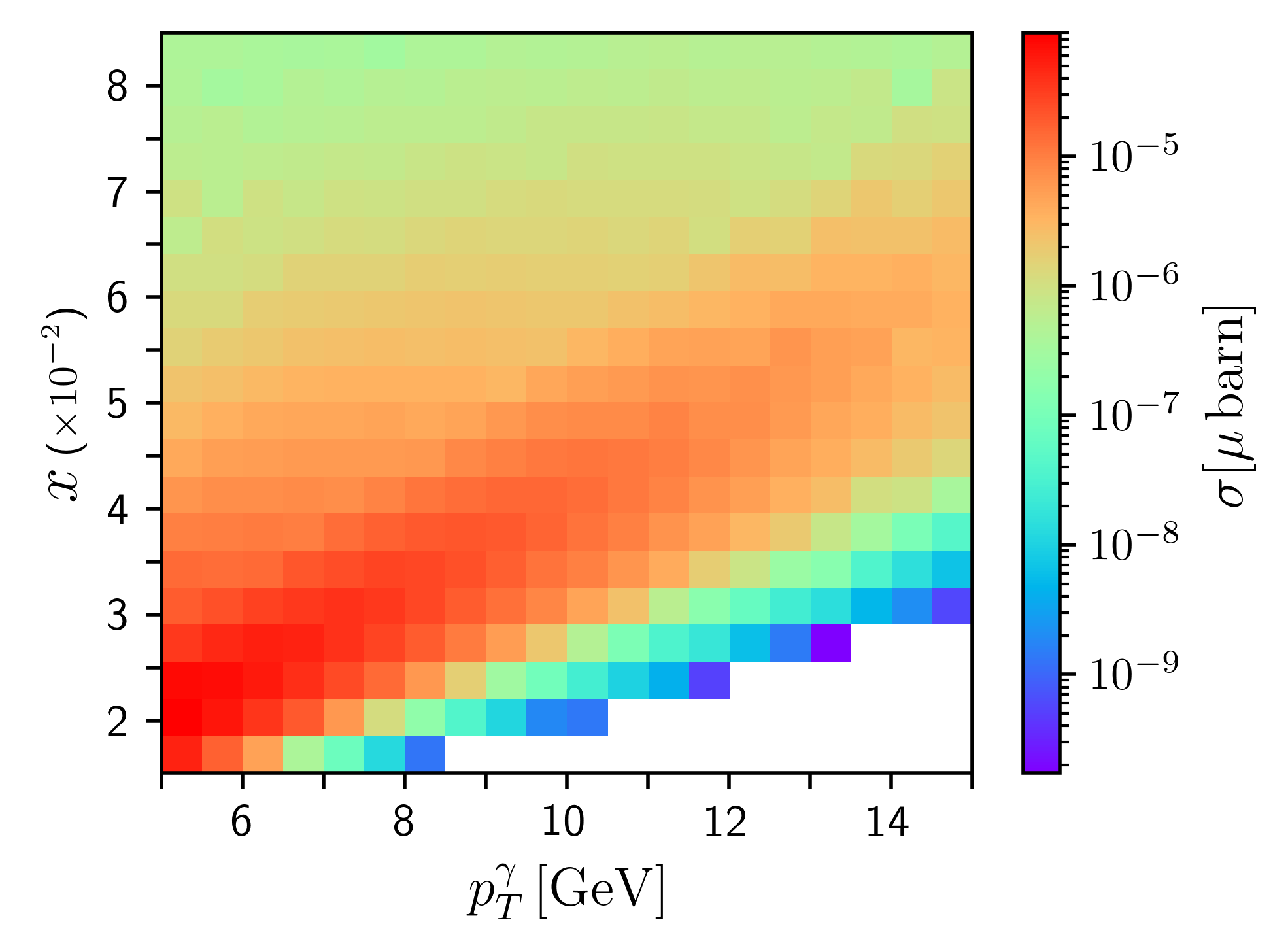}}
    \subfigure{\includegraphics[width=75mm,height=5.9cm]{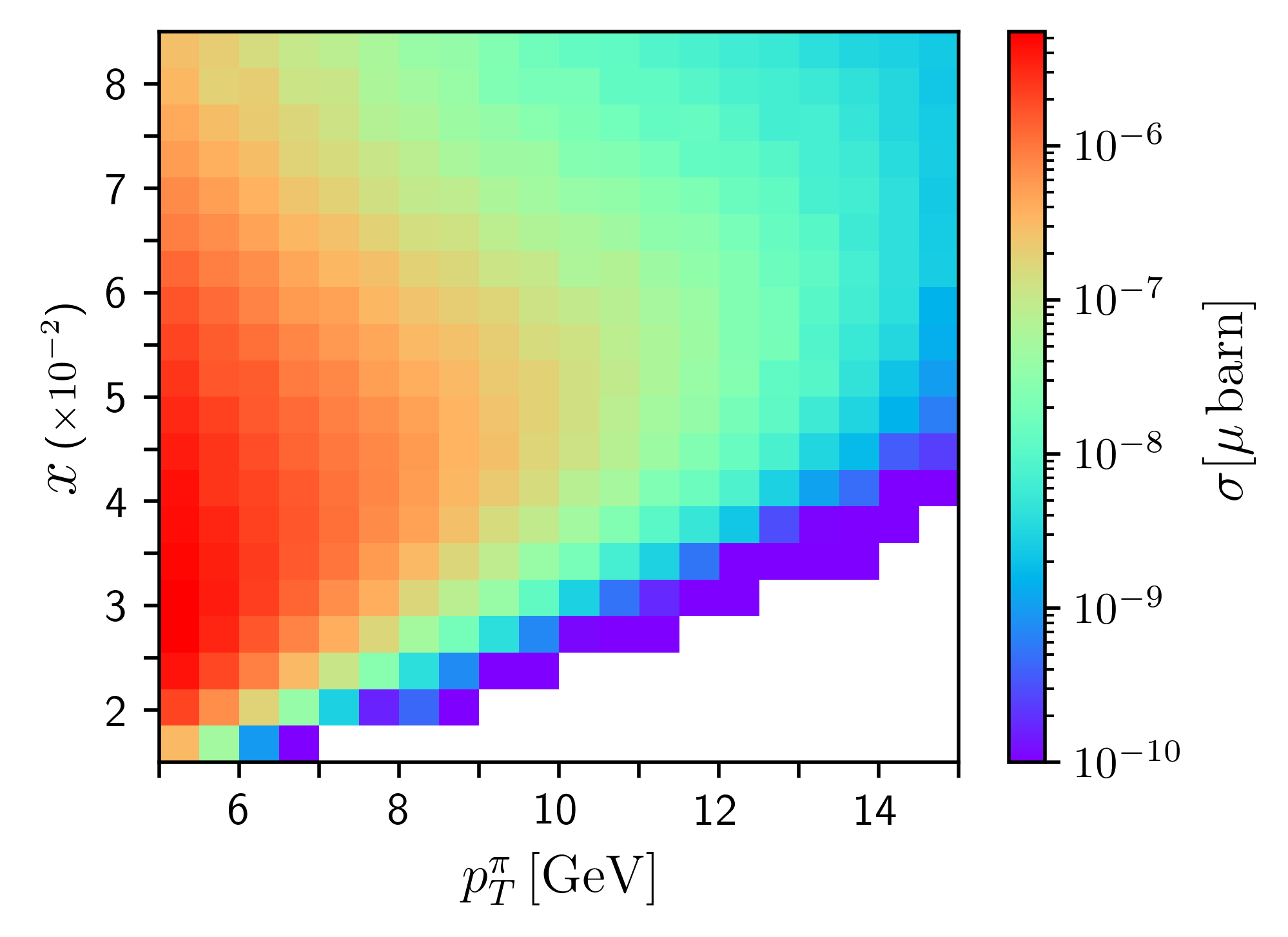}}
    \caption{The partonic momentum fraction $x=x_1$ as a function of $p_T^\gamma$ (left) and $p_T^\pi$ (right). The color scale shows the cross-section at LO QCD (upper row) and NLO QCD + LO QED (lower row).
    \label{fig:xVSptCORRELACION}}
\end{figure}

\begin{figure}[h!]
    \centering
    \subfigure{\includegraphics[width=75mm,height=5.9cm]{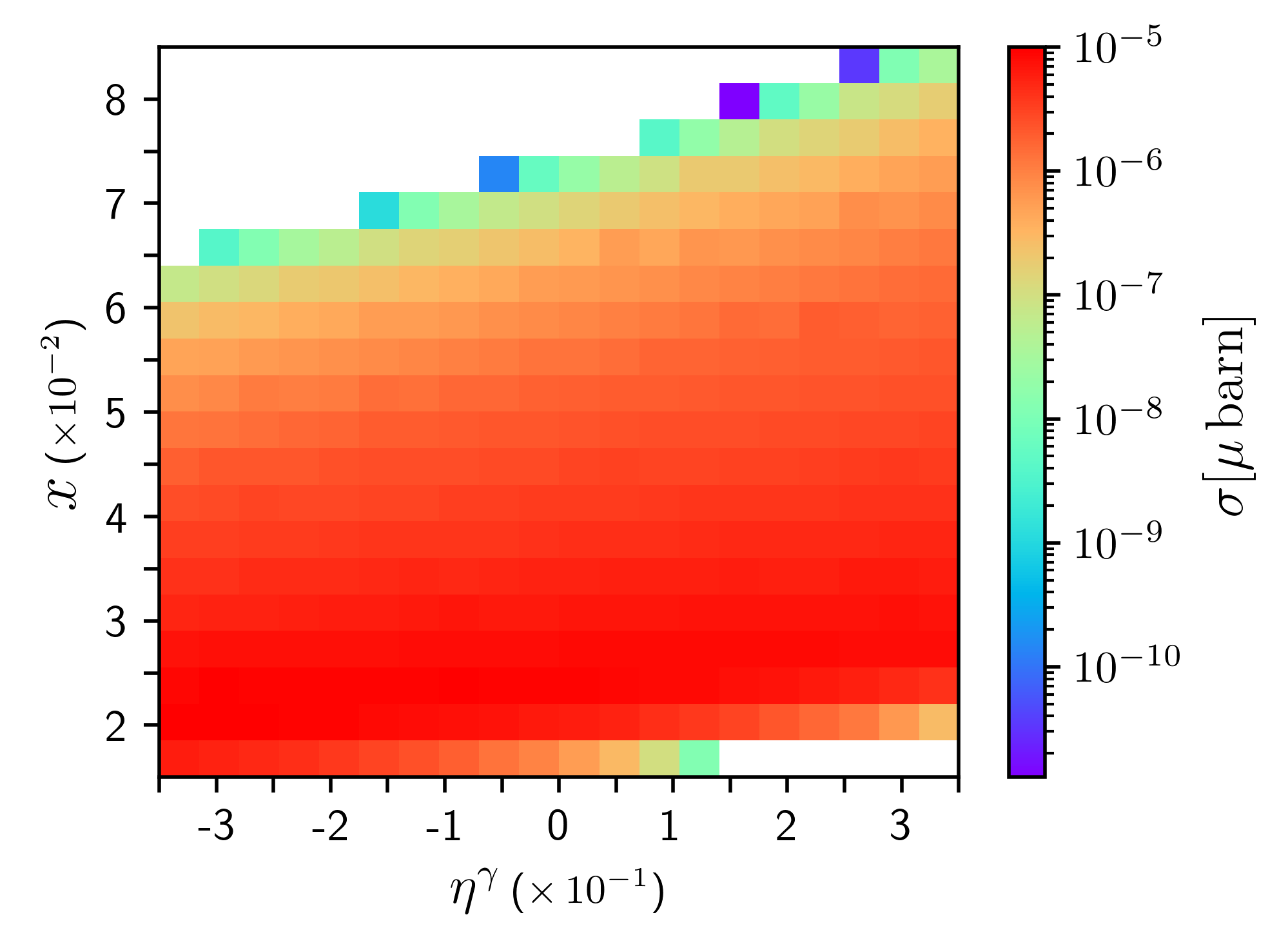}}
    \subfigure{\includegraphics[width=75mm,height=5.9cm]{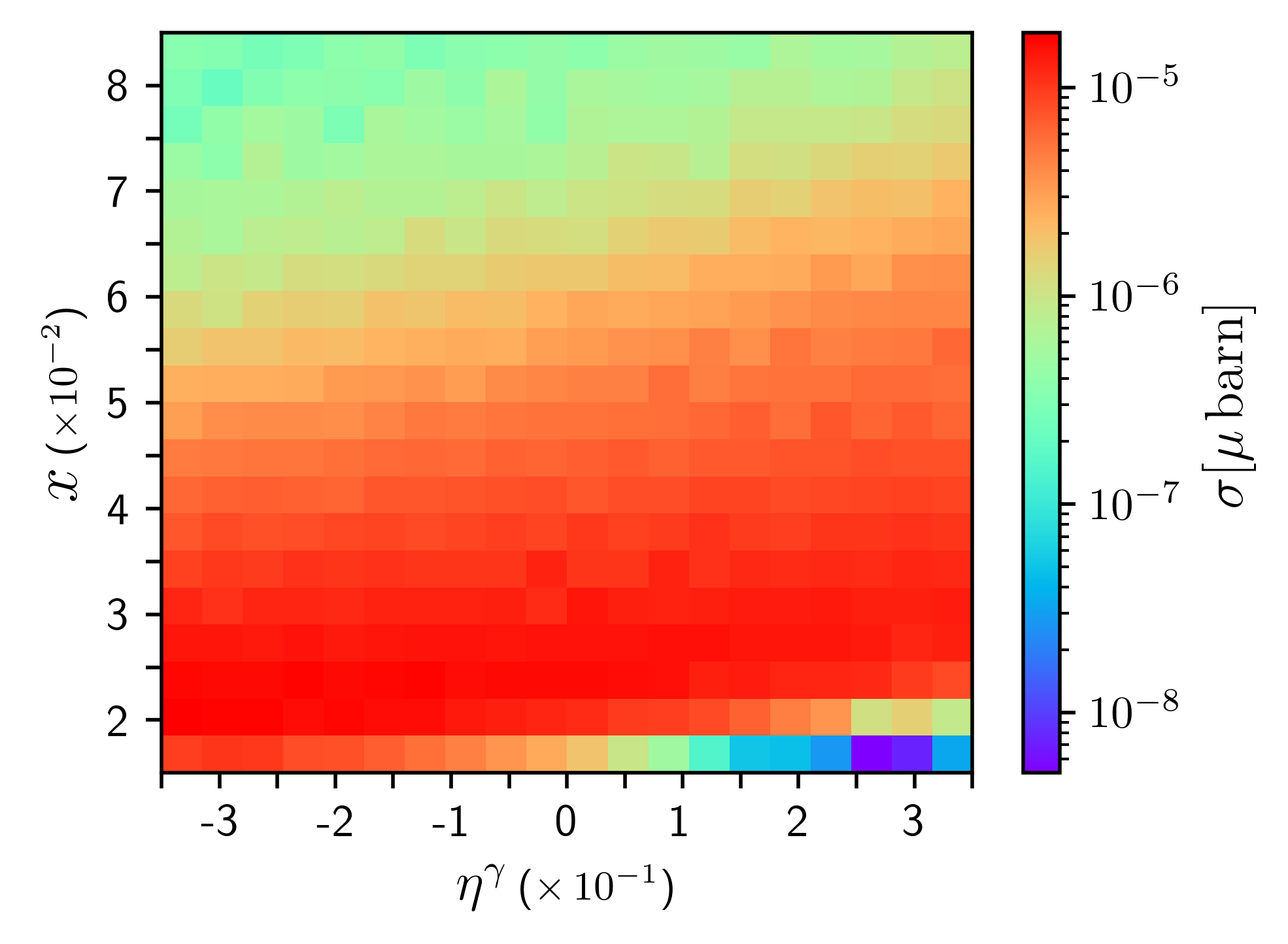}}
     \caption{The partonic momentum fraction $x$ as a function of the rapidity of the photon. The color scale shows the cross-section at LO QCD (left) and NLO QCD + LO QED (right) accuracy.
    \label{fig:xVSetaCORRELACION}}
\end{figure}

We start by considering the relation between $x=x_1$ and the transverse momentum of the particles in the final state. In Fig. \ref{fig:xVSptCORRELACION}, we present the correlation between $x_1$ and $p_T^\gamma$ (left column) and $p_T^\pi$ (right column). Each bin contains the corresponding integrated cross-section at LO QCD (upper row) and NLO QCD + LO QED (lower row) precision. Notice that the inclusion of higher-order corrections leads to a broadening of the patterns, originated by the presence of events in previously empty bins due to an extended phase-space. This is a general behaviour that also manifests when studying the correlations of other variables. Events with low $p_T^\gamma$ are associated with low $x_1$, and there is a somehow linear correlation between these variables. Events with low $p_T^\pi$ are mostly uniformly spread in the region of $x_1 \in [0.2, 0.6]$. This behaviour is expected from the fact that the photon is originated from the partonic event (its energy is directly related to the energy of the colliding partons), whilst the pion comes from an hadronization (which implies the convolution with the FF and the consequent spreading of the distributions).

\begin{figure}[h!]
    \centering
    \subfigure{\includegraphics[width=75mm,height=5.9cm]{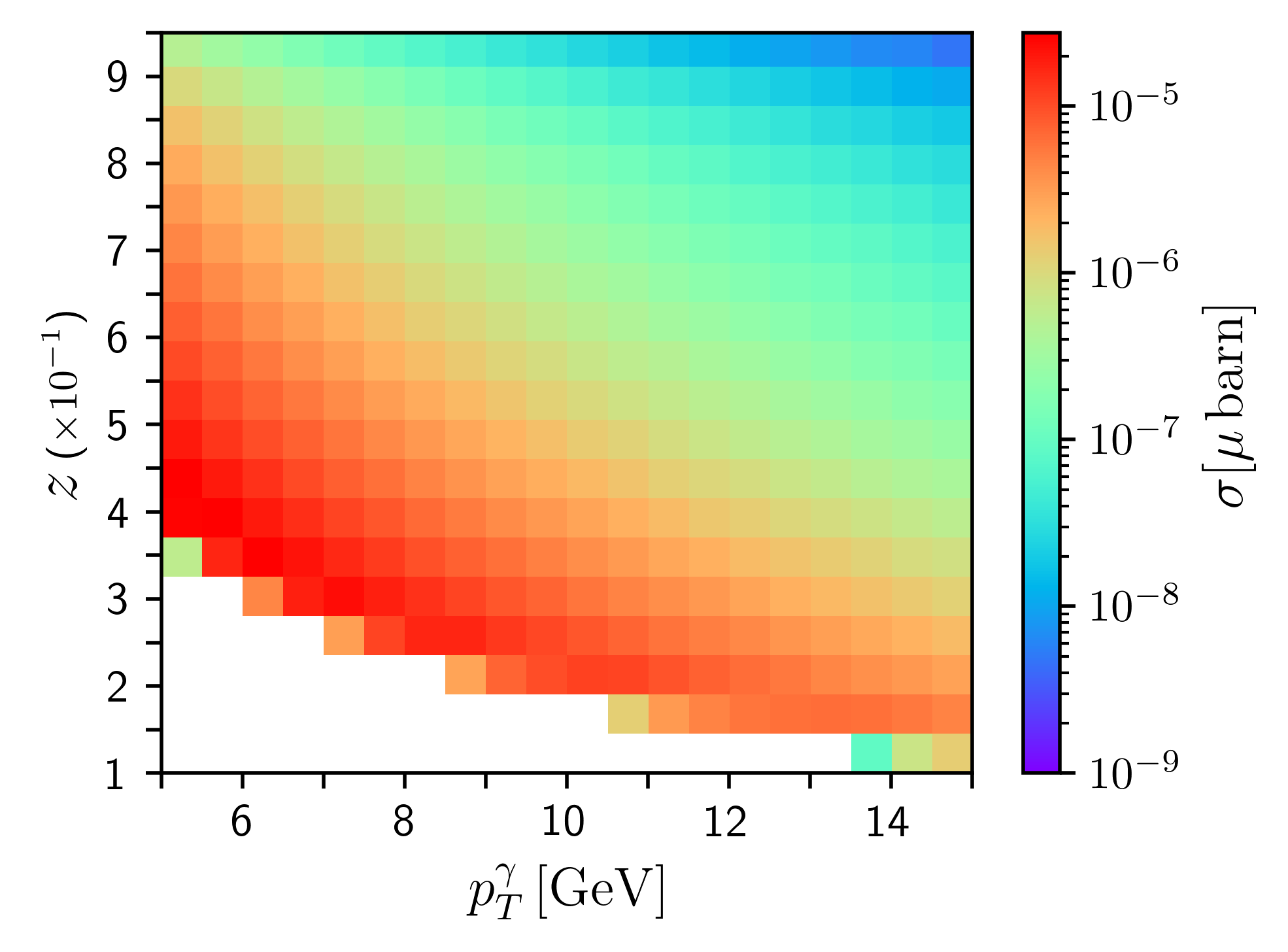}}
    \subfigure{\includegraphics[width=75mm,height=5.9cm]{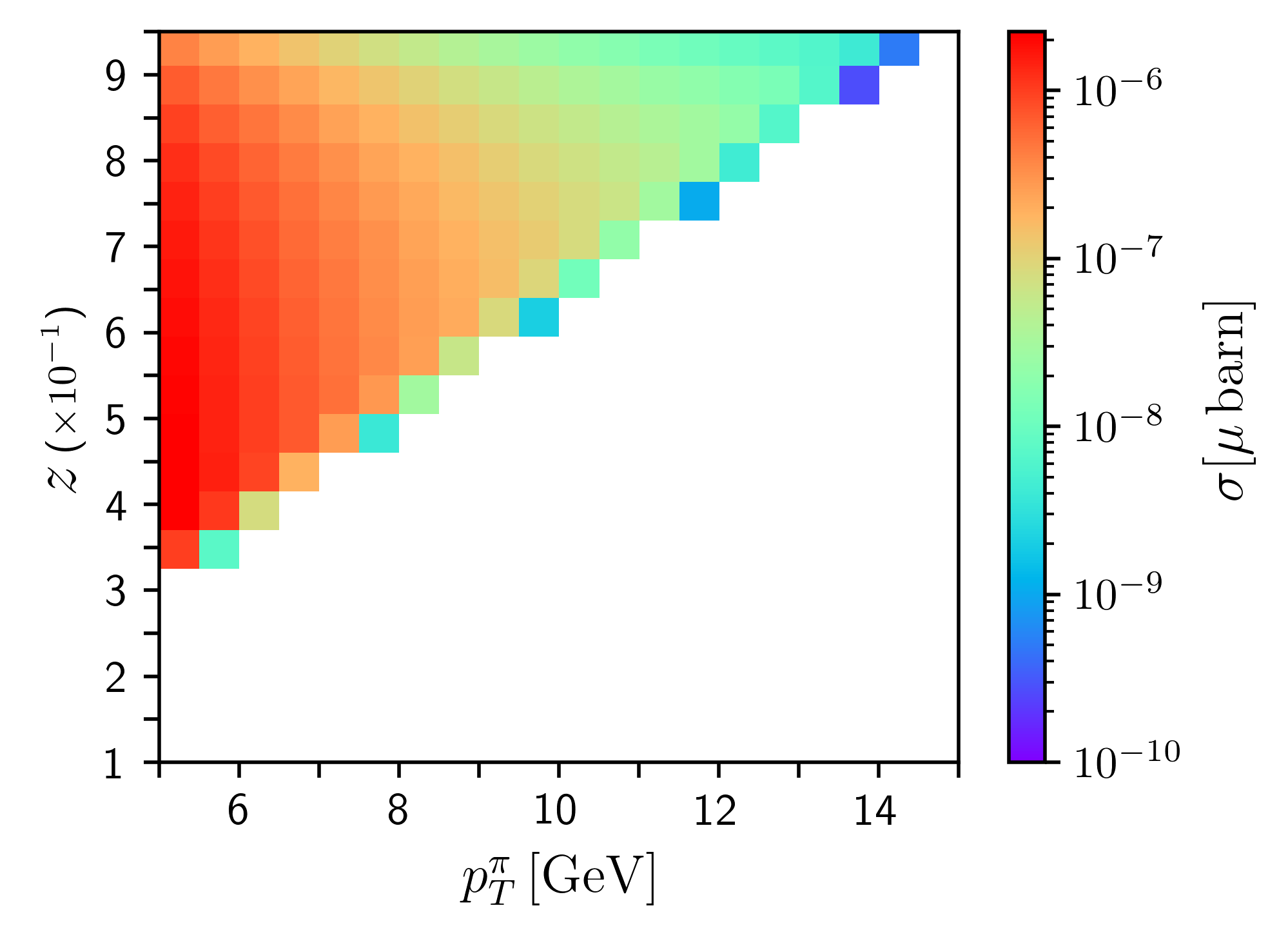}}
    \subfigure{\includegraphics[width=75mm,height=5.9cm]{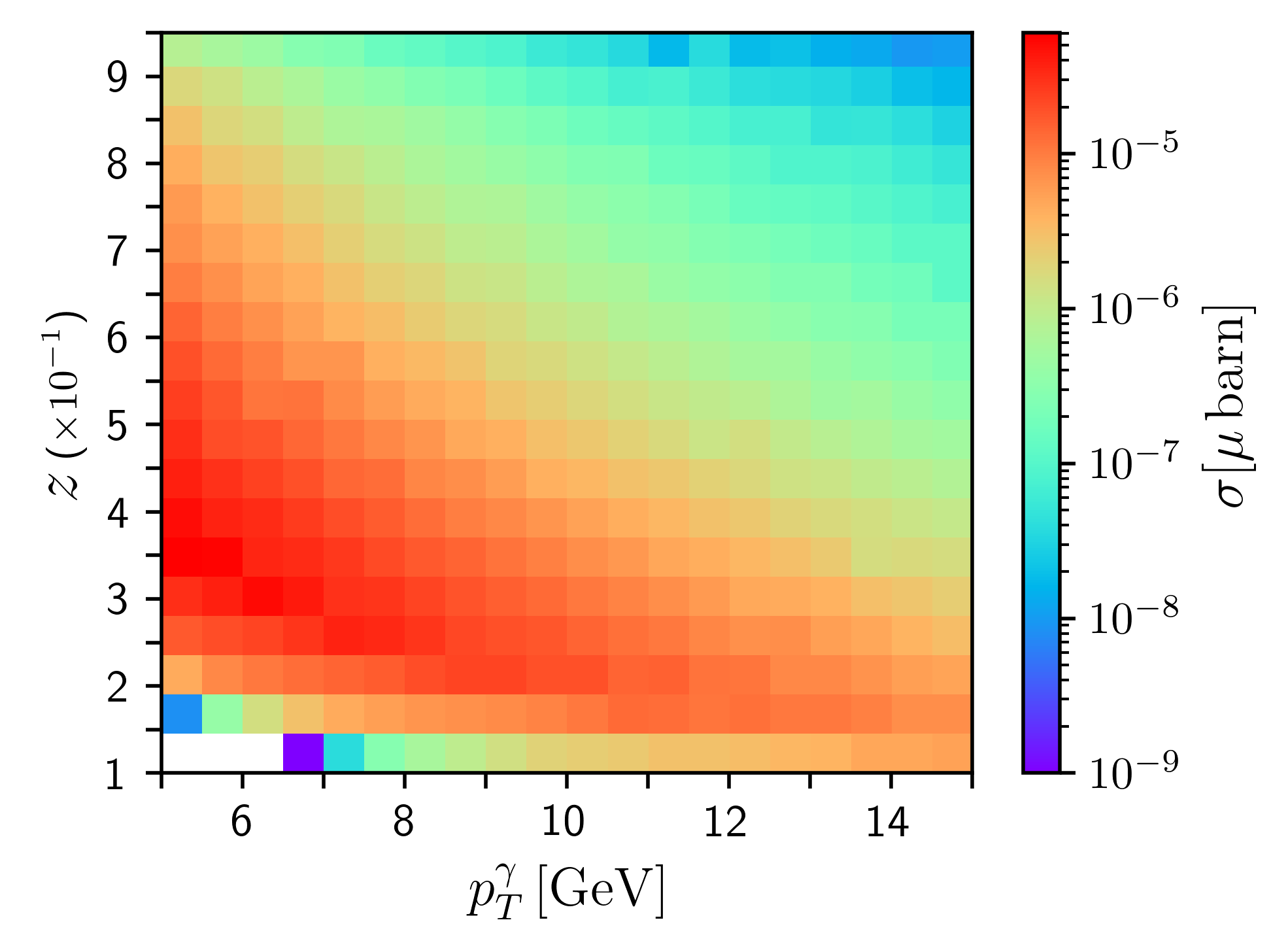}}
    \subfigure{\includegraphics[width=75mm,height=5.9cm]{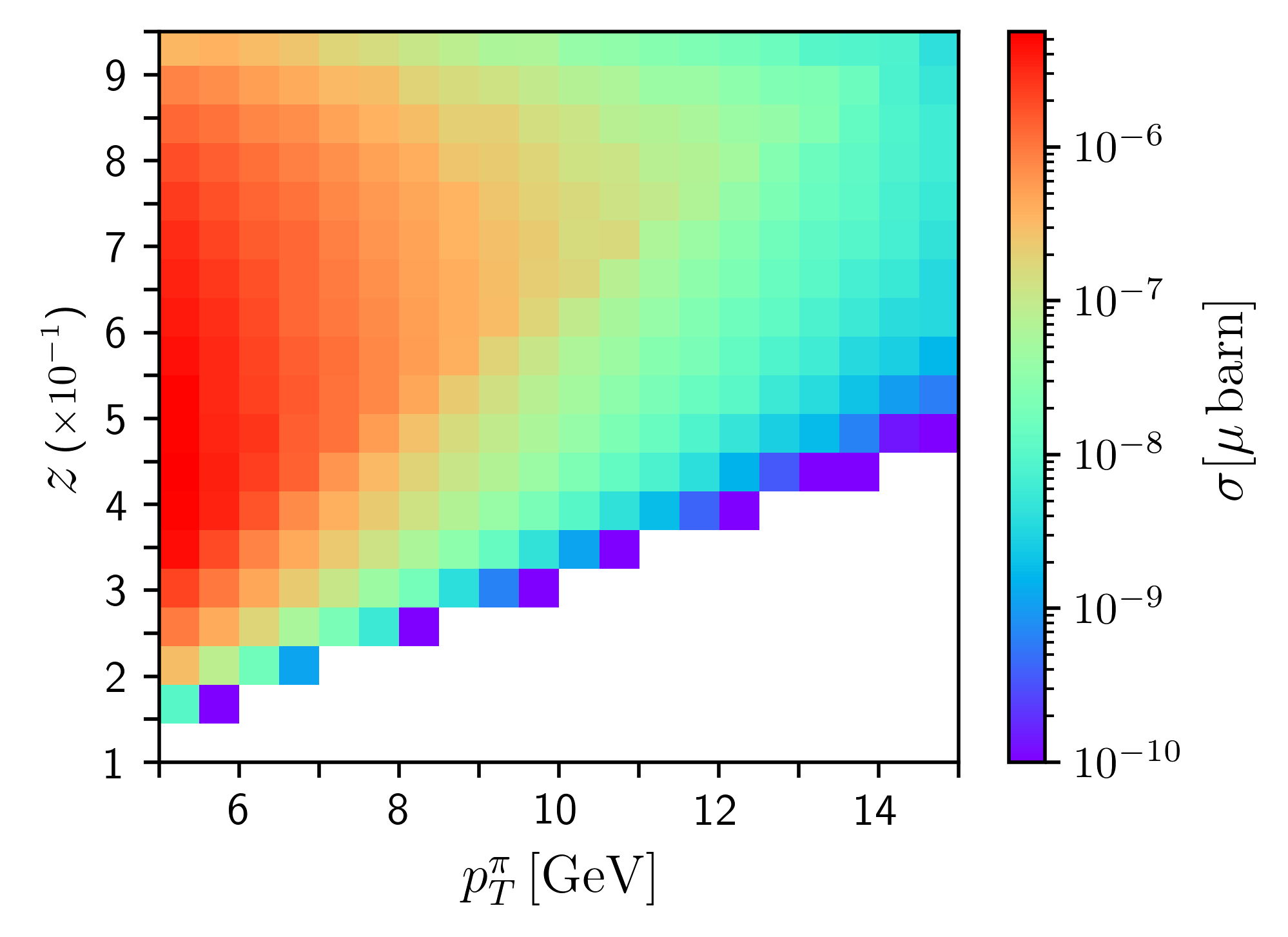}}
    \caption{Partonic momentum fraction $z$ as a function of $p_T^\gamma$ (left) and $p_T^\pi$ (right). The color scale shows the cross-section at LO QCD (upper row) and NLO QCD + LO QED (lower row).
    \label{fig:zVSptCORRELATION}}
\end{figure}

Next we move on to analyze the correlation between $x=x_1$ and the rapidities of the particles in the final state. It is important to highlight that the analysis here does depend on the momentum fraction being used, i.e. $x_1$ or $x_2$, since the rapidity introduces an asymmetry in the direction of the colliding particles. We show, in Fig. \ref{fig:xVSetaCORRELACION}, the plots of $x_1$ vs. $\eta^\gamma$ at LO QCD (left) and NLO QCD + LO QED (right), respectively. Similar results were found when considering $x_1$ vs. $\eta^\pi$ and are thus not presented here. Since the distributions are rather flat for $-0.3 \leq \eta \leq 0.3$, we find that most of the events are uniformly distributed for $x_1 \in [0.2,0.5]$. Finally, notice that below $x_1 \approx 0.2$, the cross-section falls steeply as a consequence of the imposed kinematical cuts, and the bins are empty.

\begin{figure}[h!]
    \centering
    \subfigure{\includegraphics[width=75mm,height=5.9cm]{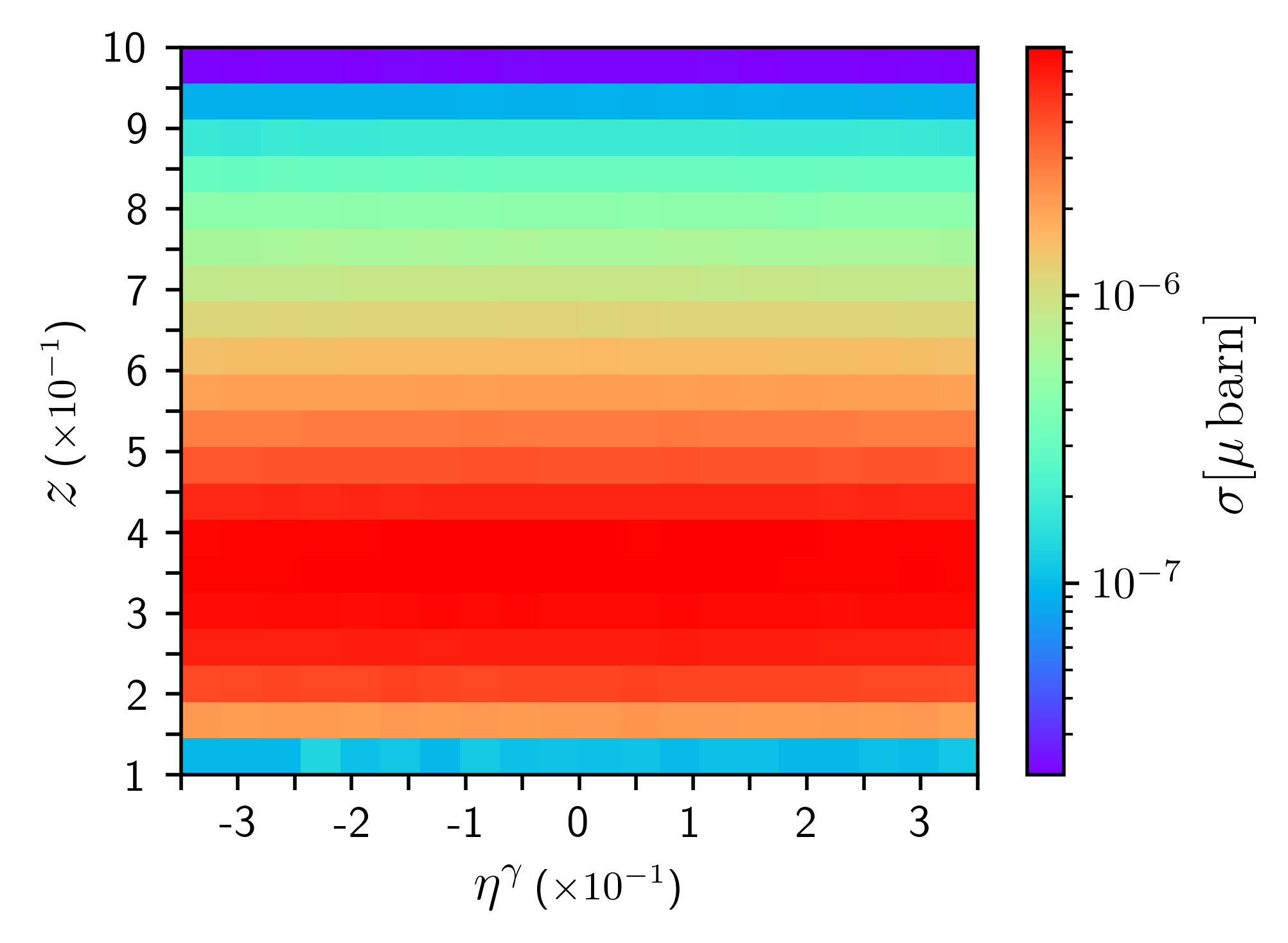}}
    \subfigure{\includegraphics[width=75mm,height=5.9cm]{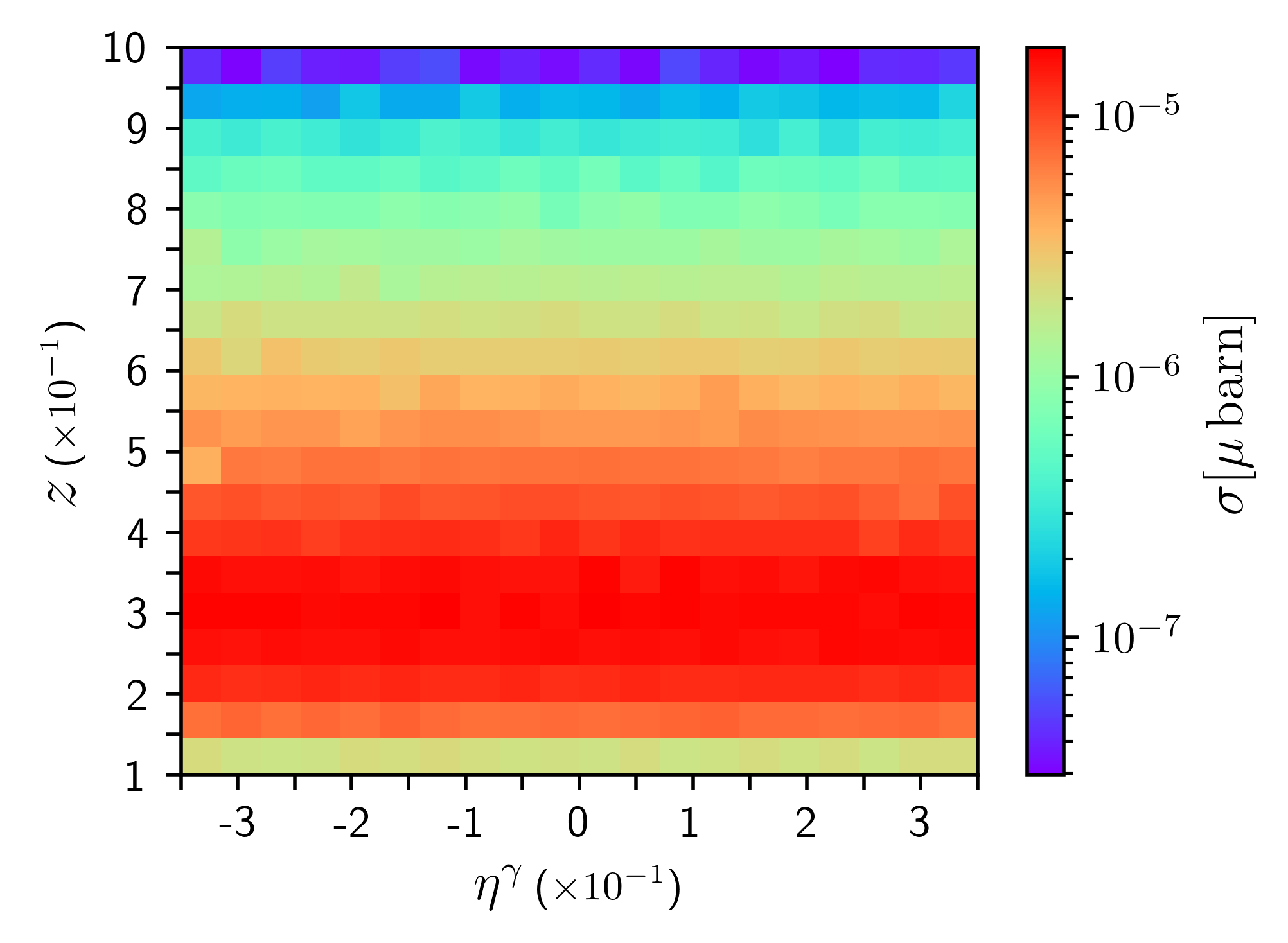}}
     \caption{Partonic momentum fraction $z$ as a function of the rapidity of the photon. The color scale shows the cross-section at LO QCD (left) and NLO QCD + LO QED (right) accuracy.
     \label{fig:zVSetaCORRELATION}}
\end{figure}

The analogous results on the $z$ dependence are presented in Fig. \ref{fig:zVSptCORRELATION}, the upper (lower) row corresponding to the LO QCD (NLO QCD + LO QED) contributions. On the left column we show the correlation between $z$ and $p_T^\gamma$, and between $z$ and $p_T^\pi$ on the right column. The former seems to be slightly negative, i.e. smaller values of $z$ tend to be favoured in events with higher $p_T^\gamma$, while the latter has a concentration of events in the low $p_T^\pi$ region with $z \geq 0.4$. Also, as expected, events with high $p_T^\pi$ require higher values of $z$ since the amount of partonic energy is limited by the cut $p_T^\gamma \leq 15$ GeV.

\begin{figure}[h!]
    \centering
    \subfigure{\includegraphics[width=75mm,height=5.9cm]{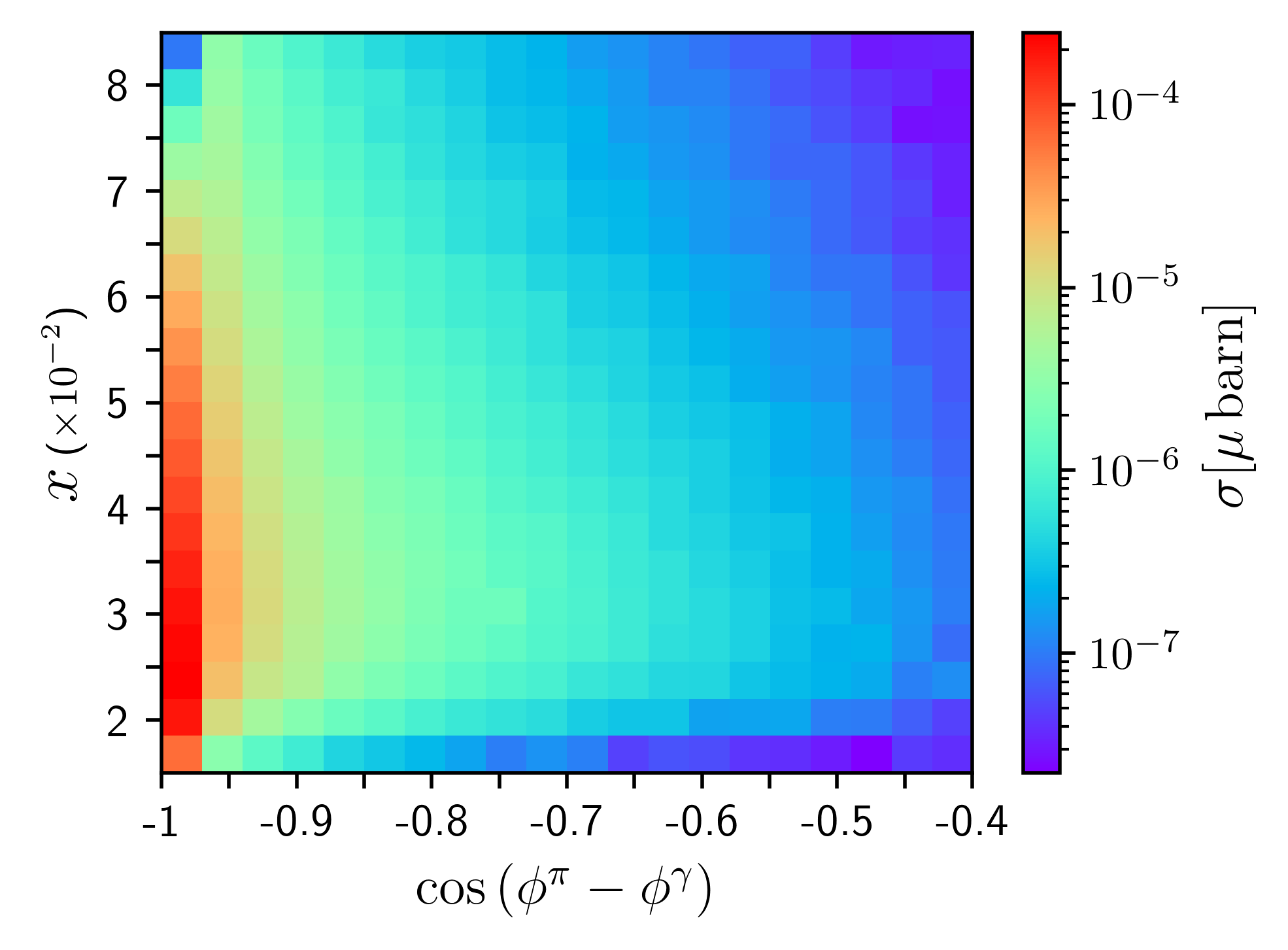}}
    \subfigure{\includegraphics[width=75mm,height=5.9cm]{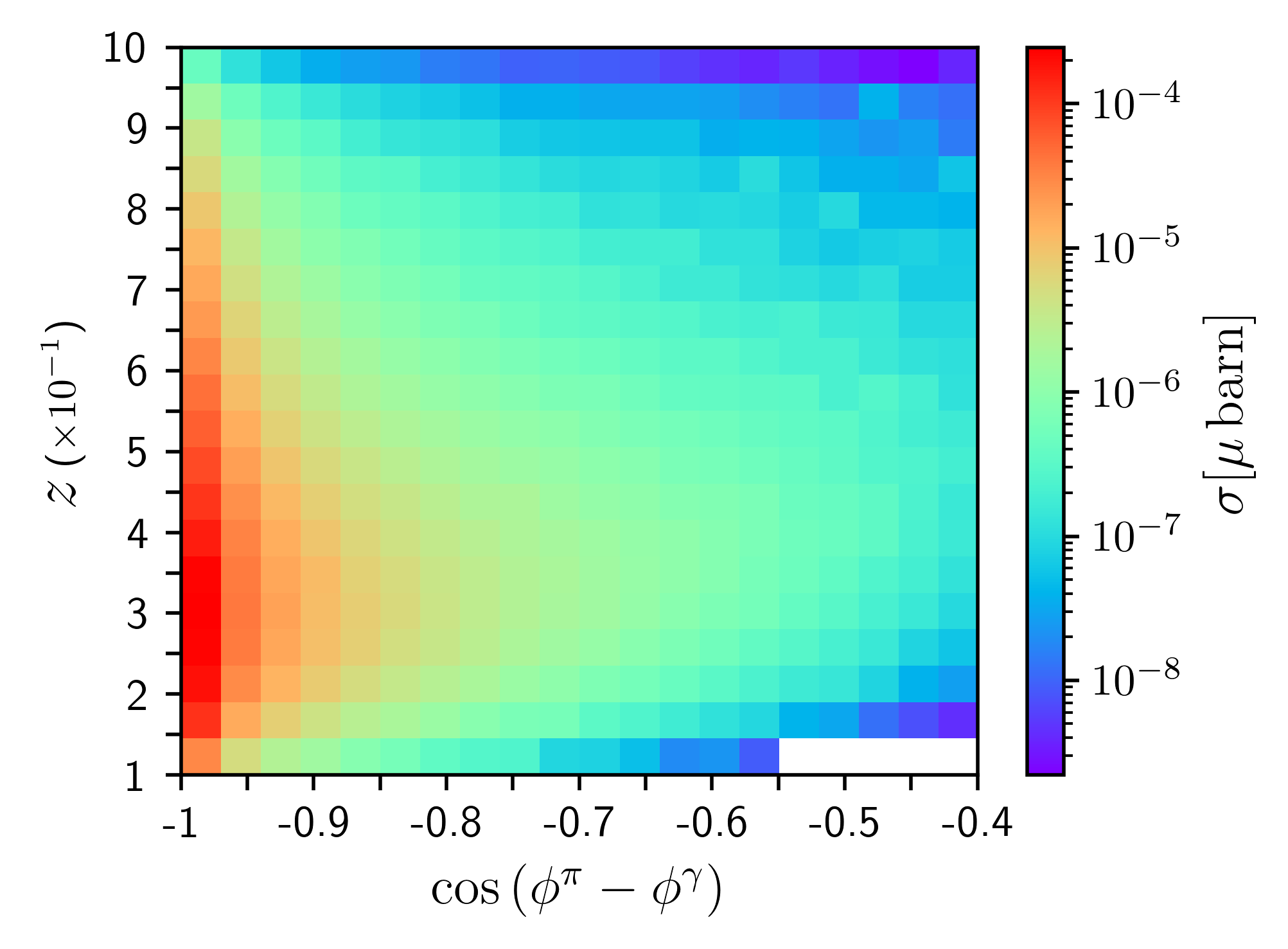}}
     \caption{Partonic momentum fractions $x_1$ (left) and $z$ (right) as a function of $\cos{(\phi^\pi-\phi^\gamma)}$. The color scale shows the integrated cross-section value per pixel with NLO QCD + LO QED accuracy.
     \label{fig:xzVScosCORRELATION} }
\end{figure}

The correlation between $z$ and the rapidities of the final state particles shows a rather flat dependence on $\eta$, as depicted in Fig. \ref{fig:zVSetaCORRELATION} for the case of $\eta^\gamma$ (similar plots were obtained when considering $\eta^\pi$).

Then, let us consider the correlations with the azimuthal variable $\cos{(\phi^\pi-\phi^\gamma)}$ in Fig. \ref{fig:xzVScosCORRELATION}. Of course, the contributions associated to the Born kinematics are restricted to the first bin because $\cos{(\phi^\pi-\phi^\gamma)}=-1$ (i.e. the pion and the photon are produced back-to-back). The remaining bins are heavily suppressed, since they only receive contributions from the real radiation. We see that the events are strongly concentrated in the medium and low-$x$ region without a clear trend or dependence w.r.t. $\cos{(\phi^\pi-\phi^\gamma)}$. For $z$, the distribution spreads over more bins, and there is a subtle trend to favour events with a bigger azimuthal separation (smaller values of $-\cos{(\phi^\pi-\phi^\gamma)}$) and slightly lower values of $z$.

\begin{figure}[h!]
    \centering
    \subfigure{\includegraphics[width=75mm,height=5.9cm]{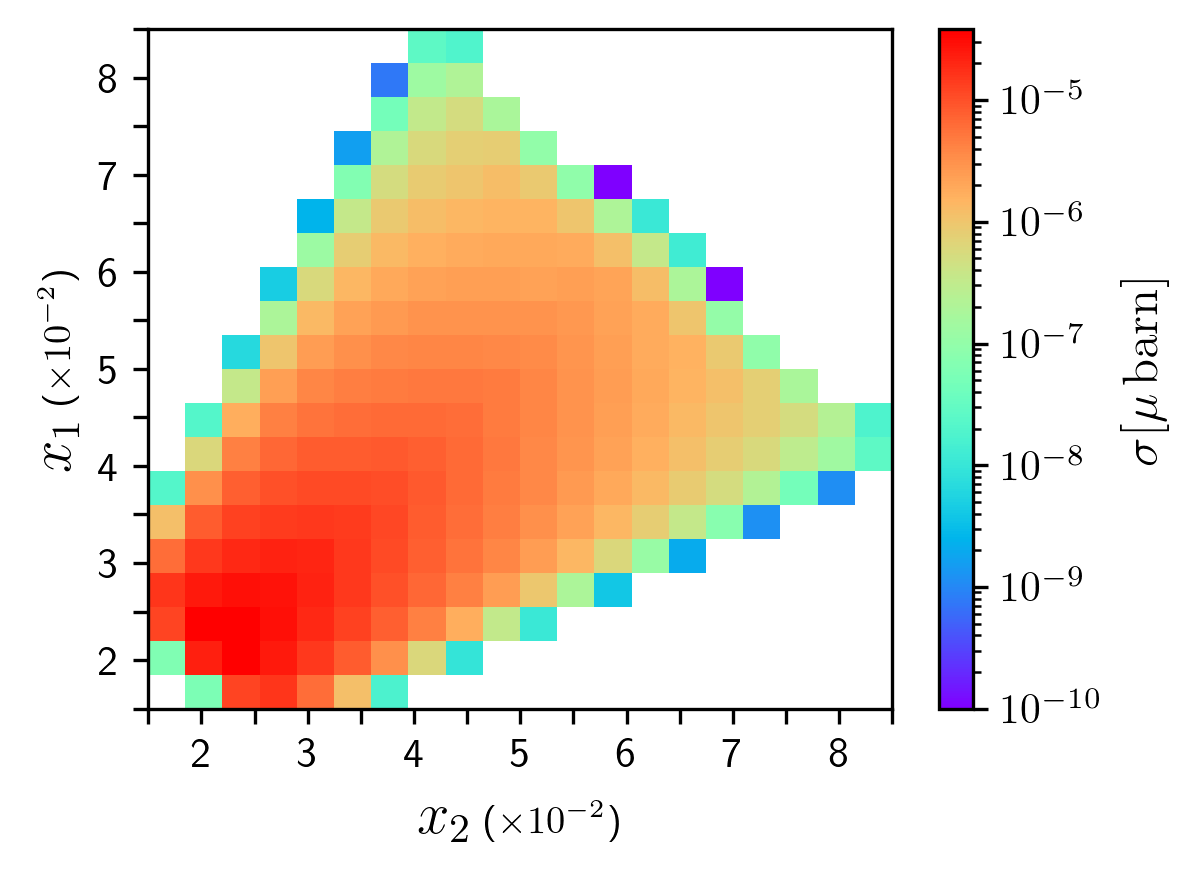}}
    \subfigure{\includegraphics[width=75mm,height=5.9cm]{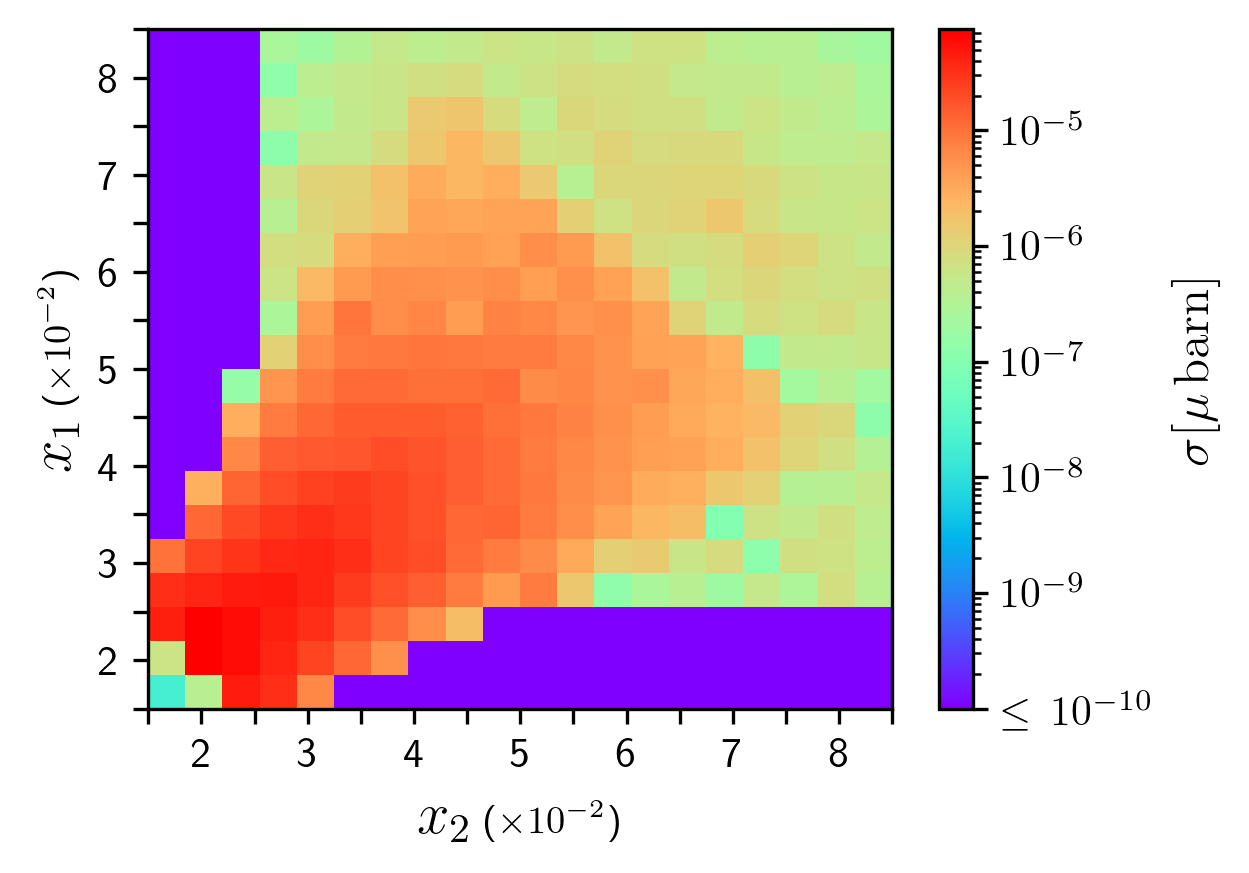}}
    \caption{Correlation between $x_1$ and $x_2$ at LO QCD (left) and NLO QCD + LO QED (right) accuracy. Even when including NLO real corrections, there is a strong suppression for those events with rather different values of $x_1$ and $x_2$.}
    \label{fig:x1VSx2CORRELATION}
\end{figure}

Finally, we analyze the correlation between $x_1$ and $x_2$ for $p+p$ collisions. In Fig. \ref{fig:x1VSx2CORRELATION}, we show the correlation plots at LO QCD (left) and NLO QCD + LO QED (right) accuracy, for RHIC kinematics. As expected, there is a compact region containing events at LO, reflecting the kinematical constraints of a $2 \to 2$ process. The events are concentrated in the low-$x$ region and show a strong positive linear correlation between $x_1$ and $x_2$: this reflects the fact that it is more probable to have events in the back-to-back region, in agreement with Fig. \ref{fig:xzVScosCORRELATION}. When introducing higher-order corrections, the real emission phase-space gets enlarged and the distributions are spread. In any case, the positive correlation between $x_1$ and $x_2$ remains, with an strong concentration of events in the middle and low-$x$ region. Also, it is worth appreciating that the NLO real corrections are not enough to enhance the number of events with rather different values of $x_1$ and $x_2$. This is, in part, a consequence of the kinematical cuts that favour central events rather than highly boosted ones.

To conclude this section, let us comment on the importance of the study of correlations. Since we want to reconstruct the partonic momentum fractions by using the measurable variables, it is important to know which ones are the most relevant. From the previous discussion, we expect that $x$ strongly depends on $p_T^\gamma$ (positive correlation) but not on the other variables. Analogously, $z$ exhibits a negative correlation with $p_T^\gamma$, a positive one with $p_T^\pi$ and a slight dependence on $-\cos{(\phi^\pi-\phi^\gamma)}$. This knowledge will be applied to the construction of a basis of functions for determining $x$ and $z$ in the next section. 

\section{Reconstruction of parton kinematics}
\label{sec:Reconstruction}
We now focus on our main goal, which is to determine the partonic variables $x_1$, $x_2$ and $z$ in terms of the measured momenta of the final state particles. At LO this is fully determined by energy-momentum conservation, and thus the LO case will serve as control. The real challenge appears at NLO, where real emissions prevent a straightforward determination of closed analytic formulae: this is what we will attempt to approximate using ML \footnote{Doing a formal description of the ML methods that we used is beyond the scope of this work, and would take much more than a simple article. Moreover much literature is available on the topic (see e.g. \cite{Murphy:2012}), so we will leave out such a discussion and mention just a few basic concepts needed in the rest of the section.}. 

In supervised ML, we have an initial set of data (the \emph{training set}) and we want to map it into another known set (the \emph{target}). Each entry in the training set is a vector of dimension $d$, with $d$ the number of variables (\emph{features}) that the target depends upon. We also assume that there is some underlying function, the so-called \emph{target} function, that connects the two; the task of a ML algorithm is to find a good estimation of this function. This estimator, in turn, depends on a set of parameters that is determined by minimising a function (the \emph{cost} function) that measures some distance between the prediction of the estimator and the actual targets. As a last step, one takes another set of data with corresponding labels (\emph{test} set) and compares how well the estimator does over it. To prevent the estimator from performing well over the training data set but poorly over the test set (\emph{overfitting}), the cost function includes also some parameters to control the trade off between a low training cost and a low test cost. The total number of \emph{regularization} parameters depend on the specific method used, and the optimal value/s have to be found by picking the one/s that minimize the test cost function. 

Armed with these basic concepts, we first discuss the generation of our input and target sets using the outputs of our MC code. After that, we present results obtained through the application of supervised ML for estimating $x \equiv x_1$ and $z$ at LO QCD and NLO QCD + LO QED accuracy. For the purpose of the present analysis, we explore three models: a Linear Model (LM), a Gaussian Regression (GR) and the Multi-Layer Perceptron (MLP) algorithm based on neural networks. These models have been implemented in Python using the \texttt{scikit-learn} library \cite{scikit-learn}.


\subsection{Construction of the training data sets}
\label{ssec:selection}
The training and test sets were generated with the MC code used and described in the previous sections. As was mentioned already, it deals independently with each term of the computation (LO, NLO real radiation, NLO virtual terms, NLO counter-terms). This poses two difficulties when generating the training set for feeding the ML algorithms. On the one hand, only the LO calculations are finite on their own; for the NLO cross-section, we have to combine all terms (real, virtual and counter-terms) to have a meaningful finite quantity. On the other hand, by the same nature of the MC integration, no two identical points are generated in the sampling, which in turn spoils the fully local cancellation of the divergences. Instead, one has to split the different variables into bins and sum over all contributions entering each of them. If a sufficient number of points is sampled, the divergences cancel and we obtain the finite cross-section per bin. This is a common feature of MC integration, and many codes provide routines that take care of this for one-dimensional binning. In our case we are interested in a more differential observable, so that we had to generate a large number of points to meet this condition. Moreover, not all sampled points pass the selection cuts, e.g. from the $10^9$ points sampled we retain $\approx 30\%$ at LO. 

For the LO we can directly use the generated points, but for the NLO case we need to discretize the differential cross-section w.r.t. the external kinematical variables defined in Eq. (\ref{eq:VARIABLES}). For this purpose, we create a five-dimensional grid by binning the variables in ${\cal V}_{\rm Exp}$. Explicitly, we define 10 bins for $p_T^{\gamma}$ and $p_T^{\pi}$, 5 bins for $\eta^{\gamma}$ and $\eta^{\pi}$, and 6 bins for $\cos(\phi^{\pi}-\phi^{\gamma})$. The set of discretized experimentally-measurable variables is denoted as  
\beq
{\bar{\cal V}}_{\rm Exp} = \{{\bar p}_T^\gamma,{\bar p}_T^\pi,{\bar\eta}^\gamma,{\bar\eta}^\pi,{\overline{\cos}}(\phi^\pi-\phi^\gamma)\} \, ,
\label{eq:VARIABLESpromedio}
\eeq
where $\bar{a}$ denotes the mean value of the variable $a$ in a given bin. In total ${\bar{\cal V}}_{\rm Exp}$ contains 15000 bins. Then, we define the cross-section per bin according to
\beqn
\nn \sigma_j({\bar p}_T^\gamma,{\bar p}_T^\pi,{\bar\eta}^\gamma,{\bar\eta}^\pi,{\overline{\cos}}(\phi^\pi-\phi^\gamma)) &=& \int_{(p_T^\gamma)_{j,\rm MIN}}^{(p_T^\gamma)_{j,\rm MAX}} \, dp_T^\gamma \, \int_{(p_T^\pi)_{j,\rm MIN}}^{(p_T^\pi)_{j,\rm MAX}} \, dp_T^\pi \, \ldots 
\\ &\times& \, \int dx_1 dx_2 dz \, d\bar\sigma \, ,
\label{eq:xSECTIONperbin}
\eeqn
with $x_{j,\rm MIN}$ ($x_{j,\rm MAX}$) the minimum (maximum) value of the variable $x$ in the $j$-th bin, $\bar{x}$ the corresponding average of $x$ over the $j$-th bin and 
\beq
d\bar\sigma = \frac{d\sigma}{d{\cal V}_{\rm Exp} \,dx_1 dx_2 dz} \, 
\eeq
is the fully-differential \emph{hadronic} cross-section as a function of the partonic momentum fractions and the experimentally-measurable variables ${\cal V}_{\rm Exp}$. At LO, $\sigma_j$ can be straightforwardly calculated since we only need to integrate the tree-level scattering amplitude in a $2 \to 2$ phase-space. However, as we explained in Sec. \ref{sec:ComputationalSetup}, the NLO corrections include several contributions calculated with different kinematics (virtual, real, counter-terms): all of these are taken into account in $d\bar\sigma$ and integrated over their corresponding phase-space to obtain $\sigma_j$~\footnote{It is worth appreciating that binning could be avoided using a fully-local framework for computing higher-order corrections \cite{Gnendiger:2017pys,TorresBobadilla:2020ekr}. One of these methods is the Four-Dimensiona Unsubtraction (FDU) \cite{Hernandez-Pinto:2015ysa,Sborlini:2016fcj,Sborlini:2016gbr,Sborlini:2016hat} based on the Loop-Tree Duality \cite{Catani:2008xa,Rodrigo:2008fp,deJesusAguilera-Verdugo:2021mvg}. Since FDU leads to a fully-differential and finite representation of the cross-section, it constitutes a perfectly suited candidate to improve the efficiency of the analysis presented in this article.}. 

Once the grid and the discretized cross-section are defined, we use the MC code to generate three histograms per each bin in the grid. These histograms corresponds to the distributions $d\sigma_j / dx_1$, $d\sigma_j / dx_2$ and $d\sigma_j / dz$, respectively. So, given a point in the grid 
\beq
p_j = \{ {\bar p}_T^\gamma,{\bar p}_T^\pi,{\bar\eta}^\gamma,{\bar\eta}^\pi,{\overline{\cos}}(\phi^\pi-\phi^\gamma) \} \in {\bar{\cal V}}_{\rm Exp} \, , 
\label{eq:pjDEF}
\eeq
we can define
\beqn 
(x_1)_j &=& \sum_i \, (x_1)_i \frac{d \sigma_j}{d x_1} (p_j;(x_1)_i) \, ,
\label{eq:x1BINj}
\\ (x_2)_j &=& \sum_i \, (x_1)_i \frac{d \sigma_j}{d x_2} (p_j;(x_2)_i) \, ,
\label{eq:x2BINj}
\\ (z)_j &=& \sum_i \, z_i \frac{d \sigma_j}{d z} (p_j;z_i) \, ,
\label{eq:zBINj}
\eeqn
which correspond to the weighted average of the partonic momentum fractions extracted from the histograms generated with the MC code.

At this stage, we can identify ${\bar{\cal V}}_{\rm Exp}$ as the training set and $\{(x_1)_j,(x_2)_j,(z)_j\}$ as the target one. Then, we can train the ML algorithms to find the target functions
\beqn
X_{1,\rm REC} &:=& {\bar{\cal V}}_{\rm Exp} \longrightarrow \bar{X}_{1,REAL} = \{(x_1)_j\} \, ,
\\ X_{2,\rm REC} &:=& {\bar{\cal V}}_{\rm Exp} \longrightarrow \bar{X}_{2,REAL} = \{(x_2)_j\} \, ,
\\ Z_{\rm REC} &:=& {\bar{\cal V}}_{\rm Exp} \longrightarrow \bar{Z}_{REAL} = \{(z)_j\} \, ,
\eeqn
that will allow us to reconstruct the MC partonic momentum fractions $\bar{X}_{1,REAL}$, $\bar{X}_{2,REAL}$ and $\bar{Z}_{REAL}$.

To conclude this discussion, notice that the definitions given in Eqs. (\ref{eq:x1BINj})-(\ref{eq:zBINj}) are crucial beyond LO. In fact, for a $2 \to 2$ process, fixing the bin $p_j \in {\bar{\cal V}}_{\rm Exp}$ leads to a unique value of the partonic-momentum fractions. Explicitly, we have
\begin{align}
X_{1,\rm REC} &=\frac{p_T^{\gamma}\exp(\eta^{\pi})+p_T^{\gamma}\exp(\eta^{\gamma})}{\sqrt{S_{CM}}} \, , \label{eq:VARIABLESLOX1}  \\
X_{2,\rm REC} &=\frac{p_T^{\gamma}\exp(-\eta^{\pi})+p_T^{\gamma}\exp(-\eta^{\gamma})}{\sqrt{S_{CM}}} \, ,\label{eq:VARIABLESLOX2}  \\
Z_{\rm REC} &=\frac{p_{T}^{\pi}} {p_T^{\gamma}}   \, ,
\label{eq:VARIABLESLOZ}
\end{align}
as explained in Ref. \cite{deFlorian:2010vy}. Due to the presence of $2 \to 3$ sub-processes contributing to the real radiation, the value of $\{x_1,x_2,z\}$ for a given $p_j$ is not unambiguously defined at NLO (and beyond). If we pick up an event with a fixed $p_j$ from our NLO MC generator, the real partonic momentum fractions might take all the kinematically-allowed values. However, the probability of the different outcomes is given by the differential-cross section of the event, which motivates the definitions introduced in Eqs. (\ref{eq:x1BINj})-(\ref{eq:zBINj}). In the following, we explain how these data sets are used with the different ML frameworks.


\subsection{Linear regression}
\label{ssec:LinReg}
Linear methods, as the name indicates, provide the estimation of the target function as a linear combination of the input set. However, the linearity occurs at the level of the parameters and one can apply prior knowledge to construct new features upon which the target dependence is simpler. Choosing a \emph{good} set of features (basis) plays an important role to achieve an accurate reconstruction.

\begin{figure}[h!]
    \centering
    \subfigure{\includegraphics[width=75mm,height=5.9cm]{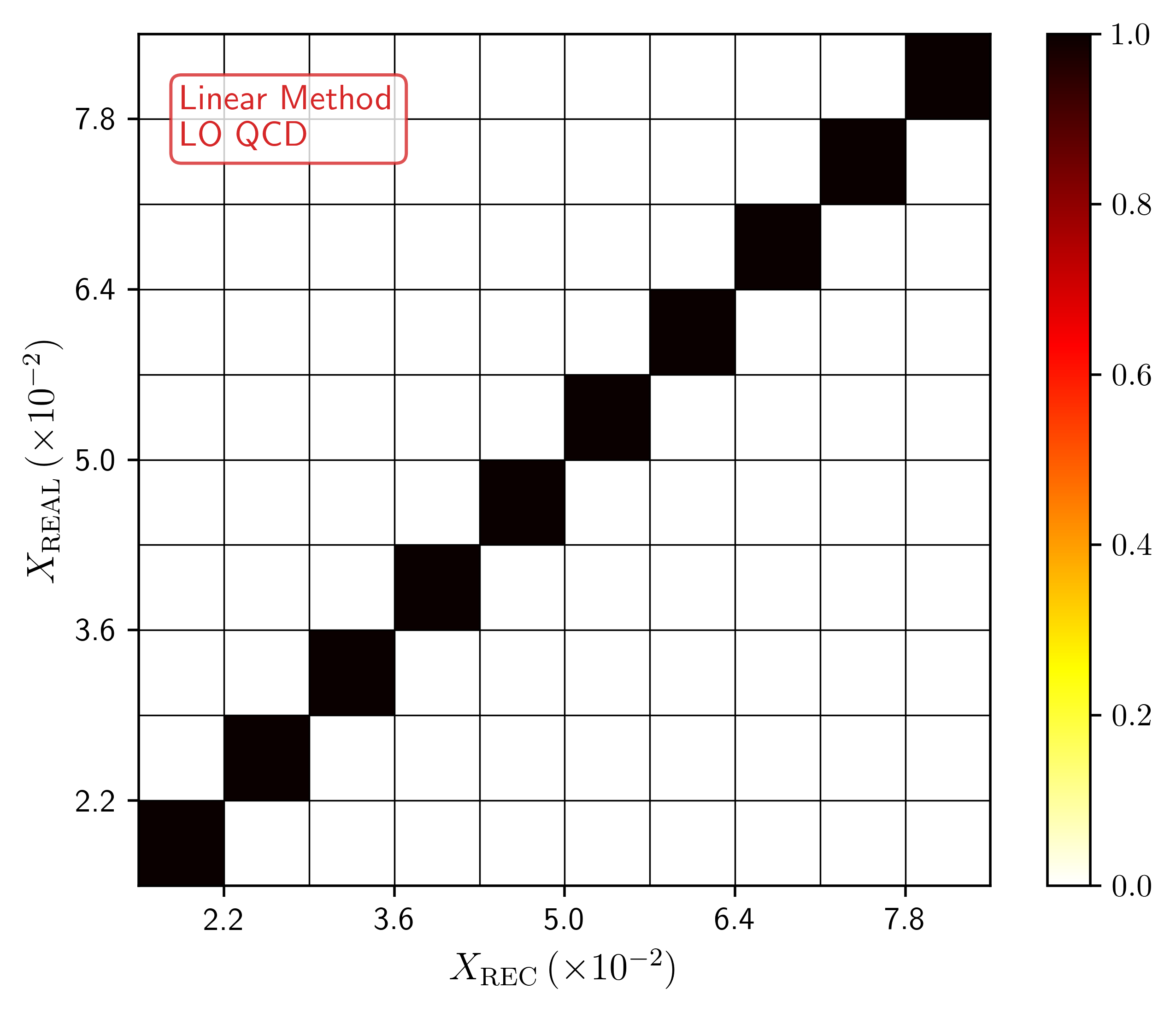}}
    \subfigure{\includegraphics[width=75mm,height=5.9cm]{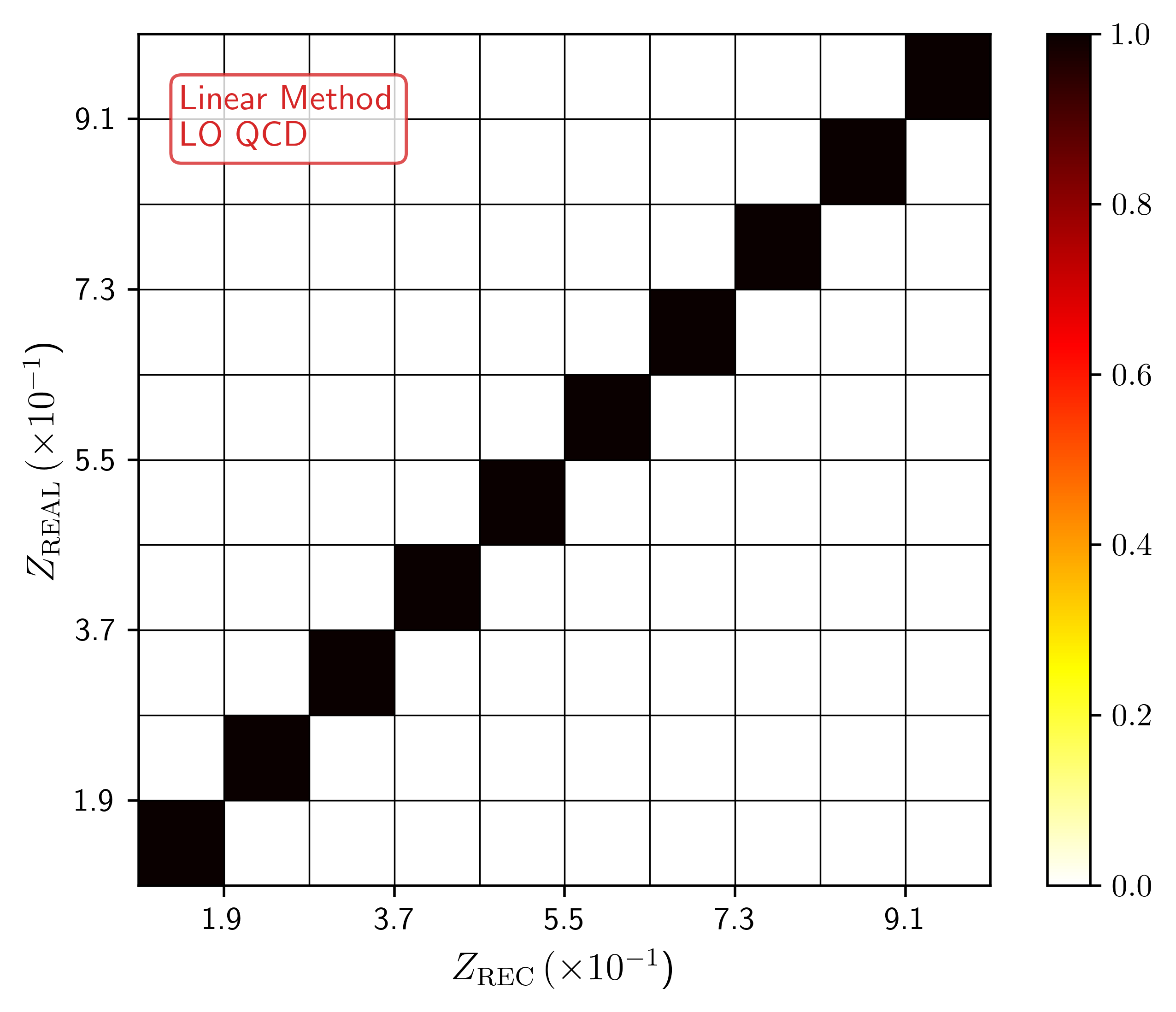}}
    \caption{Correlation between the  MC momentum fractions (i.e. $X_{\rm REAL}$ and $Z_{\rm REAL}$) versus the ones obtained at LO QCD accuracy using the LM approach. Each bin of the correlation plot is filled with the integrated cross-section.}
    \label{fig:LMreconstructionLO}
\end{figure}

For example, at LO we take inspiration from the exact analytical expressions given by Eqs. (\ref{eq:VARIABLESLOX1})-(\ref{eq:VARIABLESLOZ}) and propose the basis
\beq
{\cal B}_{\rm LO}=\{ \frac{p_T^{\gamma}}{\sqrt{S_{CM}}}\exp(\eta^{\pi}), \, \frac{p_T^{\gamma}}{\sqrt{S_{CM}}}\exp(\eta^{\gamma}), \, \frac{p_T^{\gamma}}{\sqrt{S_{CM}}}\exp(-\eta^{\pi}), \, \frac{p_T^{\gamma}}{\sqrt{S_{CM}}}\exp(-\eta^{\gamma}), \,
p_T^{\pi}/p_T^{\gamma}\, \} .
\eeq
We then expect $x_1$ to be well reconstructed by a linear combination of the first two elements of the basis (with coefficient $1$), whilst $z$ should be mainly proportional to the last element. In Fig. \ref{fig:LMreconstructionLO}, we show the correlation between the MC partonic momentum fractions (vertical axis) and the output of the linear regression (horizontal axis). Each bin contains the integrated cross-section at LO QCD accuracy. We can appreciate that the reconstruction is perfect, and the LM approach leads exactly to the Eqs. (\ref{eq:VARIABLESLOX1})-(\ref{eq:VARIABLESLOZ}).

When dealing with the NLO scenario, in principle, it should be expected an enlargement of the basis. The elements of ${\cal B}_{\rm LO}$ are not enough to fully capture the additional dependencies introduced by the NLO real kinematics. In fact, in Ref. \cite{deFlorian:2010vy} the authors proposed
\beqn
X_{1,\rm REC} &=&\frac{p_T^{\gamma}\exp(\eta^{\pi})-\cos(\phi^\pi-\phi^\gamma)\, p_T^{\gamma}\exp(\eta^{\gamma})}{\sqrt{S_{CM}}} \, , \label{eq:RECONSTRCCIONdefloX1}  
\\ X_{2,\rm REC} &=&\frac{p_T^{\gamma}\exp(-\eta^{\pi})-\cos(\phi^\pi-\phi^\gamma)\, p_T^{\gamma}\exp(-\eta^{\gamma})}{\sqrt{S_{CM}}} \, ,\label{eq:RECONSTRCCIONdefloX2}  
\\ Z_{\rm REC} &=& - \cos(\phi^\pi-\phi^\gamma)\, \frac{p_T^\pi}{p_T^\gamma} \, ,
\label{eq:RECONSTRCCIONdefloZ}
\eeqn
that agree with Eqs. (\ref{eq:VARIABLESLOX1})-(\ref{eq:VARIABLESLOZ}) at LO, but introduce an additional dependence on the azimuthal variables at higher-orders. The study of correlations performed at NLO QCD accuracy using these expressions showed a good reconstruction of the MC partonic momentum fractions.
 
With this precedent in mind, we propose here to include additional functional dependencies to have a more flexible reconstruction. We start by defining a primitive set of functions
\beq
{\cal K} = \{\frac{p_T^\gamma}{\sqrt{S_{CM}}},\frac{p_T^\pi}{\sqrt{S_{CM}}},\exp(\eta^\gamma),\exp(\eta^\pi),\cos(\phi^\pi-\phi^\gamma)\} \, ,
\eeq
in such a way that the reconstructed variables take the form
\beqn
Y_{\rm REC} &=& \sum_{i=1,i \neq 5}^9 (a^Y_i+ b^Y_i \, {\cal K}_5) \, {\cal K}_i + \sum_{i \leq j, \{i,j\} \neq 5, j-i \neq 5} \, (c^Y_{ij}+d^Y_{ij}\, {\cal K}_5) \, {\cal K}_i {\cal K}_j \, ,
\label{eq:LMgeneralNLO}
\eeqn
with $Y_{\rm REC}=\{X_{1,\rm REC},X_{2,\rm REC},Z_{\rm REC}\}$ and ${\cal K}_{i}={\cal K}_{i-5}^{-1}$ for $i= \{ 6,7,8,9 \}$. The ansatz proposed in Eq. (\ref{eq:LMgeneralNLO}) generalizes the basis ${\cal B}_{\rm LO}$ and includes products of up to three kinematical variables, which gives more flexibility to fit the data. In total, there are eighty-one functions in the basis, that we denominate \emph{general basis}. However, as we will now explicitly see, a larger basis does not imply a better reconstruction.

\begin{figure}[h!]
    \centering
    \subfigure{\includegraphics[width=75mm,height=5.9cm]{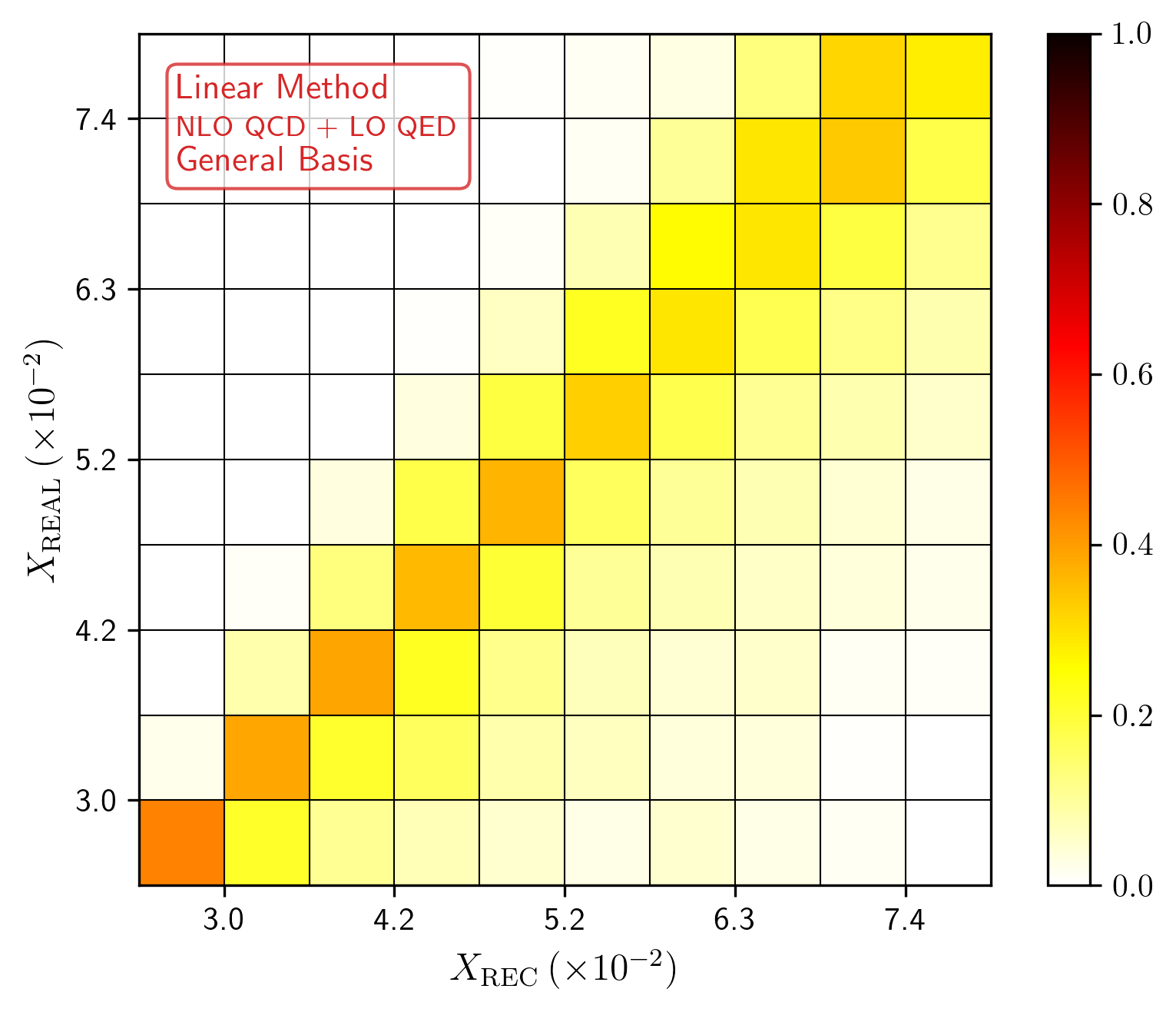}}
    \subfigure{\includegraphics[width=75mm,height=5.9cm]{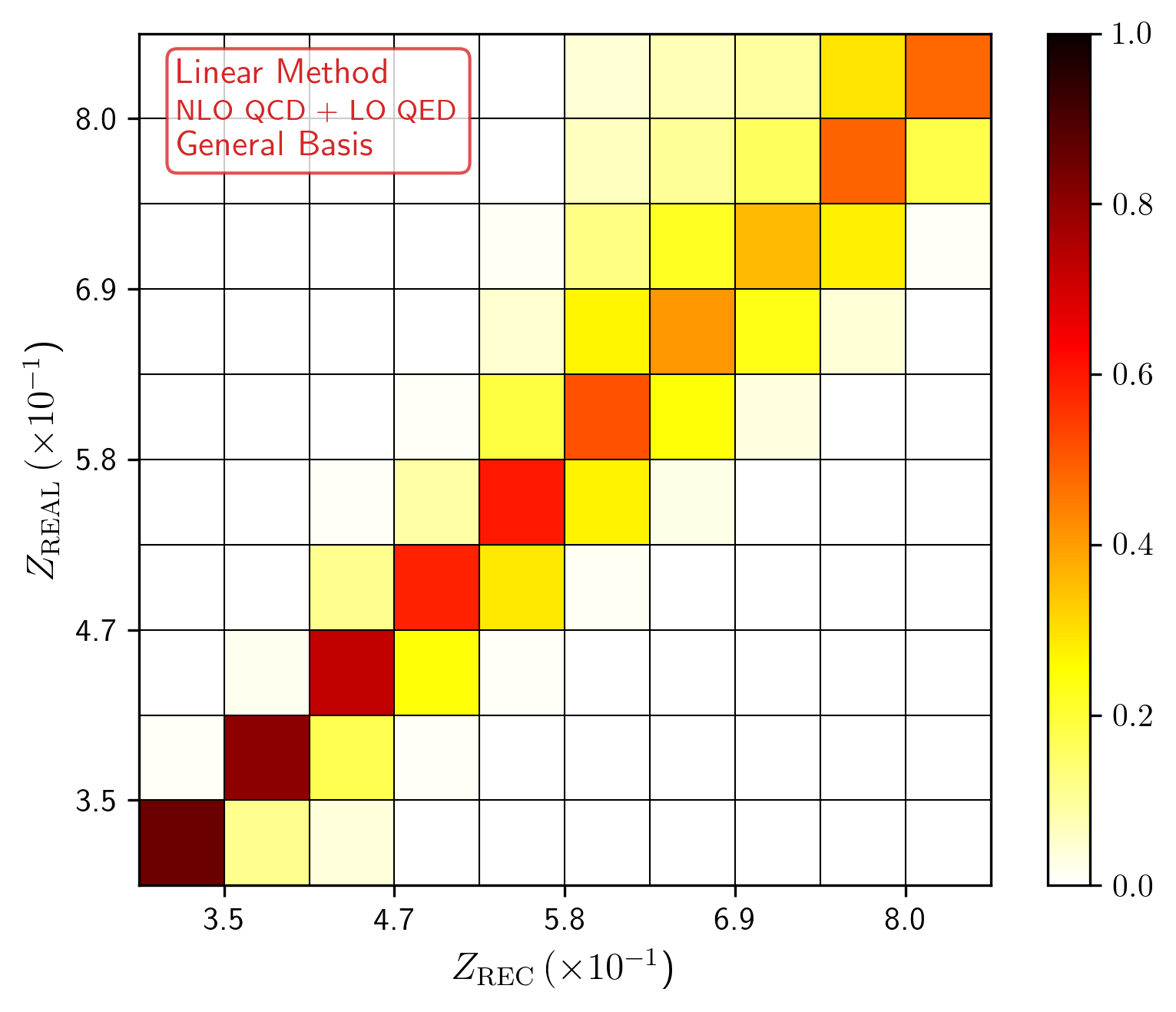}}
    \subfigure{\includegraphics[width=75mm,height=5.9cm]{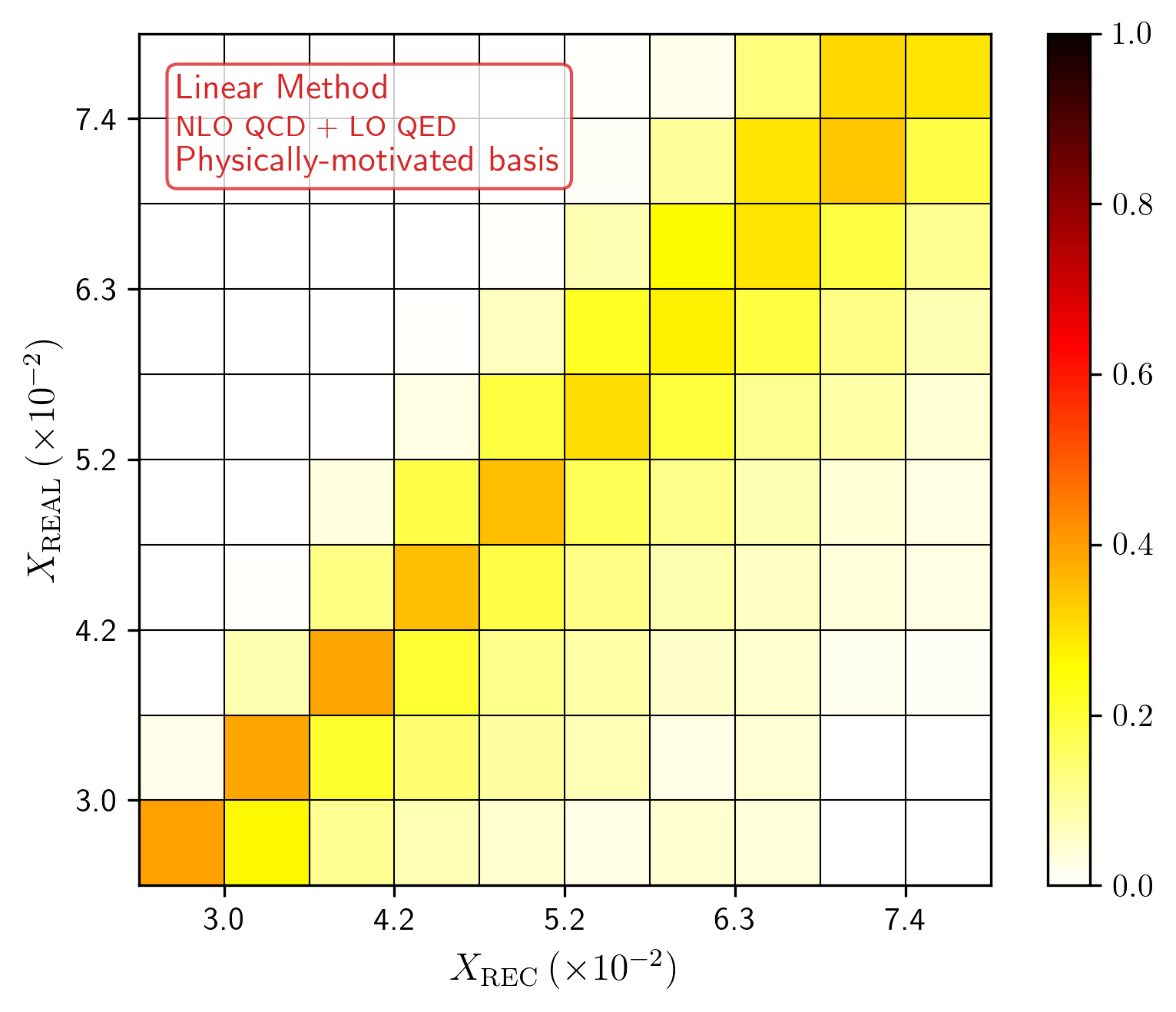}}
    \subfigure{\includegraphics[width=75mm,height=5.9cm]{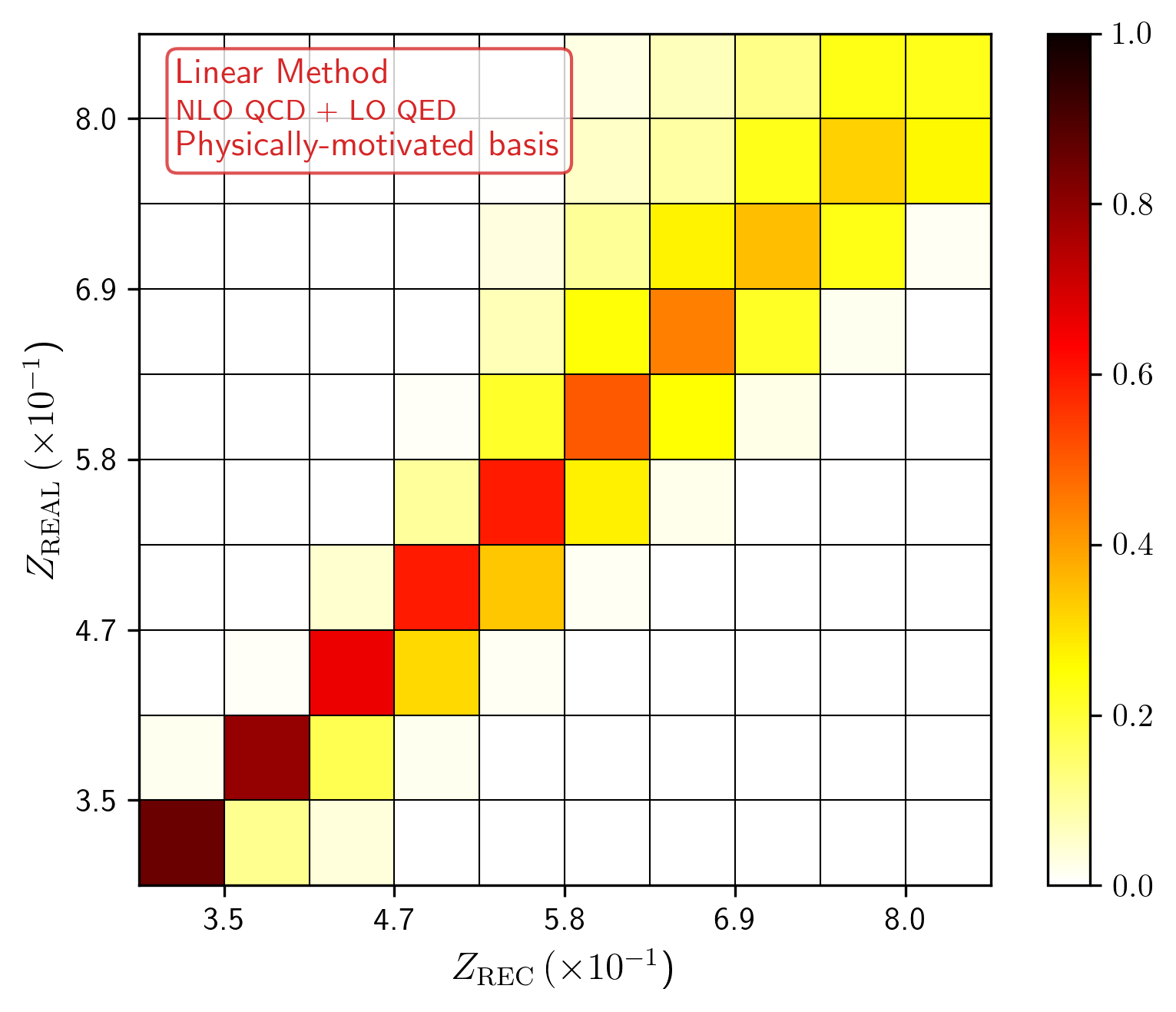}}    
    \subfigure{\includegraphics[width=75mm,height=5.9cm]{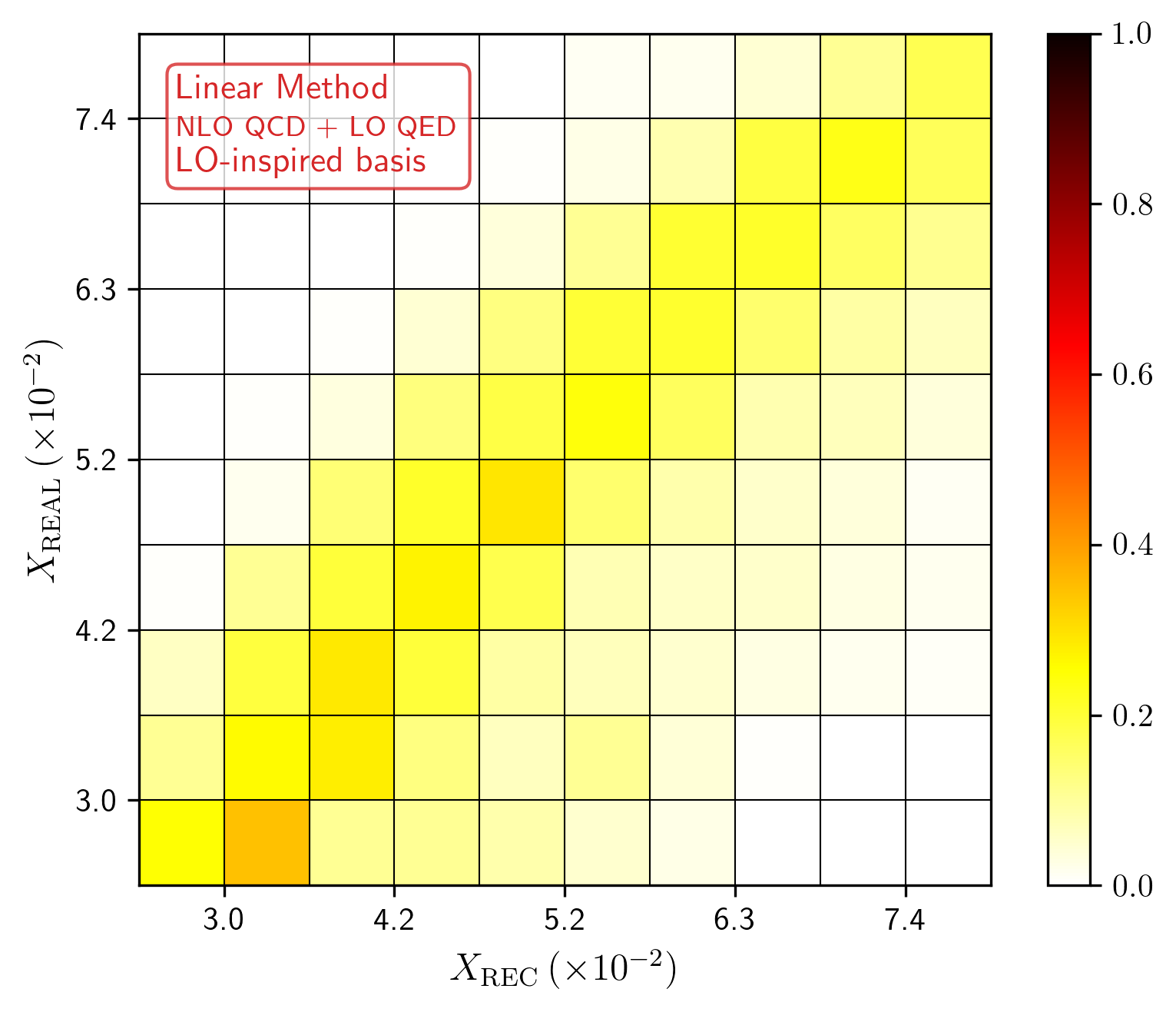}}
    \subfigure{\includegraphics[width=75mm,height=5.9cm]{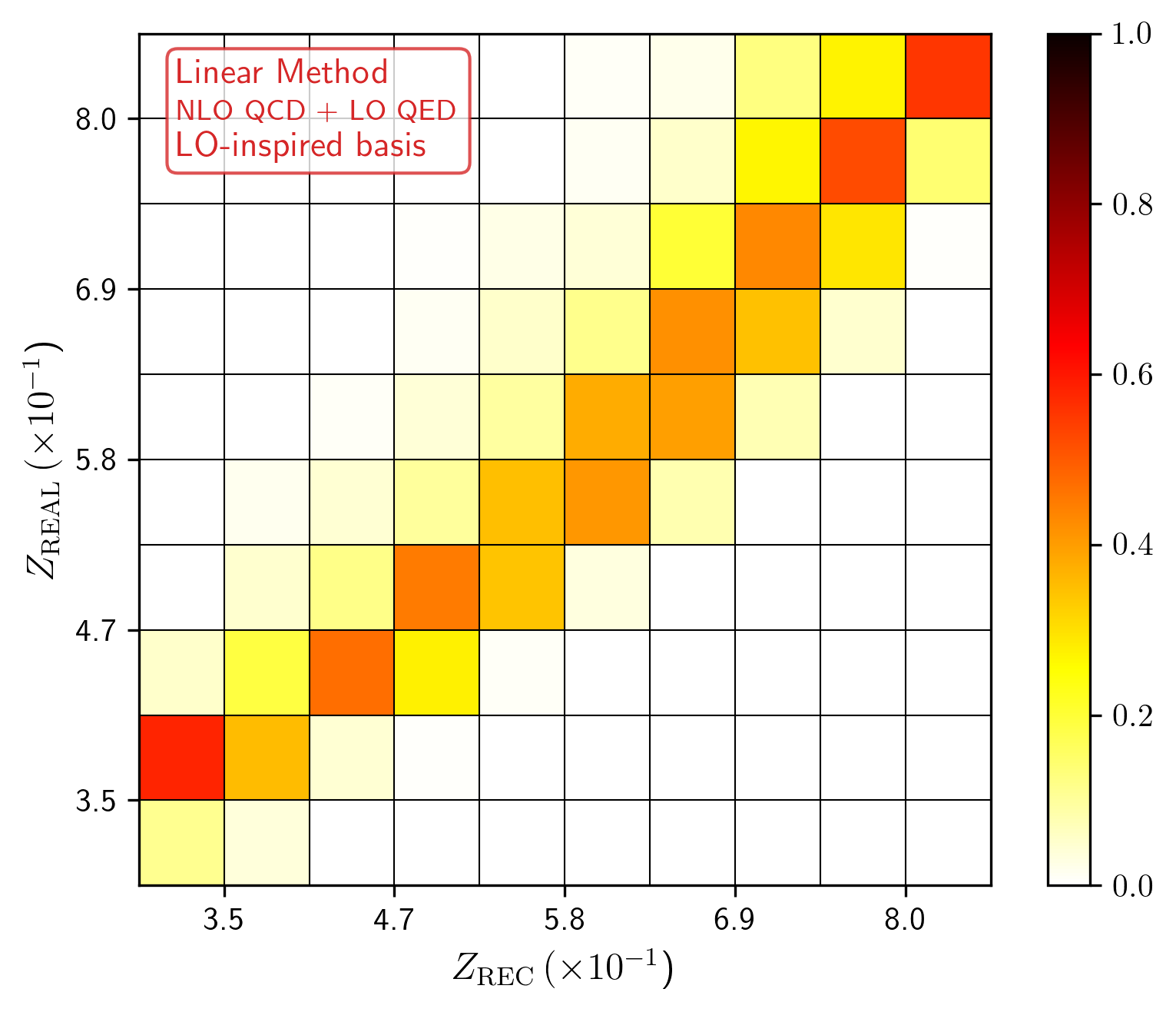}}
    \caption{Correlation between the MC momentum fractions (i.e. $X_{\rm REAL}$ and $Z_{\rm REAL}$) versus the ones obtained at NLO QCD + LO QED accuracy using the LM approach ($X_{\rm REC}$ and $Z_{\rm REC}$). Upper row: using the general basis given in Eq. (\ref{eq:LMgeneralNLO}). Middle row: \emph{physically motivated} basis. Lower row: \emph{LO-inspired} basis.}
    \label{fig:LMreconstructionNLO}
\end{figure} 

If we take Eq. (\ref{eq:LMgeneralNLO}), with $Y=\{x_{1},z\}$ we obtain the results shown in the upper row of Fig. \ref{fig:LMreconstructionNLO}. In this figure, we indicate the strength of the correlation with the integrated cross-section per bin at NLO QCD + LO QED accuracy. The coefficients $a^Y_i$, $b^Y_{ij}$, $c^Y_{ij}$ and $d^Y_{ij}$ are given in App. \ref{app:LM}. We can appreciate that the reconstruction is good in the low-$x$ and low-$z$ region. This is expected because the cross-section is larger in that region, so there are more data-points to perform the fit. However, the reconstruction becomes noisy and imprecise for higher values of the momentum fractions. The LM is unable to keep the functional dependencies that better approximate the real momentum fractions in regions with low number of events.

For this reason, we explore a second approach. We take profit from the findings in Sec. \ref{ssec:Correlation}, and distinguish different basis for $Y=x_1$ and $Y=z$. It was shown that $x_1$ exhibits a positive correlation with $p_T^\gamma$, so we remove the contributions involving ${\cal K}_6=(p_T^\gamma)^{-1}$ from Eq. (\ref{eq:LMgeneralNLO}). Regarding $z$, the conclusion of Sec. \ref{ssec:Correlation} was that it is correlated with ${\cal K}_6=(p_T^\gamma)^{-1}$, ${\cal K}_2=p_T^\pi$ and that also presents a mild correlation with ${\cal K}_5$. So, we remove the contributions that involve the primitive functions ${\cal K}_1$ and ${\cal K}_7$. As a result, we propose a \emph{physically-motivated} reconstruction by taking Eq. (\ref{eq:LMgeneralNLO}) and setting 
\beqn
\nn && b_6^{X_1} = 0  \, ,
\\ && c^{X_1}_{6,j} = d^{X_1}_{6,j} = c^{X_1}_{i,6} = d^{X_1}_{i,6} = 0 \quad  \{i,j\} \in \{1,\ldots,9\} \, ,
\label{eq:LMgeneralNLOX1prime}
\eeqn
for $x_1$ and
\beqn
\nn && b_1^{Z} = b_7^{Z} = 0  \, ,
\\ \nn && c^{Z}_{1,j} = d^{Z}_{1,j} = 0 \quad j \in \{1,\ldots,9\} \, , j \neq \{5,7\} \, ,
\\ && c^{Z}_{i,7} = d^{Z}_{i,7} = 0 \quad i \in \{1,\ldots,9\} \, , i \neq \{1,5\} \, ,
\label{eq:LMgeneralNLOZprime}
\eeqn
for $z$. The coefficients obtained with these assumptions are presented in App. \ref{app:LM}, whilst the corresponding correlations with the real MC momentum fractions are shown in the middle row of Fig. \ref{fig:LMreconstructionNLO}. We can appreciate that the correlation is slightly better for $z$, but it is worse for $x$. Even if the \emph{physically-motivated} basis includes elements that are selected according to the correlations with physical variables, it turns out that the abundance of points in a particular region of the parameter space imposes a very tight constraint in the whole fit. For $z$, it is not a big problem since it seems to be dominated by the ratio $p_T^\pi/p_T^\gamma$. However, the dependence of $x$ w.r.t. the kinematical variables is more complicated, and a linear fit is not enough to capture it. Thus, reducing the basis does not lead to an improved reconstruction of the momentum fractions.

To conclude this discussion, let us mention that we tested the LM with another basis inspired by the LO formulae. Namely, this \emph{LO-inspired} basis is given by
\beqn
\nn && {\cal B}^{X_1}_{\rm NLO}=\{ \frac{p_T^{\gamma}}{\sqrt{S_{CM}}}\exp(\eta^{\gamma}), \, \frac{p_T^{\gamma}}{\sqrt{S_{CM}}}\exp(\eta^{\pi}), \, \frac{p_T^{\pi}}{\sqrt{S_{CM}}}\exp(\eta^{\gamma}), \, \frac{p_T^{\pi}}{\sqrt{S_{CM}}}\exp(\eta^{\pi}), \,
\\  && \frac{p_T^{\gamma} {\cal K}_5 }{\sqrt{S_{CM}}}\exp(\eta^{\gamma}), \, \frac{p_T^{\gamma} {\cal K}_5}{\sqrt{S_{CM}}}\exp(\eta^{\pi}), \, \frac{p_T^{\pi} {\cal K}_5}{\sqrt{S_{CM}}}\exp(\eta^{\gamma}), \, \frac{p_T^{\pi} {\cal K}_5}{\sqrt{S_{CM}}}\exp(\eta^{\pi}) \, \} \, ,
\label{eq:BaseNLOXfisica}
\eeqn
for $x \equiv x_1$ and
\beq
{\cal B}^Z_{\rm NLO}=\{ p_T^{\pi}/p_T^{\gamma}, \, {\cal K}_5 \, p_T^{\pi}/p_T^{\gamma} , \, {\cal K}_5 \, p_T^{\pi}/\sqrt{S_{CM}} , \, {\cal K}_5 \, \sqrt{S_{CM}}/p_T^{\gamma} \, \} \, ,
\label{eq:BaseNLOZfisica}
\eeq
for $z$. In this case, the reconstruction was even worse, as can be seen in the lower row of Fig. \ref{fig:LMreconstructionNLO}. In particular, $X_{1,\rm REC}$ seems to be uncorrelated with $X_{1,\rm REAL}$. So, we can appreciate that the approach followed in Ref. \cite{deFlorian:2010vy} was more efficient than the LM. In other words, forcing a linear combination that describes the LO kinematics and then using the same formulae for higher-orders, allows to achieve a more precise reconstruction. In the next subsections, we explore other methods that will lead to a better approximation of the MC momentum fractions in a more automatized way.


\subsection{Gaussian regression}
\label{ssec:GaussReg}
While the LM method provides a good description for the LO case, at NLO the result strongly depends on the variables used to feed the algorithm. As the larger basis seems to render a slightly better reconstruction, we could use this as a motivation to further expand our basis, e.g. by including higher-powers of its elements. However this relies on deciding \emph{i)} which appropriate combinations of ${\cal K}_i$ are needed, and \emph{ii)} to which power it would be convenient to go. The first point was addressed in Subsec. \ref{ssec:LinReg} by constructing several bases, with different degree of success. Regarding the second point, we could try with different powers of a given basis, but this would be a cumbersome task. A more general and computationally efficient approach can be implemented by using the kernel trick (see e.g. \cite{kernels:2008,Rasmussen:2006}). In this method, the feature vector in the calculation is replaced by writing everything in terms of a function (kernel) of the dot product of the elements of the training set. In particular we use the radial basis function (RBF), defined as
\begin{align}
    k(x_i,x_j) =\exp\left(-\frac{d(x_i,x_j)}{2l^2}\right) \, ,
    \label{eq:GaussKernel}
\end{align}
where $x_i$, $x_j$ are two elements of the training set, $d(x_i,x_j)$ is the Euclidean distance between them, and $l$ is a distance parameter (not necessarily the same for all $\{i,j\}$). The RBF has the advantage of including all possible powers of the exponent, and therefore we expect a better reconstruction of the kinematic variables. 

\begin{figure}[htb]
    \centering
    \subfigure{\includegraphics[width=75mm,height=5.9cm]{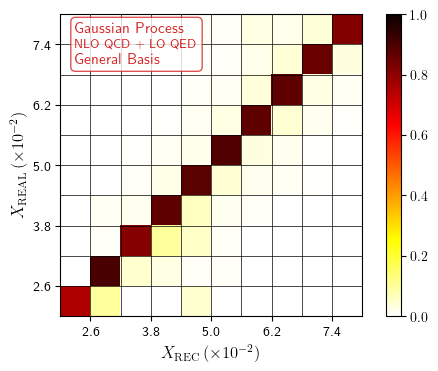}}
    \subfigure{\includegraphics[width=75mm,height=5.9cm]{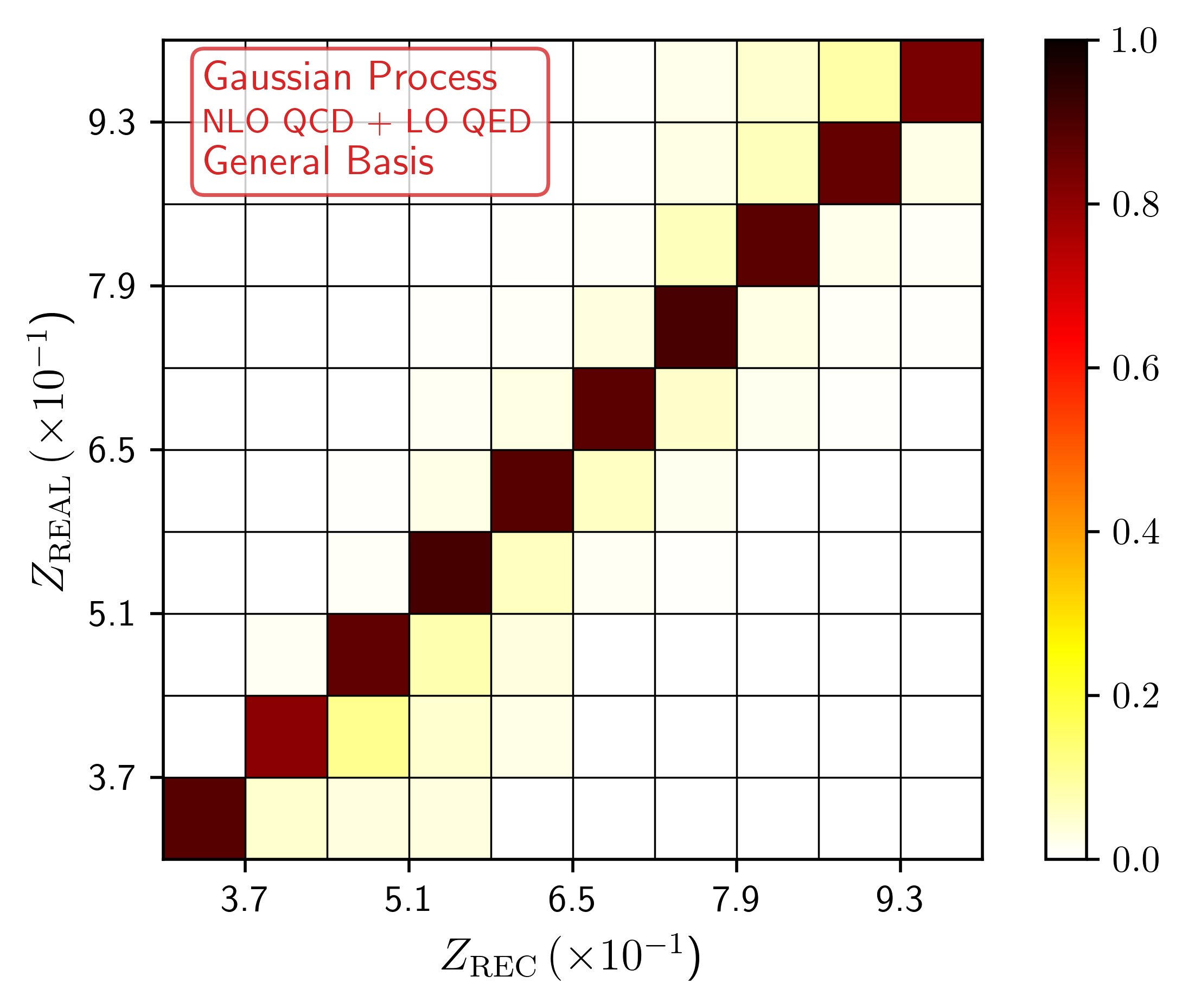}}
    \subfigure{\includegraphics[width=75mm,height=5.9cm]{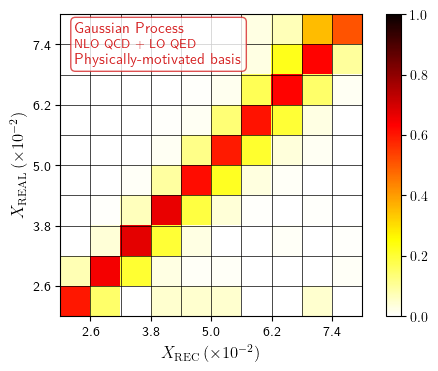}}
    \subfigure{\includegraphics[width=75mm,height=5.9cm]{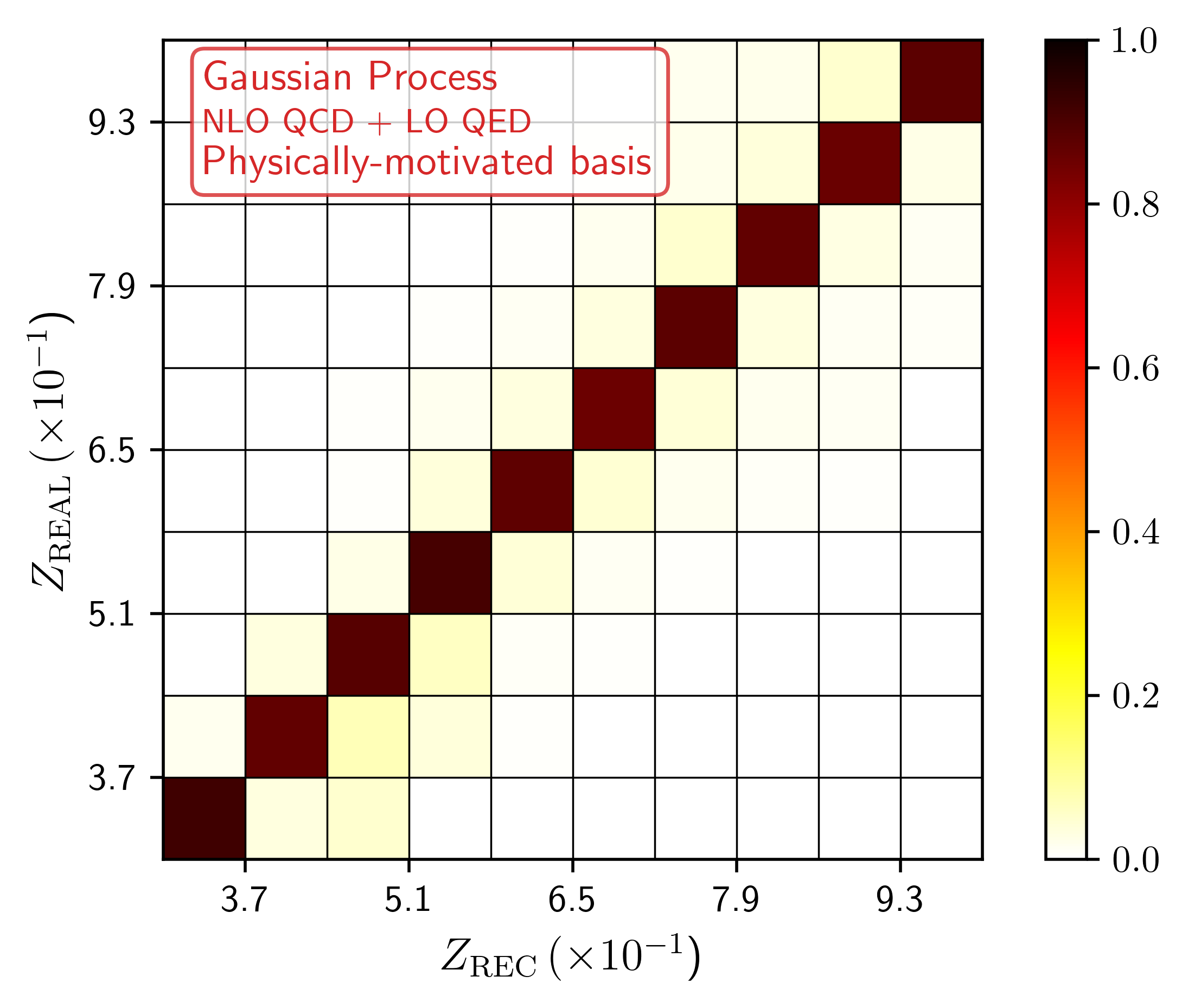}}    
    \subfigure{\includegraphics[width=75mm,height=5.9cm]{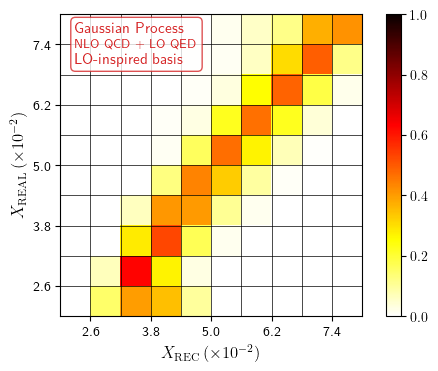}}
    \subfigure{\includegraphics[width=75mm,height=5.9cm]{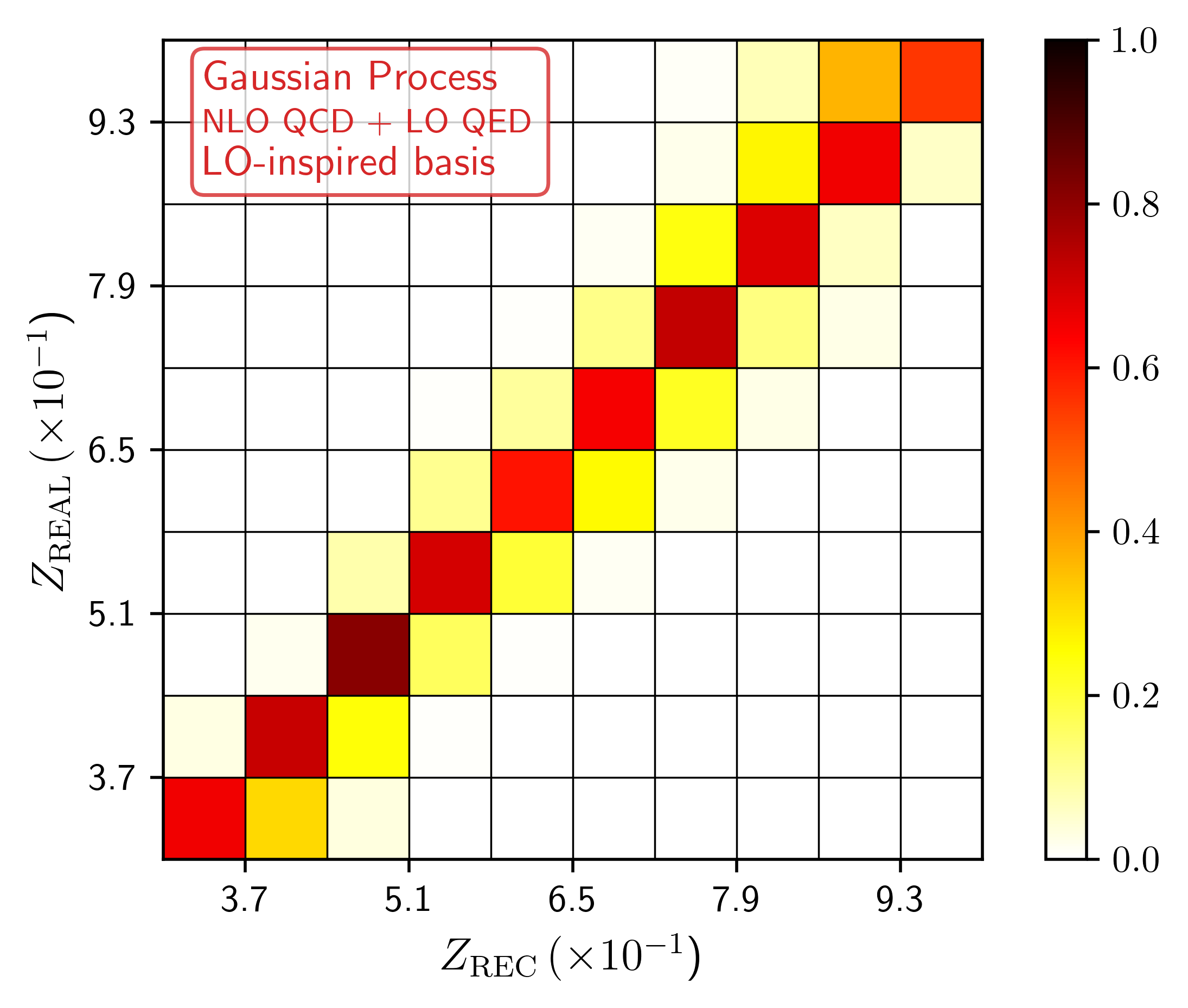}}     
    \caption{Correlation between the MC momentum fractions (i.e. $X_{\rm REAL}$ and $Z_{\rm REAL}$) versus the ones obtained at NLO QCD + LO QED accuracy ($X_{\rm REC}$ and $Z_{\rm REC}$). We show the results corresponding to the GR approach, using the general basis (upper row), the \emph{physically-motivated} basis (middle row) and the \emph{LO-inspired} basis (lower row).}
    \label{fig:NLOGM}
\end{figure}

Similarly to the LM, the GR requires a set of input variables. In order to properly compare the methods, we take the same bases for both. The GR also needs the user to select the {\it width} of each Gaussian function, $l$, which is by default $l=1$. In principle it could be different for each feature of the input set, but for simplicity we keep it feature-independent. However we did find better reconstructions when using different $l$ for $x$ and $z$. The optimal values of $l$ for each basis can be found in Table \ref{tab:GMell}.

\begin{table}[h!]
    \centering
    \begin{tabular}{|c|c|c|c|}
    \hline
    \hline
       Reconstructed & General  & Physically-motivated & LO-inspired  \\
       quantity & basis &  basis & basis \\
       \hline
       $x$ & 26 & 1 & 30 \\
       $z$ & 21 & 1.5 & 25 \\
    \hline
    \hline
   \end{tabular}
    \caption{Values of the $l$-parameter to reconstruct the $x$ and $z$ momentum fractions in three different basis used within the GR framework.}
    \label{tab:GMell}
\end{table}

We find that, when using the most general basis, a better agreement between the reconstructed and the real data sets requires {\it broad} Gaussian functions. In addition, if we reduce the basis the GR tends to require {\it wider} Gaussian functions to achieve a good description of the data sets. Finally, we find that in the {\it physically-motivated} basis, the GR finds the best agreement by choosing $l=1$ for the prediction of $x$ and $l=1.5$ for $z$, i.e. {\it sharp} Gaussian functions are needed meaning that a combination of these variables is enough to reproduce the full data sets.

These facts can appreciated in Fig. \ref{fig:NLOGM} where we present the results obtained at NLO QCD + LO QED accuracy. As expected, the inclusion of higher-order terms (higher non-linearity) in the training set brings a significant improvement with respect to the LM, in particular for the reconstruction of $x$. In addition, we point out that among the three basis, in general, the reconstruction of $x$ is harder than the $z$ momentum fraction. The general basis can extract the information to almost determine completely a function for the prediction of the momentum fractions but with wide Gaussian functions. In contrast, the \emph{physically-motivated} basis makes a good job in the determination of $z$ but is not that accurate on the extraction of $x$, although it requires {\it sharp} Gaussian functions, meaning that they are well localized and determined. 

To conclude this section, we appreciate that the GR method leads to a more reliable reconstruction of the MC momentum fractions, compared to the LM. The best results are obtained with a larger basis, in order to have more flexibility. Moreover, the non-linearity inherent to the GR allows to overcome the limitation of the overfitting in the low-$x$ and low-$x$ region that we observed in the LM, leading to a very accurate reconstruction in a wider range.


\subsection{Neural Networks}
\label{ssec:NeuNet}
Before jumping into the results of this section, let us briefly remind the reader of what is a neural network (NN). The building blocks of a NN are algorithms (called \emph{Perceptrons}) used in supervised learning to decide if an input belongs into a class or not (binary classifier). They consist of a set of input values $X$, that will be linearly combined by weights ($W$) and independent terms $B$ (\emph{biases}), after which the sum will be transformed by the (usually non-linear) activation function $f$, giving an output $Y$: $Y=f(z)$ with $z=X*W+B$. Each Perceptron mimics a neuron, and a combination of them makes a NN. The standard nomenclature labels the inputs and outputs as input and output \emph{layers}, respectively. To increase the capabilities of the NN (and its complexity) one can add more neurons in between, organised in \emph{hidden} layers. The activation functions connecting one layer to the next do not need to be the same, neither the number of neurons in each hidden layer. The learning proceeds in two steps. First, the NN computes the output from the inputs (feed-forward). In a second step (back-propagation), it calculates the cost and then minimizes it. This can be implemented in different ways, one of the most popular being stochastic gradient descent\footnote{This procedure depends on the size of a parameter called the \emph{learning rate}, that also requires adjustment. For more details about the implementation of NN in \texttt{scikit-learn} and specifics of the MPL algorithm we refer the reader to Ref. \cite{scikit-learn}.}.  

\begin{table}[h!]
    \centering
    \begin{tabular}{|l|c|c|c|c|}
    \hline
    \hline
    &\small{$X_{REC}$ (LO)} & \small{$Z_{REC}$ (LO)} & \small{$X_{REC}$ (NLO)} & \small{$Z_{REC}$ (NLO)}\\
    \hline
    $\#$ of hidden layers & \small{2}& \small{1} &\small{5} & \small{5} \\    
    $\#$ of neurons/layer & \small{200}& \small{100} &\small{300} & \small{300}\\
    activation function & ReLU& ReLU& ReLU& ReLU\\
    $\#$ iterations & $1\times\,10^{5}$ & 200 & $1\times\,10^{12}$ & $1\times\,10^{12}$ \\
    learning rate & $1\times\,10^{-3}$& $1\times\,10^{-3}$& $1\times\,10^{-4}$& $1\times\,10^{-4}$\\ 
    \hline
    \hline
   \end{tabular}
    \caption{Architecture for the MLP best fit parameters for the reconstruction of the momentum fractions at LO in QCD: $X_{\rm REC}$(LO) and $Z_{\rm REC}$(LO) (second and third columns), and for the momentum fractions at NLO QCD + LO QED: $X_{\rm REC}$(NLO) and $Z_{\rm REC}$(NLO) (fourth and fifth columns).}
    \label{tab:paramMLP}
\end{table}

The choice of the activation function/s and relevant parameters is highly non-trivial, and trial-and-error was required to find a configuration that could reproduce the momentum fractions. A non-exhaustive comparison of different combinations is presented in App. \ref{app:NNarchitectu}, but here we limit ourselves to present the results corresponding to the parameters summarised in Table \ref{tab:paramMLP}, which are used within the \texttt{scikit-learn} framework. 

\begin{figure}[h]
    \centering
    \subfigure{\includegraphics[width=75mm,height=5.9cm]{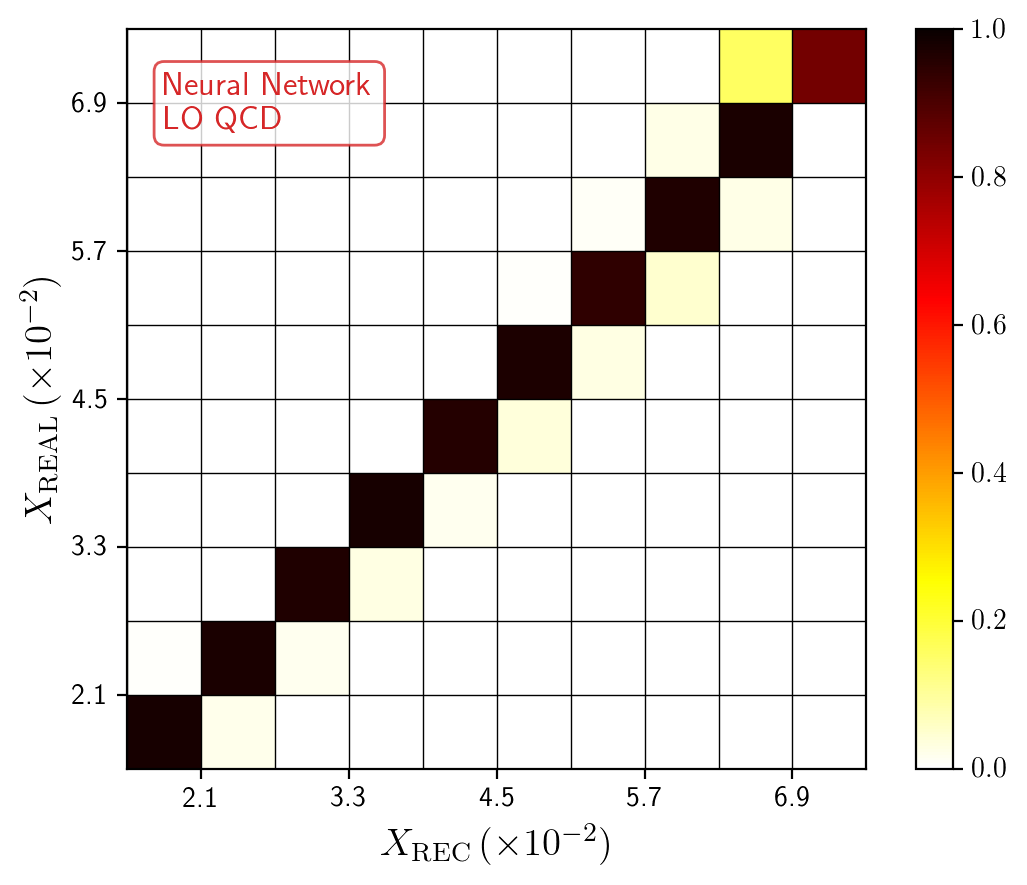}}
    \subfigure{\includegraphics[width=75mm,height=5.9cm]{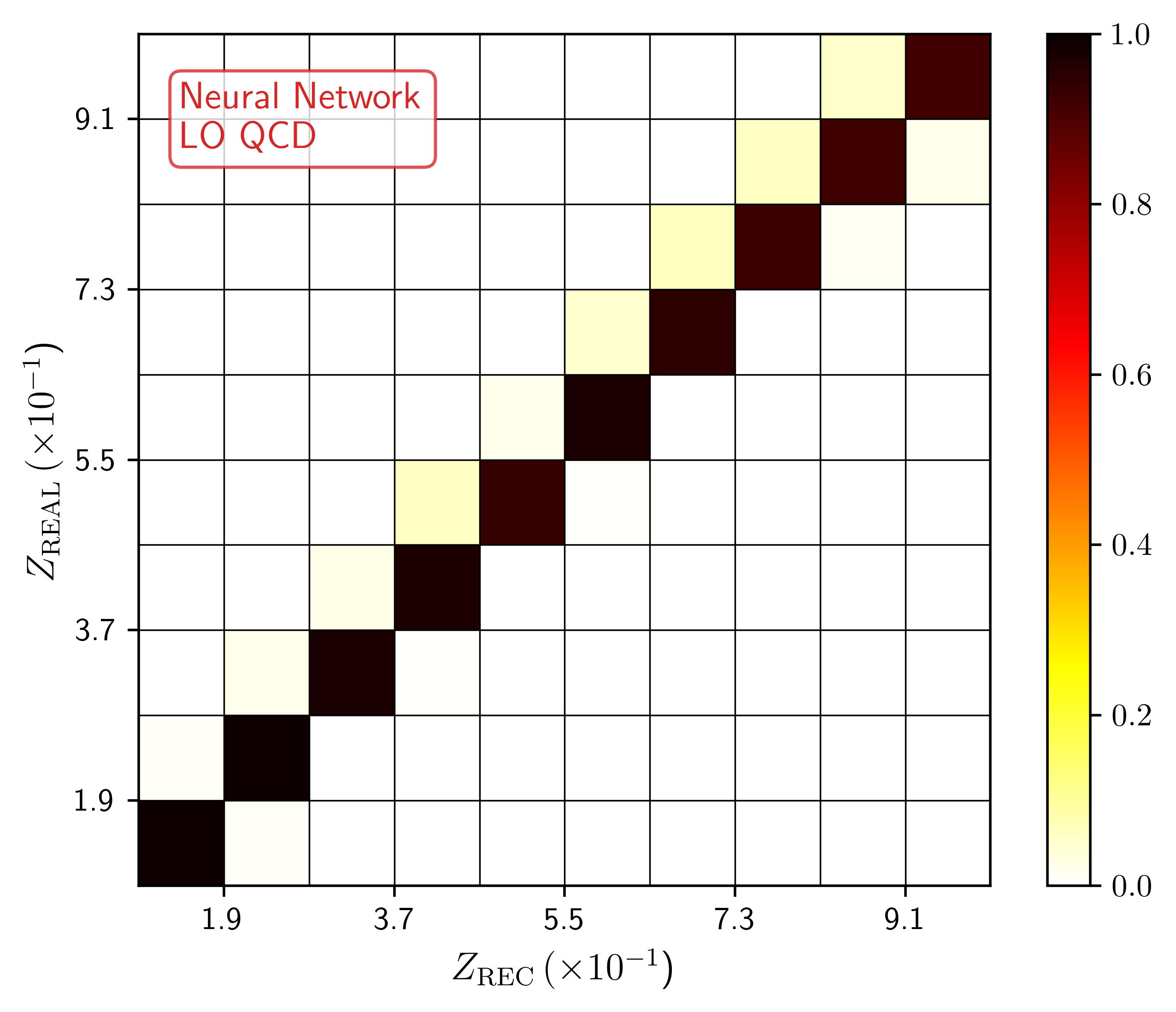}}
    \subfigure{\includegraphics[width=75mm,height=5.9cm]{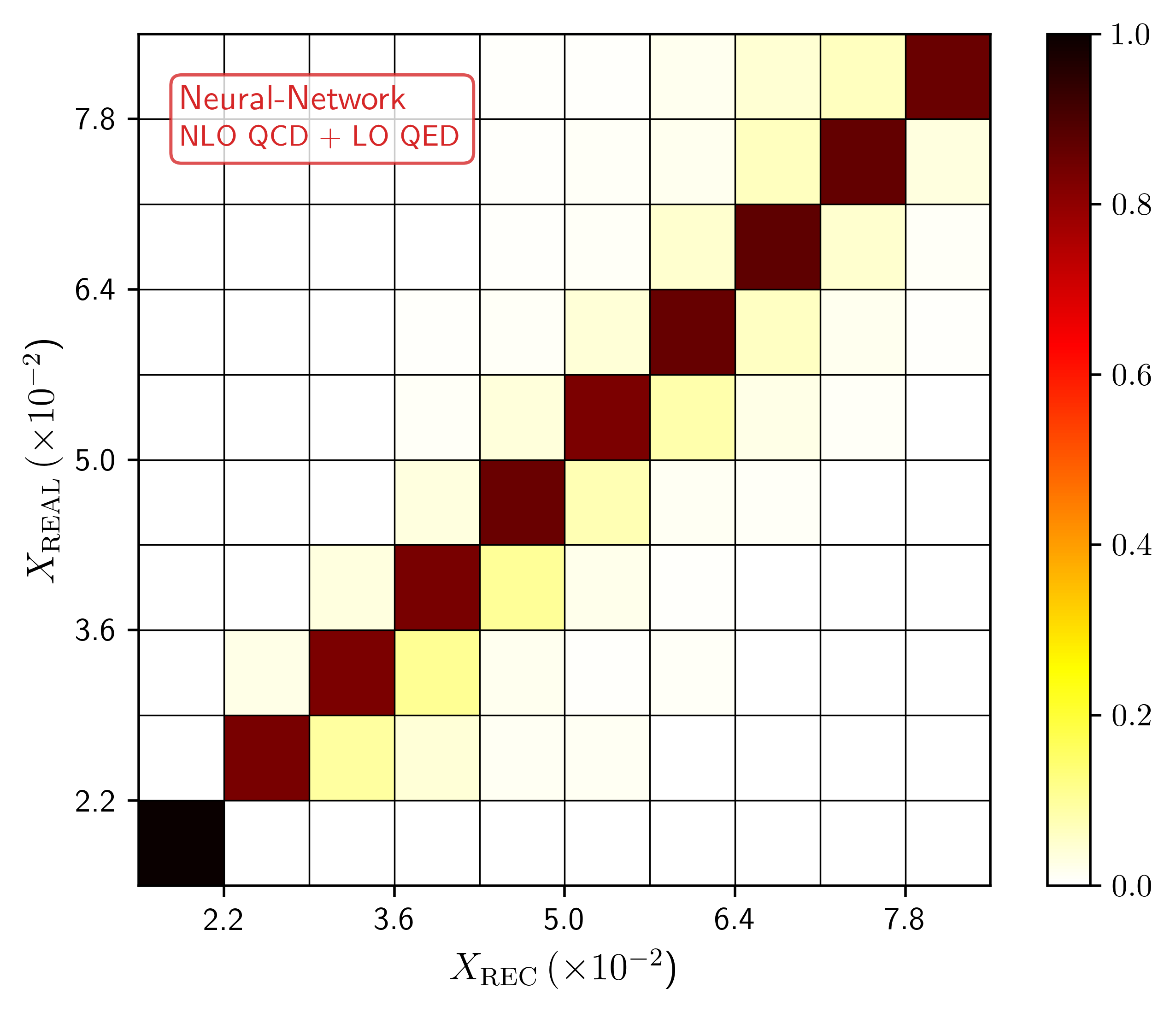}}
    \subfigure{\includegraphics[width=75mm,height=5.9cm]{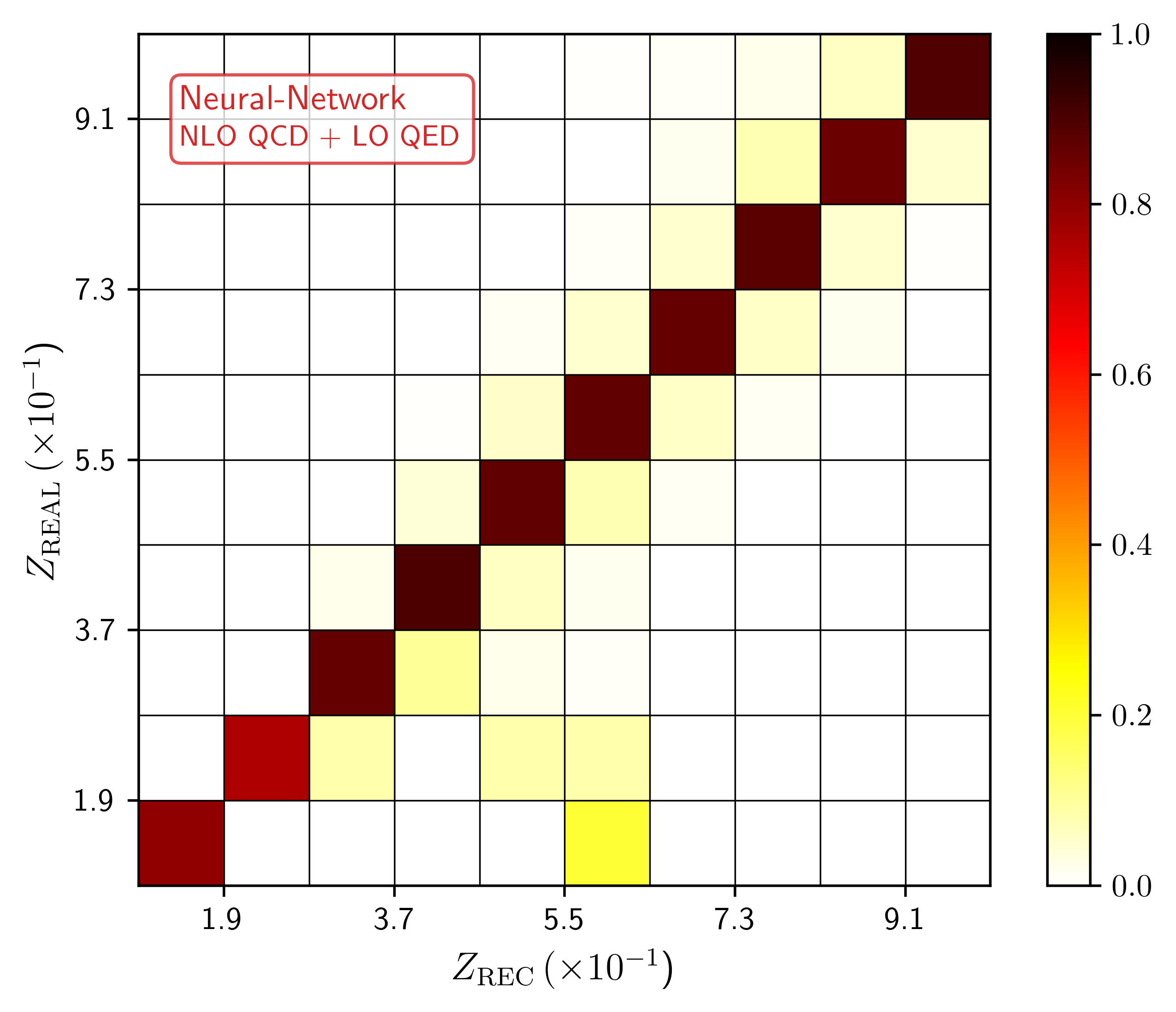}}
    \caption{Left: Comparison of the momentum fractions $X_{\rm REAL}$ and $X_{\rm REC}$ obtained with MPL neural networks with the parameters given in Table \ref{tab:paramMLP}. The upper (lower) row corresponds to the LO QCD (NLO QCD + LO QED) data set. Right: same as the r.h.s but for $Z_{\rm REAL}$ and $Z_{\rm REC}$.}
    \label{fig:LO+NLOMPL}
\end{figure}

The results of the MPL algorithm are presented in Fig.\ref{fig:LO+NLOMPL} for the LO QCD contribution (upper row) and the NLO QCD + LO QED correction (lower row). In the LO case the reconstruction is quite good, without reaching the level of accuracy of the LM or GR. This is a strong evidence that the complexity of the NN machinery greatly exceeds that of the task to be solved. In the NLO case, on the contrary, the reconstruction is much better than the one obtained with the LM using any basis, and similar to the GR one with the general basis (upper row of Fig. \ref{fig:NLOGM}). The plots show an almost perfect agreement in all bins for both $x$ and $z$. The largest discrepancy appears for $x$, which can be partially due to the higher complexity of the target function for $x$ than for $z$, already suggested by the analytic LO expressions. Indeed, almost all trials performed with different methods and configurations arrive to reasonable relations between $Z_{\rm REAL}$ and $Z_{\rm REC}$. However for $x$, we have to either increase the number of elements in our basis (GR) or the number of layers/nodes (NN).

In any case, we can highlight that the MPL algorithm does not require to choose any particular basis: the complexity is translated into defining the proper architecture. This task is more suitable for automation, thus more appropriate for tackling generic physical processes regardless of the number or kind of particles involved. Whereas LM or GR could take advantage from physically-motivated parameter's choice to speed-up an accurate reconstruction, the NN framework relies mainly on computational power to reduce the problem to a \emph{black-box} function. 


\section{Conclusions and outlook}
\label{sec:Conclusions}
In this work we have explored the reconstruction of the parton-level kinematics for the process $p+p \to \gamma + h$ using Machine-Learning (ML) tools. In first place, we implemented the calculation in a Monte-Carlo (MC) code with NLO QCD and LO QED accuracy. We relied on the FKS algorithm to cancel the infrared singularities, and the smooth cone isolation criteria to select those events with direct photons. This prescription is crucial to have access to cleaner information from the hard process.

Then, we studied different kinematical distributions with the purpose of identifying the regions with the largest number of events. After imposing selection cuts similar to those used by experimental collaborations, dynamical cuts were induced in the $x$ and $z$ distributions. These restrictions were taken into account when selecting events for analysing the correlations between experimentally-accessible quantities ($p_T$, $\eta$ and $\phi$ for the photon and pion) and the partonic momentum fraction. We realized that $x$ strongly depends on $p_T^\gamma$ (positive correlation) but not on the other variables, whilst $z$ exhibits a negative correlation with $p_T^\gamma$, a positive one with $p_T^\pi$ and a mild dependence with $\cos(\phi^\pi-\phi^\gamma)$. 

After that, we applied ML algorithms to reconstruct the partonic variables $x_1$, $x_2$ and $z$. We started by introducing a proper discretization of the multi-differential cross-section w.r.t. the set of variables $\{p_T^\pi,p_T^\gamma,\eta^\pi,\eta^\gamma,\cos(\phi^\pi-\phi^\gamma)\}$, in order to have a reliable estimation of the higher-order corrections in each bin. For these distributions, we generated the data sets and explored three different ML reconstruction strategies: linear methods (LM), Gaussian Regression (GR) and Multi-Layer Perceptron (MLP). For the first two approaches, we introduced three bases of functions inspired by the results obtained from the analysis of two-dimensional correlations in Sec. \ref{ssec:Correlation}. 
In all the cases, the reconstruction at LO QCD accuracy was very successful, and in agreement with the known analytical expressions. When dealing with the NLO QCD + LO QED corrections, the flexibility of the MLP approach leads to a very reliable reconstruction, achieving a better performance than the LM and comparable to the GR when using a sufficiently large basis. In particular, the LM results were highly-influenced by the abundance of data in the low-$x$ and low-$z$ region, leading to an unreliable fit when extrapolated outside these regions. 

It is worth appreciating that the number of assumptions related to the setup of the MLP framework is rather limited, compared to the ones done for linear and Gaussian regression. In particular, we want to highlight that there was no need to introduce an specific basis of functions, which makes this approach fully process-independent and suitable for other analysis.

In conclusion, the application of ML-inspired methods (and Neural Networks in particular) is suitable to unveil the partonic kinematics at hadron colliders, including also higher-order corrections. In this way, ML-assisted event reconstruction might allow to achieve a highly-precise description of the deepest constituents of matter and their interactions, complementing the current developments in other areas of theoretical particle physics.


\section*{Acknowledgements}
We would like to thank Germ\'an Rodrigo and Andreas Sch\"afer for fruitful comments about the manuscript. This research was partially supported by COST Action CA16201 (PARTICLEFACE) and MCIN/AEI/10.13039/501100011033, Grant No.
PID2020-114473GB-I00. The work of D. F. R.-E. and R. J. H.-P. is supported by CONACyT (M\'exico) through the Project No. A1- S-33202 (Ciencia Basica) and Ciencia de Frontera 2021-2042. Besides, R. J. H.-P. is also funded by Sistema Nacional de Investigadores from CONACyT and PROFAPI 2022 (Universidad Aut\'onoma de Sinaloa). P.Z. acknowledges support from the Deutsche Forschungs-gemeinschaft (DFG, German Research Foundation) - Research Unit FOR 2926, grant number 409651613.

\appendix

\section{Coefficients for the Linear Method}
\label{app:LM}
For completeness, we present the coefficients associated to the linear regression for each of the three bases studied in Subsec. \ref{ssec:LinReg}. We restrict our attention to the fit of the data sets at NLO QCD + LO QED accuracy, since the LO contributions were perfectly in agreement with the analytical LO formulae. In Tab. \ref{tab:coefBASEcompleta} we present the coefficients of the most general basis, Eq. (\ref{eq:LMgeneralNLO}), that reproduce the plots in the upper row of Fig. \ref{fig:LMreconstructionNLO}. The parameters of the \emph{physically-motivated} basis, given by Eq. (\ref{eq:LMgeneralNLO}) with the constraints of Eqs. (\ref{eq:LMgeneralNLOX1prime})-(\ref{eq:LMgeneralNLOZprime}), are in Tabs. \ref{tab:coefBASEXred} and \ref{tab:coefBASEZred} for $x$ and $z$, respectively. The corresponding correlation with the real MC variables can be seen in the middle row of Fig. \ref{fig:LMreconstructionNLO}. Finally, the coefficients for the \emph{LO-inspired} basis, associated to the constraints in Eqs. (\ref{eq:BaseNLOXfisica})-(\ref{eq:BaseNLOZfisica}), can be found in Tab. \ref{tab:coefBASEphys}. These fall short in the quality of the fit, as we can appreciate from the lower row of Fig. \ref{fig:LMreconstructionNLO}.

\section{Comparison of different NN architectures}
\label{app:NNarchitectu}
We summarize here some results that were obtained before the {\it optimal} architecture described in Subsec. \ref{ssec:NeuNet} was found. In Tab. \ref{tab:NNTest} we present the parameters corresponding to three different tests implemented.

\begin{table}[h!]
    \centering
    \begin{tabular}{|l|c|c|c|}
    \hline
    \hline
    Parameters & TEST 1 & TEST 2 & TEST 3 \\
    \hline
    $\#$ hidden layers & \small{2}& \small{4} &\small{3}  \\    
    $\#$ neurons/layer & \small{50}& \small{100} &\small{100}  \\    
    tolerance & \small{$ 10^{-2}$}& \small{$ 10^{-2}$} &\small{$ 10^{-3}$} \\
    max. number of iterations & $10^{8}$ & $ 10^{8}$ & $10^{9}$  \\
    $\#$ iterations w/o change & $14,000$& $21,000$& $100,000$\\ 
    \hline
    \hline
   \end{tabular}
    \caption{Architectures for the MLP of three different tests for the reconstruction of the momentum fractions at NLO in QCD. All parameters are taken to be the same for $X_{\rm REC}$ and $Z_{\rm REC}$.}
    \label{tab:NNTest}
\end{table}

\begin{figure}
    \centering
    \subfigure{\includegraphics[width=75mm,height=5.8cm]{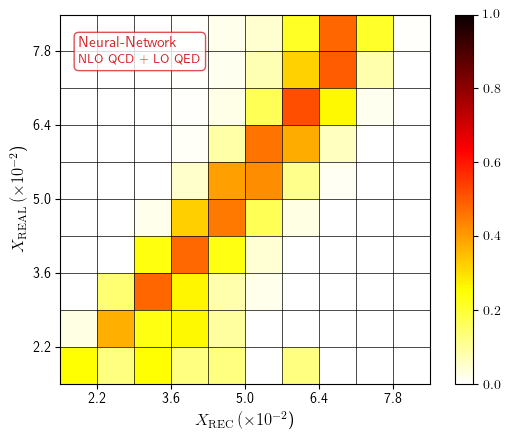}}
    \subfigure{\includegraphics[width=75mm,height=5.8cm]{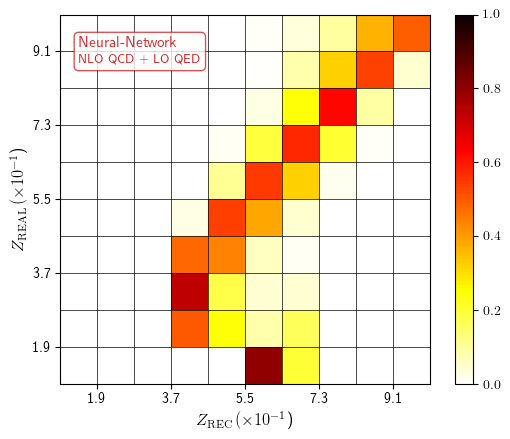}}
    \subfigure{\includegraphics[width=75mm,height=5.8cm]{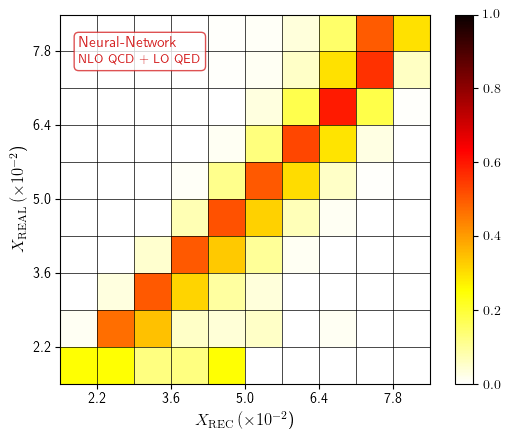}}
    \subfigure{\includegraphics[width=75mm,height=5.8cm]{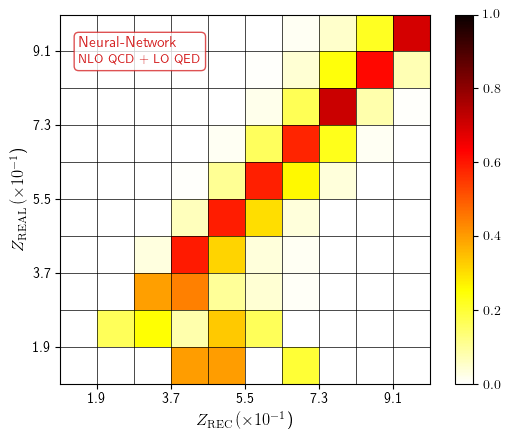}}    
    \subfigure{\includegraphics[width=75mm,height=5.8cm]{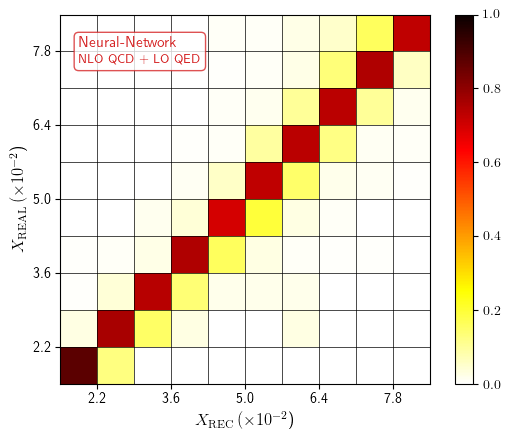}}
    \subfigure{\includegraphics[width=75mm,height=5.8cm]{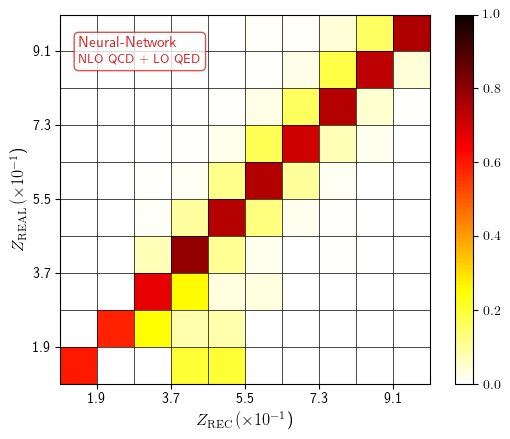}}    
   \caption{Comparison of the momentum fractions $X_{\rm REAL}$ vs. $X_{\rm REC}$ (left) and $Z_{\rm REAL}$ vs. $Z_{\rm REC}$ (right) obtained with MPL at NLO QCD + LO QED accuracy. The parameters for TEST1 (upper row), TEST2 (middle row) and TEST3 (lower row) are given in Tab. \ref{tab:NNTest}.}
    \label{fig:Tests}
\end{figure}

In TEST1 (upper row of Fig. \ref{fig:Tests}), we use a lower number of neurons/layer and less layers than for obtaining the results in Fig. \ref{fig:LO+NLOMPL}. We find a poor agreement between the real and reconstructed quantities, in particular for low-$z$ bins. An improvement is achieved by increasing the number of layers and neurons/layer (TEST2), while simultaneously requiring the NN to see no variation of the cost function (within a given tolerance) through a larger number of iterations. As seen in Fig. \ref{fig:Tests} (middle row), this gives a better reconstruction, thought it is still far from ideal. A third example, TEST3, reinforces the conditions for convergence and returns a significantly improved result (lower row of Fig. \ref{fig:Tests}). Each step towards a more complex architecture and more stringent requirements for convergence is translated into an increase of the computational time required for the training. These, and other trials, have guided us to the selection of the \emph{best} architecture for our task, summarised in Tab. \ref{tab:paramMLP}.

\begin{table}[t!]
    \centering
    \begin{tabular}{|ccc|ccc|}
    \hline
    \hline
    Coefficient &\small{$X_{REC}$ (NLO)} & \small{$Z_{REC}$ (NLO)} & Coefficient &\small{$X_{REC}$ (NLO)} & \small{$Z_{REC}$ (NLO)} \\
    \hline
$a^Y_1$ & $-5.7\times\,10^{1}$ & $-1.1\times\,10^{3}$ & $c^Y_{68}$ & $-5.7\times\,10^{1}$ & $7.3\times\,10^{-2}$ \\
$a^Y_2$ & $7.2\times\,10^{1}$ & $3.3\times\,10^{2}$ & $c^Y_{69}$ & $7.2\times\,10^{1}$ & $6.6\times\,10^{-2}$ \\
$a^Y_3$ & $5.4\times\,10^{0}$ & $-5.6\times\,10^{1}$ & $c^Y_{77}$ & $5.4\times\,10^{0}$ & $-1.9\times\,10^{-4}$ \\
$a^Y_4$ & $-2.4\times\,10^{0}$ & $2.7\times\,10^{0}$ & $c^Y_{78}$ & $-2.4\times\,10^{0}$ & $3.6\times\,10^{-2}$ \\
$a^Y_5$ & $-4.9\times\,10^{0}$ & $8.0\times\,10^{1}$ & $c^Y_{79}$ & $-4.9\times\,10^{0}$ & $5.5\times\,10^{-3}$ \\
$a^Y_6$ & $-2.8\times\,10^{-2}$ & $-1.2\times\,10^{-1}$ & $c^Y_{88}$ & $-2.8\times\,10^{-2}$ & $1.1\times\,10^{1}$ \\
$a^Y_7$ & $3.8\times\,10^{-2}$ & $1.6\times\,10^{-2}$ & $c^Y_{89}$ & $3.8\times\,10^{-2}$ & $3.8\times\,10^{-2}$ \\
$a^Y_8$ & $5.2\times\,10^{0}$ & $-5.6\times\,10^{1}$ & $c^Y_{99}$ & $5.2\times\,10^{0}$ & $-2.5\times\,10^{0}$ \\
$a^Y_9$ & $-2.1\times\,10^{0}$ & $9.4\times\,10^{-1}$ & $d^Y_{11}$ & $-2.1\times\,10^{0}$ & $4.4\times\,10^{3}$ \\
$b^Y_1$ & $-6.8\times\,10^{1}$ & $-1.2\times\,10^{3}$ & $d^Y_{12}$ & $-6.8\times\,10^{1}$ & $-1.3\times\,10^{4}$ \\
$b^Y_2$ & $5.8\times\,10^{1}$ & $5.2\times\,10^{2}$ & $d^Y_{13}$ & $5.8\times\,10^{1}$ & $2.3\times\,10^{2}$ \\
$b^Y_3$ & $4.9\times\,10^{0}$ & $-5.6\times\,10^{1}$ & $d^Y_{14}$ & $4.9\times\,10^{0}$ & $3.8\times\,10^{2}$ \\
$b^Y_4$ & $-2.2\times\,10^{0}$ & $-1.6\times\,10^{-1}$ & $d^Y_{17}$ & $-2.2\times\,10^{0}$ & $1.7\times\,10^{0}$ \\
$b^Y_6$ & $-3.1\times\,10^{-2}$ & $-9.1\times\,10^{-2}$ & $d^Y_{18}$ & $-3.1\times\,10^{-2}$ & $2.6\times\,10^{2}$ \\
$b^Y_7$ & $3.5\times\,10^{-2}$ & $3.2\times\,10^{-2}$ & $d^Y_{19}$ & $3.5\times\,10^{-2}$ & $3.4\times\,10^{2}$ \\
$b^Y_8$ & $4.7\times\,10^{0}$ & $-5.7\times\,10^{1}$ & $d^Y_{22}$ & $4.7\times\,10^{0}$ & $-3.2\times\,10^{3}$ \\
$b^Y_9$ & $-1.9\times\,10^{0}$ & $-2.2\times\,10^{0}$ & $d^Y_{23}$ & $-1.9\times\,10^{0}$ & $1.4\times\,10^{2}$ \\
$c^Y_{11}$ & $-4.9\times\,10^{2}$ & $2.4\times\,10^{3}$ & $d^Y_{24}$ & $-4.9\times\,10^{2}$ & $1.3\times\,10^{1}$ \\
$c^Y_{12}$ & $1.9\times\,10^{3}$ & $-9.8\times\,10^{3}$ & $d^Y_{26}$ & $1.9\times\,10^{3}$ & $-5.5\times\,10^{0}$ \\
$c^Y_{13}$ & $1.7\times\,10^{0}$ & $2.4\times\,10^{2}$ & $d^Y_{28}$ & $1.7\times\,10^{0}$ & $1.5\times\,10^{2}$ \\
$c^Y_{14}$ & $6.2\times\,10^{0}$ & $3.6\times\,10^{2}$ & $d^Y_{29}$ & $6.2\times\,10^{0}$ & $7.9\times\,10^{0}$ \\
$c^Y_{17}$ & $1.8\times\,10^{-1}$ & $1.6\times\,10^{0}$ & $d^Y_{33}$ & $1.8\times\,10^{-1}$ & $1.1\times\,10^{1}$ \\
$c^Y_{18}$ & $1.4\times\,10^{-1}$ & $2.6\times\,10^{2}$ & $d^Y_{34}$ & $1.4\times\,10^{-1}$ & $-1.3\times\,10^{0}$ \\
$c^Y_{19}$ & $9.9\times\,10^{0}$ & $3.1\times\,10^{2}$ & $d^Y_{36}$ & $9.9\times\,10^{0}$ & $6.7\times\,10^{-2}$ \\
$c^Y_{22}$ & $-7.2\times\,10^{2}$ & $-3.0\times\,10^{3}$ & $d^Y_{37}$ & $-7.2\times\,10^{2}$ & $3.4\times\,10^{-2}$ \\
$c^Y_{23}$ & $-3.1\times\,10^{1}$ & $1.5\times\,10^{2}$ & $d^Y_{39}$ & $-3.1\times\,10^{1}$ & $-3.6\times\,10^{-1}$ \\
$c^Y_{24}$ & $-1.4\times\,10^{1}$ & $2.5\times\,10^{1}$ & $d^Y_{44}$ & $-1.4\times\,10^{1}$ & $-2.4\times\,10^{0}$ \\
$c^Y_{26}$ & $5.3\times\,10^{-1}$ & $-4.3\times\,10^{0}$ & $d^Y_{46}$ & $5.3\times\,10^{-1}$ & $8.1\times\,10^{-2}$ \\
$c^Y_{28}$ & $-2.5\times\,10^{1}$ & $1.5\times\,10^{2}$ & $d^Y_{47}$ & $-2.5\times\,10^{1}$ & $3.0\times\,10^{-3}$ \\
$c^Y_{29}$ & $-1.1\times\,10^{1}$ & $1.9\times\,10^{1}$ & $d^Y_{48}$ & $-1.1\times\,10^{1}$ & $-1.3\times\,10^{0}$ \\
$c^Y_{33}$ & $-1.2\times\,10^{0}$ & $1.0\times\,10^{1}$ & $d^Y_{66}$ & $-1.2\times\,10^{0}$ & $-2.1\times\,10^{-4}$ \\
$c^Y_{34}$ & $3.6\times\,10^{-1}$ & $-7.7\times\,10^{-1}$ & $d^Y_{67}$ & $3.6\times\,10^{-1}$ & $-1.0\times\,10^{-3}$ \\
$c^Y_{36}$ & $-6.6\times\,10^{-4}$ & $6.9\times\,10^{-2}$ & $d^Y_{68}$ & $-6.6\times\,10^{-4}$ & $7.2\times\,10^{-2}$ \\
$c^Y_{37}$ & $-9.4\times\,10^{-3}$ & $3.5\times\,10^{-2}$ & $d^Y_{69}$ & $-9.4\times\,10^{-3}$ & $7.0\times\,10^{-2}$ \\
$c^Y_{39}$ & $4.8\times\,10^{-1}$ & $5.3\times\,10^{-2}$ & $d^Y_{77}$ & $4.8\times\,10^{-1}$ & $-1.3\times\,10^{-4}$ \\
$c^Y_{44}$ & $5.6\times\,10^{-1}$ & $-3.3\times\,10^{0}$ & $d^Y_{78}$ & $5.6\times\,10^{-1}$ & $3.6\times\,10^{-2}$ \\
$c^Y_{46}$ & $2.8\times\,10^{-3}$ & $7.6\times\,10^{-2}$ & $d^Y_{79}$ & $2.8\times\,10^{-3}$ & $1.2\times\,10^{-3}$ \\
$c^Y_{47}$ & $-6.3\times\,10^{-3}$ & $7.8\times\,10^{-3}$ & $d^Y_{88}$ & $-6.3\times\,10^{-3}$ & $1.1\times\,10^{1}$ \\
$c^Y_{48}$ & $4.4\times\,10^{-1}$ & $-8.0\times\,10^{-1}$ & $d^Y_{89}$ & $4.4\times\,10^{-1}$ & $-3.8\times\,10^{-1}$ \\
$c^Y_{66}$ & $2.2\times\,10^{-5}$ & $-2.2\times\,10^{-4}$ & $d^Y_{99}$ & $2.2\times\,10^{-5}$ & $-1.6\times\,10^{0}$ \\
$c^Y_{67}$ & $1.4\times\,10^{-4}$ & $-7.9\times\,10^{-4}$ &  &  & \\
    \hline
    \hline
   \end{tabular}
    \caption{Coefficients for the LM with the general basis expressed in Eq. (\ref{eq:LMgeneralNLO}) for both $x$ and $z$ momentum fractions.}
    \label{tab:coefBASEcompleta}
\end{table}

\begin{table}[h!]
    \centering
    \begin{tabular}{|cccc|}
    \hline
    \hline
    Coefficient &\small{$X_{REC}$ (NLO)} & Coefficient &\small{$X_{REC}$ (NLO)}  \\
    \hline
    $a^Y_1$ & $5.5\times\,10^{1}$ & $c^Y_{48}$ & $4.2\times\,10^{-1}$ \\
    $a^Y_2$ & $1.4\times\,10^{2}$ & $c^Y_{77}$ & $-1.0\times\,10^{-4}$ \\
    $a^Y_3$ & $5.4\times\,10^{0}$ & $c^Y_{78}$ & $-8.0\times\,10^{-3}$ \\
    $a^Y_4$ & $-2.3\times\,10^{0}$ & $c^Y_{79}$ & $-5.4\times\,10^{-3}$ \\
    $a^Y_5$ & $-8.4\times\,10^{0}$ & $c^Y_{88}$ & $-1.3\times\,10^{0}$ \\
    $a^Y_7$ & $5.6\times\,10^{-2}$ & $c^Y_{89}$ & $5.3\times\,10^{-1}$ \\
    $a^Y_8$ & $5.2\times\,10^{0}$ & $c^Y_{99}$ & $2.5\times\,10^{-1}$ \\
    $a^Y_9$ & $-1.8\times\,10^{0}$ & $d^Y_{11}$ & $-4.1\times\,10^{2}$ \\
    $b^Y_1$ & $6.3\times\,10^{1}$ & $d^Y_{12}$ & $-6.4\times\,10^{2}$ \\
    $b^Y_2$ & $1.4\times\,10^{2}$ & $d^Y_{13}$ & $3.9\times\,10^{0}$ \\
    $b^Y_3$ & $4.9\times\,10^{0}$ & $d^Y_{14}$ & $-7.4\times\,10^{0}$ \\
    $b^Y_4$ & $-2.1\times\,10^{0}$ & $d^Y_{17}$ & $-5.6\times\,10^{-1}$ \\
    $b^Y_7$ & $5.8\times\,10^{-2}$ & $d^Y_{18}$ & $2.5\times\,10^{0}$ \\
    $b^Y_8$ & $4.7\times\,10^{0}$ & $d^Y_{19}$ & $-8.0\times\,10^{0}$ \\
    $b^Y_9$ & $-1.6\times\,10^{0}$ & $d^Y_{22}$ & $-6.5\times\,10^{2}$ \\
    $c^Y_{11}$ & $-3.2\times\,10^{2}$ & $d^Y_{23}$ & $-3.2\times\,10^{1}$ \\
    $c^Y_{12}$ & $-6.0\times\,10^{2}$ & $d^Y_{24}$ & $-1.4\times\,10^{1}$ \\
    $c^Y_{13}$ & $4.1\times\,10^{0}$ & $d^Y_{28}$ & $-2.5\times\,10^{1}$ \\
    $c^Y_{14}$ & $-7.3\times\,10^{0}$ & $d^Y_{29}$ & $-1.0\times\,10^{1}$ \\
    $c^Y_{17}$ & $-4.8\times\,10^{-1}$ & $d^Y_{33}$ & $-1.1\times\,10^{0}$ \\
    $c^Y_{18}$ & $2.6\times\,10^{0}$ & $d^Y_{34}$ & $3.8\times\,10^{-1}$ \\
    $c^Y_{19}$ & $-7.8\times\,10^{0}$ & $d^Y_{37}$ & $-9.6\times\,10^{-3}$ \\
    $c^Y_{22}$ & $-6.3\times\,10^{2}$ & $d^Y_{39}$ & $5.1\times\,10^{-1}$ \\
    $c^Y_{23}$ & $-3.1\times\,10^{1}$ & $d^Y_{44}$ & $5.5\times\,10^{-1}$ \\
    $c^Y_{24}$ & $-1.4\times\,10^{1}$ & $d^Y_{47}$ & $-5.9\times\,10^{-3}$ \\
    $c^Y_{28}$ & $-2.5\times\,10^{1}$ & $d^Y_{48}$ & $4.6\times\,10^{-1}$ \\
    $c^Y_{29}$ & $-1.0\times\,10^{1}$ & $d^Y_{77}$ & $-1.1\times\,10^{-4}$ \\
    $c^Y_{33}$ & $-1.2\times\,10^{0}$ & $d^Y_{78}$ & $-8.1\times\,10^{-3}$ \\
    $c^Y_{34}$ & $3.3\times\,10^{-1}$ & $d^Y_{79}$ & $-5.3\times\,10^{-3}$ \\
    $c^Y_{37}$ & $-9.5\times\,10^{-3}$ & $d^Y_{88}$ & $-1.2\times\,10^{0}$ \\
    $c^Y_{39}$ & $4.6\times\,10^{-1}$ & $d^Y_{89}$ & $5.9\times\,10^{-1}$ \\
    $c^Y_{44}$ & $6.3\times\,10^{-1}$ & $d^Y_{99}$ & $1.7\times\,10^{-1}$ \\
    $c^Y_{47}$ & $-6.1\times\,10^{-3}$ &   &\\
    \hline
    \hline
   \end{tabular}
    \caption{Coefficients for the LM with the \emph{physically-motivated} basis expressed in Eq. (\ref{eq:LMgeneralNLO}) with the constraints given in Eq. (\ref{eq:LMgeneralNLOX1prime}) and Eq. (\ref{eq:LMgeneralNLOZprime}) for the $x$ momentum fraction.}
    \label{tab:coefBASEXred}
\end{table}

\begin{table}[h!]
    \centering
    \begin{tabular}{|cccc|}
    \hline
    \hline
    Coefficient &\small{$Z_{REC}$ (NLO)} & Coefficient &\small{$Z_{REC}$ (NLO)}  \\
    \hline
    $a^Y_2$ & $5.5\times\,10^{1}$ & $c^Y_{67}$ & $3.3\times\,10^{-1}$ \\
    $a^Y_3$ & $1.4\times\,10^{2}$ & $c^Y_{68}$ & $-9.5\times\,10^{-3}$ \\
    $a^Y_4$ & $5.4\times\,10^{0}$ & $c^Y_{69}$ & $4.6\times\,10^{-1}$ \\
    $a^Y_5$ & $-2.3\times\,10^{0}$ & $c^Y_{88}$ & $6.3\times\,10^{-1}$ \\
    $a^Y_6$ & $-8.4\times\,10^{0}$ & $c^Y_{89}$ & $-6.1\times\,10^{-3}$ \\
    $a^Y_8$ & $5.6\times\,10^{-2}$ & $c^Y_{99}$ & $4.2\times\,10^{-1}$ \\
    $a^Y_9$ & $5.2\times\,10^{0}$ & $d^Y_{11}$ & $-1.0\times\,10^{-4}$ \\
    $b^Y_2$ & $-1.8\times\,10^{0}$ & $d^Y_{22}$ & $-8.0\times\,10^{-3}$ \\
    $b^Y_3$ & $6.3\times\,10^{1}$ & $d^Y_{23}$ & $-5.4\times\,10^{-3}$ \\
    $b^Y_4$ & $1.4\times\,10^{2}$ & $d^Y_{24}$ & $-1.3\times\,10^{0}$ \\
    $b^Y_6$ & $4.9\times\,10^{0}$ & $d^Y_{26}$ & $5.3\times\,10^{-1}$ \\
    $b^Y_8$ & $-2.1\times\,10^{0}$ & $d^Y_{28}$ & $2.5\times\,10^{-1}$ \\
    $b^Y_9$ & $5.8\times\,10^{-2}$ & $d^Y_{29}$ & $-4.1\times\,10^{2}$ \\
    $c^Y_{11}$ & $4.7\times\,10^{0}$ & $d^Y_{33}$ & $-6.4\times\,10^{2}$ \\
    $c^Y_{22}$ & $-1.6\times\,10^{0}$ & $d^Y_{34}$ & $3.9\times\,10^{0}$ \\
    $c^Y_{23}$ & $-3.2\times\,10^{2}$ & $d^Y_{36}$ & $-7.4\times\,10^{0}$ \\
    $c^Y_{24}$ & $-6.0\times\,10^{2}$ & $d^Y_{39}$ & $-5.6\times\,10^{-1}$ \\
    $c^Y_{26}$ & $4.1\times\,10^{0}$ & $d^Y_{44}$ & $2.5\times\,10^{0}$ \\
    $c^Y_{28}$ & $-7.3\times\,10^{0}$ & $d^Y_{46}$ & $-8.0\times\,10^{0}$ \\
    $c^Y_{29}$ & $-4.8\times\,10^{-1}$ & $d^Y_{48}$ & $-6.5\times\,10^{2}$ \\
    $c^Y_{33}$ & $2.6\times\,10^{0}$ & $d^Y_{66}$ & $-3.2\times\,10^{1}$ \\
    $c^Y_{34}$ & $-7.8\times\,10^{0}$ & $d^Y_{67}$ & $-1.4\times\,10^{1}$ \\
    $c^Y_{36}$ & $-6.3\times\,10^{2}$ & $d^Y_{68}$ & $-2.5\times\,10^{1}$ \\
    $c^Y_{39}$ & $-3.1\times\,10^{1}$ & $d^Y_{69}$ & $-1.0\times\,10^{1}$ \\
    $c^Y_{44}$ & $-1.4\times\,10^{1}$ & $d^Y_{88}$ & $-1.1\times\,10^{0}$ \\
    $c^Y_{46}$ & $-2.5\times\,10^{1}$ & $d^Y_{89}$ & $3.8\times\,10^{-1}$ \\
    $c^Y_{48}$ & $-1.0\times\,10^{1}$ & $d^Y_{99}$ & $-9.6\times\,10^{-3}$ \\
    $c^Y_{66}$ & $-1.2\times\,10^{0}$ &   &   \\
    \hline
    \hline
   \end{tabular}
    \caption{Same as Tab. \ref{tab:coefBASEXred}, now for the $z$ momentum fraction.}
    \label{tab:coefBASEZred}
\end{table}

\begin{table}[h!]
    \centering
    \begin{tabular}{|cc|cc|}
    \hline
    \hline
    Coefficient &\small{$X_{REC}$ (NLO)} & Coefficient &\small{$Z_{REC}$ (NLO)}  \\
    \hline
    $c^Y_{13}$ & $3.8\times\,10^{0}$ & $c^Y_{26}$ & $3.8\times\,10^{0}$ \\
    $c^Y_{14}$ & $4.7\times\,10^{-1}$ & $d^Y_{26}$ & $4.7\times\,10^{-1}$\\
    $c^Y_{23}$ & $2.0\times\,10^{-1}$ & $b^Y_{6}$ & $2.0\times\,10^{-1}$\\
    $c^Y_{24}$ & $1.6\times\,10^{0}$ & $b^Y_{2}$ & $1.6\times\,10^{0}$ \\
    $d^Y_{13}$ & $3.6\times\,10^{0}$ &  &\\
    $d^Y_{14}$ & $1.7\times\,10^{-1}$ & &\\
    $d^Y_{23}$ & $-5.4\times\,10^{-1}$ & &\\
    $d^Y_{24}$ & $9.1\times\,10^{-1}$  &  &\\
    \hline
    \hline
   \end{tabular}
    \caption{Coefficients for the LM with the \emph{LO-inspired} basis expressed in Eqs. (\ref{eq:BaseNLOXfisica}) and (\ref{eq:BaseNLOZfisica}) for both $x$ and $z$ momentum fractions.}
    \label{tab:coefBASEphys}
\end{table}



\providecommand{\href}[2]{#2}\begingroup\raggedright\endgroup

\end{document}